\documentclass[12pt,preprint]{aastex}
\begin{document}

\title{Catalog of 93 Nova Light Curves: Classification and Properties}
\author{Richard J. Strope, Bradley E. Schaefer\affil{Physics and Astronomy, Louisiana State University, Baton Rouge, LA 70803}}
\author{Arne A. Henden\affil{American Association of Variable Star Observers, 49 Bay State Road, Cambridge MA 02138}}

\begin{abstract}

We present a catalog of 93 very-well-observed nova light curves.  The light curves were constructed from 229,796 individual measured magnitudes, with the median coverage extending to 8.0 mag below peak and 26\% of the light curves following the eruption all the way to quiescence.  Our time-binned light curves are presented in figures and as complete tabulations.  We also calculate and tabulate many properties about the light curves, including peak magnitudes and dates, times to decline by 2, 3, 6, and 9 magnitudes from maximum, the time until the brightness returns to quiescence, the quiescent magnitude, power law indices of the decline rates throughout the eruption, the break times in this decline, plus many more properties specific to each nova class.  We present a classification system for nova light curves based on the shape and the time to decline by 3 magnitudes from peak ($t_3$).  The designations are ÔSÕ for smooth light curves (38\% of the novae), ÔPÕ for plateaus (21\%), ÔDÕ for dust dips (18\%), ÔCÕ for cusp-shaped secondary maxima (1\%), ÔOÕ for quasi-sinusoidal oscillations superposed on an otherwise smooth decline (4\%), ÔFÕ for flat-topped light curves (2\%), and ÔJÕ for jitters or flares superposed on the decline (16\%).  Our classification consists of this single letter followed by the $t_3$ value in parentheses; so for example V1500 Cyg  is S(4), GK Per is O(13), DQ Her is D(100), and U Sco is P(3).

\end{abstract}

\section{Introduction}

	Nova eruptions (Payne-Gaposhkin 1964) occur on the surface of white dwarf stars in interacting binaries, generically called `cataclysmic variables' (Warner 1995), usually with a red dwarf companion star, where material accumulates on the surface until the pressure and temperature are high enough that a thermonuclear runaway reaction occurs (Bode \& Evans 2008).  During a nova eruption, a normally faint star is observed to brighten by 8-20 magnitudes to peak within a few days, to fade initially fairly quickly over a timescale of weeks to months, and then slowly fade back to the quiescent level.  The light curves of novae are usually the best-measured feature of the events, and are seen to vary substantially from system to system.  The light curves carry much information about each nova.
	
	The literature is full of nova magnitudes and fragmentary light curves.  Only a handful of the brightest novae (e.g., V1500 Cyg, T CrB, RS Oph, V603 Aql, DQ Her) have well observed published light curves.  Of these five examples of the best, only two were followed to quiescence in the professional literature, while all were followed to quiescence by amateur observers reporting to the American Association of Variable Star Observers (AAVSO).  Even amongst the 13 classical novae presented as ``First Class'' by Payne-Gaposhkin (1964), that is the best-of-the-best up until 1964, only five have professional light curves going to quiescence and only two of those appear in the book.  In modern times, little has changed, with RS Oph having wonderful coverage in 2006 but most novae having only a handful of magnitudes near peak published.  In one of the better cases, V5114 Sgr, one of us helped push for many professional observations, yet still our omnibus paper (Ederoclite et al. 2006) only has 46 V-band magnitudes not even going to 5.5 mag below the peak, and the position is mis-identified in the Downes \& Shara Catalog (Downes \& Shara 1993; Downes et al. 1997; 2001; henceforth referred to collectively as D\&SCat).  For another recent example, Figure 1a displays all the V-band magnitudes published in the professional literature for V705 Cas, and we see only 220 magnitudes with no coverage after the end of the prominent dip in the light curve.  And this is for one of the {\it best} observed events in recent decades.
	
	For a set of light curves with the best coverage of all novae, one of us (Schaefer 2010, henceforth S2010) has made an exhaustive compilation of essentially all photometry of all ten galactic recurrent novae (RNe, RN singular).  This comprises 37 known eruptions.  Of these, 16 are known with fewer than 20 magnitudes, while 8 have 20-100 points on their light curves.  Such coverage is largely useless for classifying or understanding the light curves.  Only 6 of these nova light curves (for T CrB and RS Oph) have been followed to quiescence.  Excepting these six eruptions, the median magnitude below peak of this coverage is just six magnitudes.  Even the famous U Sco has never been followed past 30 days after maximum.  Of the 13 eruptions with more than 100 magnitudes, all of them have the majority (typically 80-98\%) of their data from the AAVSO data base.  This sampling of nova light curves shows a wide range of coverage quality, mostly poor yet with a useful fraction of good coverage provided only by the AAVSO.  These RNe have attracted much attention (at least after their first eruption), so this sampling of light curve coverage is better than for ordinary novae.
	
	Several prior works have collected large numbers of nova light curves (McLaughlin 1939; Payne-Gaposhkin 1964; van den Bergh \& Younger 1987).  McLaughlin (1939) displayed only 17 light curves, with median coverage going to 7 magnitudes below peak and none going to quiescence.  Payne-Gaposhkin (1964) presented 43 light curves, 15 of which are copied from McLaughlin; in total, the coverage goes to a median of 5 magnitudes below peak.  van den Bergh \& Younger (1987) collected all nova light curves with UBV photometry and could only find 32 novae, for which their light curves have a median coverage of only 26 measures extending just 2.7 mag below peak.
	
	The classification of classical nova light curves is in a similar primitive state.  McLaughlin (1939) merely divided the light curves into `fast' and `slow'.  Payne-Gaposhkin (1964) made five finer divisions (`very fast', `fast', \ldots) based solely on the time it took the light curve to decline by two magnitudes from peak ($t_2$).  Payne-Gaposhkin never used her classification system for any purpose, nor has anyone else.  Duerbeck (1981) largely kept the speed classes as the primary classification parameter, but added a new class for those novae that showed `dust dips' (the sharp drop to a faint level followed by a fast recovery of brightness like seen for DQ Her) caused by the formation of dust in the expanding nova shell that hides the interior light.  His five classes are fast/smooth, irregulars, dust dip events, slows, and very slows.  The only later attempt to classify nova light curves that we are aware of is by Rosenbush (1999a; 1999b) who plots Òoutburst amplitudeÓ versus the logarithm of the main shell radius.  The primary reason why none of these classification schemes has been usefully adopted is that the classes are so broad that greatly different physical settings are lumped together and they have not been found to correlate with anything else.
	
	The measure of nova light curve properties is still fairly primitive.  The only properties that are known for a significant number of novae are the peak magnitude and the time it takes to decline by two or three magnitudes below peak ($t_2$ and $t_3$ respectively).  The most complete compilation from the literature, a heroic task in itself, is given in Shafter (1997).  Burlak \& Henden (2008) used the AAVSO data base to derive the peak magnitude, $t_2$, and $t_3$ for 64 novae, with no other properties being measured.  But this work was seriously flawed because no attempt was made to include observations from before the AAVSO data started, so the compilation missed the peak of about half the novae, and the decline times are correspondingly wrong.  Recently, theorists and observers have been starting to ask questions about nova light curves that need measures from many events.  Hachisu \& Kato (2006; H\&K06) have proposed a `universal decline law', where they make specific predictions as to the power law slopes at late times and early times, and the time when the break occurs is related closely to the mass of the white dwarf.  They have been systematically collecting many light curves and fitting them to broken power laws to find white dwarf masses for many individual systems (Hachisu \& Kato 2007; 2008; Hachisu et al. 2009a).  Independently, they have also been pointing to the existence of plateau phases as being a hallmark that a nova is recurrent (Hachisu et al. 2000; 2002; 2003; 2009a).  This has caused one of us to measure plateau parameters for the known RNe, and to ask what fraction of classical novae show similar plateaus (S2010).  Pejcha (2009) has been looking at a mixture of oscillations and flares (jitters) in seven novae trying to find whether the peak spacing is uniform in logarithmic or linear time, so as to test his theoretical model.
	
	In this paper, we present a catalog of 93 well-sampled nova light curves.  Almost all our light curves are all for the V-band as taken from the AAVSO database.  This sample solves the problems described above, in that we have a large number of light curves with extensive time coverage (often to quiescence) that go deep and have many observations each day.  With this large sample, we have created a new classification for nova light curves, with the categories more in line with modern ideas.  We also have measured many light curve properties for all of the novae, with our measures hopefully being useful to theorists.  We present plots and tabulated light curves for all novae.  Finally, we have sought and found a variety of correlations between our nova classes and our measured properties.  These new correlations and fits now form a challenge to theorists to explain their physical origin.
	
\section{AAVSO Data}

	The data we used for this catalog come from the AAVSO, with only a few exceptions.  The AAVSO database has photometric magnitude estimates from amateur and professional astronomers all around the world.  This database has magnitudes for roughly 200 novae, with some of the stars having tenÕs of thousands of observations from the eruption to several decades later.  For about half of the novae, the total light curve does not have enough coverage to allow for a confident light curve over most of the eruption, and these novae will not be considered in this paper.  But this still leaves us with 93 novae with good coverage throughout the eruption plus a high time density of observations.  
	
	The AAVSO light curves have some tremendous advantages over almost all nova light curves that have appeared in the professional literature in that the time coverage is often all the way until the nova is in quiescence (and often with continuing coverage for decades after).  Another tremendous advantage is that AAVSO coverage usually has a very high density of measures, typically ranging from 1-10 magnitudes per day.  To quantify this, our catalog of 93 nova light curves comprises 229,796 individual observations.  Of these novae, the median depth of coverage goes to 8.0 magnitudes below peak.  We have 24 novae for which the coverage goes completely to the quiescent level, and so we can determine the number of days after peak at which the light curve goes essentially flat ($T_Q$).  These numbers are large improvements on all previous nova light curve studies.  The large number of novae, the extremely large number of observations, and the deep coverage all mean that the AAVSO light curve data reveals many results that had been previously hidden.  This includes detailed statistics on light curve shapes and measured light curve properties for large numbers of events.
	
	The AAVSO data are primarily visual data, made by experienced amateur astronomers looking through a telescope eyepiece and comparing the brightness of the nova against the brightnesses of nearby comparison stars.  In the past twenty years, many of the AAVSO astronomers have become equipped with CCD cameras and the AAVSO data base now has CCD magnitudes for many novae with many filters.  For this catalog, with only three exceptions (DO Aql, T Aur, and BT Mon) we used only observations in the V-band, whether they are from visual observations or from CCD images.
	
	Some might think that a disadvantage of the AAVSO data is that the magnitudes are estimated by eye and they would worry about reliability.  This is wrong.  The reliability of the AAVSO magnitudes is about the same as for observations in the professional literature.  Indeed, with the AAVSO having many independent observers providing measures at effectively the same time, the rejection of systematic errors can be more reliably done with AAVSO data than with professional data which is generally sparse.  An alternative worry is that the visual magnitude estimates have a one-sigma uncertainty that is so large as to be unusable.  Indeed, from extensive experience and control studies, we find a typical one-sigma error for single visual magnitude estimates to be 0.15 mag.  For professional astronomers used to photoelectric or CCD magnitudes with real uncertainties of 0.015 mag for a single measure even for good photon statistics (Landolt 1992; 2009), the visual accuracy sounds poor.  But this worry is naive for two reasons.  First, the AAVSO light curves generally have many observations, and the usual measurement errors improve by a factor of the square root of that number, so the typical real uncertainty is more like 0.05 mag.  This is not so far from the 0.015 mag attainable with modern instrumentation.  Second, the required accuracy for nova light curves is much poorer than 0.15 mag, simply due to the features being sought having a much larger amplitude than 0.15 mag.  That is, for a light curve that fades by, say, ten magnitudes, even half-magnitude accuracy is adequate for defining the shape and features of a light curve, while the usual amplitude of flares and oscillations and dips is always much larger than the uncertainty in even one visual estimate.  It would be `nice' to have CCD magnitudes over visual magnitudes, but the improved accuracy does not translate into anything useable or any improved science.  For the purposes at hand, the naive worries are not realized and the AAVSO light curves are comparable in quality to those from professional observatories (while having vastly better time coverage).
	
	Of the $\sim$250 known galactic novae, we narrowed our selection for this catalog based on a number of criteria.  Our base list of novae came from D\&SCat, and therefore consisted only of novae that erupted before 2006.  The cut-off date from D\&SCat is reasonable as most later novae cannot yet have light curves that cover the late tails of the eruption.  With the purpose of this paper being that we are aiming to classify novae based on their light curve shapes, we include novae with good coverage in the AAVSO database.  We do not include symbiotic novae (such as RR Tel or PU Vul) in our catalog because they are so different from classical novae.  We are left with 85 nova light curves in a sample which has no apparent bias based on light curve type.
	
	Our base catalog is comprised of novae with peaks before 2006 for which there is excellent coverage from the AAVSO database.  We have supplemented this selection in a few cases for a variety of reasons.  (1) V2362 Cyg and V2491 Cyg peak in 2006 and 2008 respectively, yet they are two of the three known novae with prominent cusps in their light curve.  As their AAVSO coverage is excellent, they have been included so that this light curve class has the best possible representation.  (2)  V2467 Cyg peaked in 2007, yet this light curve displays a good example of oscillations, for which few other examples are known, so this nova is included anyway, as based on the AAVSO light curve.  (3) RS Oph peaked in 2006, has an incredibly well observed light curve and has already returned to its normal quiescent level, so it is included.  (4) DO Aql and BT Mon are included, despite little AAVSO data, because they are two of the four known flat-topped light curves.  The light curves in both cases were taken from the literature and are in the B-band.  The data for DO Aql came from Vogt (1928), Beyer (1929), Hopman (1926), Steavenson (1926; 1927), and Zikeev (1927).  For BT Mon, we took B-band magnitudes from Schaefer \& Patterson (1983), Wachmann (1968), Bertiau (1954), and Whipple (1940); for which we have used modern measures of their quoted comparison stars to convert their reported magnitudes to the modern B-band magnitude scale.  (5) GK Per erupted in 1901, long before the founding of the AAVSO (in 1911), yet it has a well-observed light curve that represents the best example of a rare type of nova light curve (displaying oscillations in brightness somewhat below the peak), and thus we have included a V-band light curve taken from the literature (Pickering 1903).  (6) T Aur erupted in 1891, again long before the foundation of the AAVSO, and has a B-band light curve that we have taken from Leavitt (1920).  We added this to extend the number of novae with dust dips to improve tests of possible correlations.  In all, we have added eight novae (DO Aql, T Aur, V2362 Cyg, V2467 Cyg, V2491 Cyg, BT Mon, RS Oph, and GK Per) to our unbiased sample of 85 novae, for a total of 93 well-sampled light curves in our catalog.  Table 1 contains a complete list of these novae, along with their year of eruption.  The novae are identified and ordered in the usual manner as adopted from the General Catalog of Variable Stars (Kukarkin 1985).
	
	For the AAVSO data, we download the observations from the AAVSO website into a spreadsheet.  Each magnitude estimate includes the Julian Date (JD) of observation, the observed magnitude, the band (such as visual or B), and the observer's identification.  We immediately remove observations that are not in the visual or V-band and those that are just brightness limits.  
	
	As discussed above, a small fraction of the AAVSO visual measures are outliers in the light curve and are likely errors in some way.  These are recognized as outliers that differ significantly (by more than three-sigma) from the consensus of many other observers with effectively simultaneous observations.  (This is a big advantage of the AAVSO data in that such observations are available to recognize outliers.)  When we recognize such outliers, we conservatively reject all observations by that observer.  In all, we reject $\sim$0.1\% of the data.  In practice, this procedure is purely cosmetic because the outliers are swamped by the normal data.
	
	 We have binned the AAVSO magnitudes in time to create our final light curves.  This binning makes for a substantial reduction in the light curve statistical uncertainty and allows for a reduction in the quantity of data to a manageable size.  The time bin sizes range from 0.5 to 256 days, depending on the rate of change of the novae.  We picked a Ôstop dateÕ for when the nova had reached quiescence or when the observations had stopped.  We chose bin sizes that averaged as many observations as possible without losing any of the fine detail of the fluctuations in the nova light curve.  If the light curve had relatively quick changes in brightness, such as near the peak, it was necessary to use short binning on the order of 0.5 to 4 days.  In cases where there is relatively little fluctuation in the brightness, such as late in the tail of the light curve, we found it acceptable to use long binning on the order of 64 to 256 days.  This process is illustrated in Figure 1 with an example of an unbinned light curve (Fig. 1b) and binned light curve (Fig. 1c).   A comparison between the panels of Figure 1 shows convincingly that the AAVSO light curves are substantially better than those available in all the professional literature, even though this is one of the best observed novae in recent decades as based on the published light curves.
	
	To calculate the binned magnitude, we performed a weighted average of all estimates in the bin.  The weight of the visual measures was taken as 1, while the weight of the CCD and photoelectric V-band measures was taken as 10.  The uncertainty in the magnitudes ($\sigma_V$) is taken as the larger of 0.02 mag,  $0.15N^{-0.5}$ mag where $N$ is the number of observations, and $\sigma_{obs}N^{-0.5}$ where $\sigma_{obs}$ is the observed RMS scatter of the magnitudes in the bin.
	
	 Almost all our data is from the AAVSO, but we did supplement this with small but important sets of data from the literature:  (1) Most importantly, we selectively added in published magnitudes that cover the light curve {\it before} the start of the AAVSO light curve.  Typically, this is pre-discovery or discovery observations from the IAU Circulars.  The inclusion of these magnitudes is often vital as it shows either a rise to peak or a peak that is brighter than the earliest AAVSO magnitudes.  We have supplemented one magnitude or a few magnitudes as such for about half of the novae in our catalog.  The total number of added observations is 1,269.  (2) For V838 Her, we have added a small number of late-time V-band magnitudes from the literature simply to fill out the tail of the light curve.  We use the late time data from Table 1 of Ingram et al. (1992) to describe the late tail.  (The late AAVSO points fainter than 13 mag were rejected as being measures of a star 1.7 arc-seconds from the nova.)  (3) For V445 Pup, we use the late V-band magnitudes from Woudt et al. (2009), and we also have our own V-band magnitudes from 2003 at McDonald Observatory.  (4) As mentioned previously, we took all the light curves for DO Aql, BT Mon, GK Per, and T Aur from the literature.  The magnitudes for DO Aql, BT Mon, and T Aur are all exclusively in the B-band.

	Our binned light curves are presented in Table 2 for 93 novae.  The version of this table in the printed paper is only a stub of 15 lines, while the whole table, all 11,843 lines of observations, is presented in the electronic version available on-line.  The first column gives the name of the nova.  Column 2 gives the Julian date of the middle of each time bin.  The start and end times of each bin are halfway between the preceding and following bin times.  The third column gives the time ($T$) of the middle of the bins with respect to the fiducial time ($T_0$).  These fiducial times are intended to be the time of the explosion, that is when the light curve starts to rise from its quiescent level.  These explosion times cannot be known with any certainty as they are always before any observations.  We generally select $T_0$ to be within a few days of the first observation showing the nova in outburst.  Occasionally, we use observational evidence (for example, based on spectral `dating') or other information to adopt an earlier explosion date.  The choice of $T_0$ will matter only for determining the power law slope of the early decline, and even then the uncertainties from our choice are generally negligible.  The reason for using $T_0$ for the power law slopes (as opposed to, say, the better measured time of peak brightness) is because this is the quantity required by the universal decline law and we must make our derived slopes directly comparable to the model.  Our $T_0$ values are all tabulated in Table 1.  The fourth and fifth columns are our binned and averaged V-band magnitude along with the one-sigma uncertainty ($\sigma_V$).  The final column lists the number of individual observations going into each bin ($N$).
	
	For each of our binned light curves, we determined the bin with the peak brightness.  The Julian date of this peak ($T_{peak}$) and the brightest V-band magnitude ($V_{peak}$)  are listed in the fourth and fifth columns of Table 1.  With this, we have also calculated the times from the peak until the {\it last} time that the light curve falls below various magnitude thresholds.  We choose to report the times after the peak for which the light curve has fallen by 2, 3, 6, and 9 magnitudes below the peak magnitude ($t_2$, $t_3$, $t_6$, and $t_9$ respectively) in columns six to nine of Table 1.  
	
	We choose to define $V_{peak}$ as the brightest observed V-band magnitude.  This has the strong advantages that it is what the name implies, is what people generally quote, and has an unambiguous value.  Nevertheless, other possibilities might be reasonable.  For example, Burlak \& Henden (2008) defined the peak to be for a smoothed light curve that ignores `short flares'.  In most cases, these differences in definition do not change $V_{peak}$.  But for many novae with large amplitude jitters, the effects can be large.  A prominent example highlighted by Burlak \& Henden (2008) is for V723 Cas, for which they quote $V_{peak}=8.9$ as appropriate for the day of discovery.  This requires subjective judgment as to whether the hundred day rise after discovery is important, as well as subjective judgment as to whether the prominent jitter with a 1.8 mag amplitude and 23 day duration is regarded as a `short flare'.  We chose to take the `peak magnitude' at its face value of $V_{peak}=7.1$ because this has no subjective judgments.  Another universal problem is that many novae are only caught fading down from some maximum, which might be somewhat brighter than observed.  There is nothing that can be done about this, as blind extrapolation to some guessed peak date will only add systematic noise.  This all goes to point out that peak magnitudes can sometimes have substantial uncertainties that are hard to quantify.
	
	We have similar problems in defining the decline times.  With a given peak and a smooth light curve, $t_2$ and $t_3$ values are straight-forward to calculate from binned light curves.  However, when the usual jitters and cusps and dips are included in the light curves, the original verbal definitions (i.e, the number of days required to decline from the peak by a given magnitude) become ambiguous.  For a light curve that goes up and down, there can be multiple times when the light curve has declined by the given magnitude.  Alternatively, perhaps the declines could be calculated from some fitted power law decline.  We have chosen to use the observed light curve (i.e., not a model fit) and to use the {\it last} time in the light curve for which the brightness is above the threshold.  To be specific, we take $t_3=T_{t3}-T_{peak}$, where $T_{t3}$ is the {\it last} time for which the light curve has $V<V_{peak}+3$.  If we had chosen to take the {\it first} time fainter than threshold, then we would get many instances like for V723 Cas where $t_2=17$ days for an apparently fast decline in what is manifestly a slow nova light curve.  Our choice of the {\it last} time is a selection for the ambiguous cases of novae with cusps and dips.
	
	For the RNe, we have chosen to adopt the template values as reported by Schaefer (2010).  These templates were constructed with data from many sources and from many eruptions, so the composite light curve is the best information that we have for these systems.  (The light curves from all eruptions of a specific RN are proven to be consistent with one unchanging template, see S2010.)  These templates provide much of the information on each line in Table 1 for the RNe.  A typical example of this improvement is that the 1866 eruption of T CrB was discovered on the rise and seen to come to a peak of 2.5 mag, whereas the 1946 eruption was discovered about one day after peak when the light curve had already declined to 3.5 mag, so we chose to include the best values in Table 1.
	
	The quiescent V-band magnitude ($V_Q$) is generally taken from the AAVSO data from long after the eruption is over.  $V_Q$ for the RNe are taken from Schaefer (2010), while some other values are taken from D\&SCat or from the literature.  For the well observed light curves, the late tail of the eruption has the light curve asymptotically approaching the quiescent level.  Given that the quiescent level varies on all time scales with typical amplitudes of 0.5 mag (Collazzi et al. 2009; S2010; Honeycutt et al. 2006; Kafka \& Honeycutt 2004), the time when the eruption light has gone to zero is ill-defined.  Nevertheless, we can derive the stop date for the eruption by plotting the magnitude versus the logarithm of $T-T_0$ and seeing when a straight line fit to the late tail passes through $V_Q$.  The time of this stop date, with respect to the peak date, is notated at $T_Q$.    For some novae, the light curve flattens out (i.e., the eruption ends) with the nova substantially brighter than the pre-eruption level (Schaefer \& Collazzi 2010), with the post-eruption magnitude being listed.  In Table 1, the tenth and eleventh columns give our values for $V_Q$ and $T_Q$.
	
\section{Light Curve Classification}

	The shape of nova light curves carries substantial information on the physics of the eruptions.  The morphology of these shapes is a necessary first step for pulling out information and correlations.  Past nova light curve classification systems barely went beyond looking at the $t_2$ value.  With our set of 93 light curves, we can finally see a large sample of well observed novae, and the morphology becomes apparent.  Here, we propose a new classification system for nova light curves.  
	
	We define seven classes of nova light curves, based solely on their shapes.  Each class is designated by a single capital letter.  The letters are S (smooth), P (plateau), D (dust dips), C (cusps), O (oscillations), F (flat-topped), and J (jitters).  These classes have definitions and examples presented in Table 3.  To add a quantitative time dimension to the classification, the $t_3$ value in days can be added after the letter within parentheses.  Examples from famous novae would be that V1500 Cyg is S(4), GK Per is O(13), DQ Her is D(100), and U Sco is P(4).  For some purposes, it is fine to not include the $t_3$ value in parentheses.  Prototype novae light curves are displayed in Figure 2 for all seven classes.

	S (smooth) light curves are the stereotypical case.  Indeed, the S class comprises the single largest fraction.  In general, light curves after the peak are well described as a broken power law.  That is, the early decline can be represented as a straight line on a magnitude versus log-time plot, and the late decline is also a power law of steeper slope.  This very general behavior is well described and predicted by the Ôuniversal decline lawÕ of H\&K06, which is a simple consequence of radiative transfer within an expanding shell.  Largely, all other classes of light curves are S shapes with some complication superposed.

	P (plateau) light curves have relatively flat intervals superposed on the otherwise smooth declines.  The plateaus need not be perfectly flat, and generally there is some fading during the plateau.  Hachisu et al. (2009a) distinguish a Ôtrue plateauÕ where the continuum radiation (for example, measured with the y-band filter) has the plateau from a Ôfalse plateauÕ where a broad band filter which includes substantial flux from emission lines (like the V-band) creates a flat portion due to the rise of emission line flux relative to the continuum.  They point to the true plateau as being uniquely from RNe (relating to the extended supersoft source illuminating the accretion disk).
	
	D (dust dip) light curves have the characteristic sudden cutoff, minimum, and recovery in brightness.  These dips are well known to be caused by the formation of dust particles in the expanding shell when the material reaches such a distance from the nova that the gas cools down to roughly 1400$\degr$ K and that refractory dust can form.  This dust will effectively block out the light from the photosphere interior to it, causing the sudden drop of light.  The recovery to the previous decline is caused by the geometric dilution of the dust as the shell expands, allowing the photosphere to be seen again with little extinction.
	
	C (cusp) light curves have the characteristic secondary maxima 1-8 months after the primary peak with the distinctive shape of an additive component (superposed on an apparently normal S light curve shape) that rises at an accelerating rate until some time when the additive component turns off sharply.  This cusp phenomenon was first recognized in 1999 with V1493 Aql.  No further examples were identified until 2006 (V2362 Cyg) and 2008 (V2491 Cyg), so now with three examples we have a distinctive class.  
	
	O (oscillation) light curves have a time interval somewhat after the peak where the brightness varies with moderate amplitude in an apparently periodic modulation and with similar shapes for each cycle.  These oscillations might continue for 5-25 cycles.  Outside of the oscillations, the light curve looks like an S class event.  These variations appear to be both brighter and dimmer than the interpolation from before and after the oscillations.  The only famous or obvious cases are V603 Aql and GK Per.
	
	F (flat-topped) light curves have an extended interval from 2-8 months of nearly constant magnitude at peak.  Other than this flat-top, the light curves appear to be ordinary smooth power law declines.  We have only one well-observed example in our basic AAVSO sample (V849 Oph), so it would have been easy to just call it an anomalous `S' and go no further.  But the light curves of BT Mon and DO Aql are sufficiently similar and distinct from all the other S light curves, that we are making this into a separate class.  A further motivation for thinking that the F novae have some distinct physics creating the flat-top is the light curve for the symbiotic nova PU Vul which has a remarkably constant peak that lasts for ten years (with a spectacular dip superposed).
	
	J (jitter) light curves have the basic smooth power law declines except that there is substantial jitter about that basic shape.  These jitters are variations of substantial amplitude (typically greater than half a magnitude) that generally extend above a smoothed light curve as often isolated brightenings.  J light curves have more than one such flicker and these flares occur only in the first part of the light curve.  
	
	Any such classification scheme must have cases where the class identification is ambiguous.  This could be due to the light curve not being adequate to distinguish between the classes.  A typical example is the S class nova V4643 Sgr, where the small bump at 19 days might be either a normal fluctuation or it might represent a weak cusp.  Another typical example is the S class nova V533 Her, where the significant but small fluctuations at 30 and 60 days might point to a J class light curve.  And we wonder whether the 8 day long nearly-constant maximum for the S class nova V2295 Oph might really be a very short flat top F class event.  We accept this reality (especially for the many poorly observed novae), and we can notate such cases with a question mark.  For example, a fragmentary light curve that looks to be a smooth power law decline can be given the classification of ``S?'', while a more poorly observed nova might be simply classed as ``?''.  
	
	Ambiguity can also arise for cases near the threshold between classes.  This is expected since the various physical phenomena that create the classes likely form a continuum, so it will always be somewhat arbitrary as to where the threshold is drawn.  So for example, perhaps all S light curves show oscillations and we can only define the O class for those events where the oscillation amplitude exceeds some standardized limit.  For other examples, in principle, all light curves will have some sort of a plateau (from the supersoft phase) and some sort of a dust dip (as dust formation is inevitable to some degree), but we will only identify the light curve as P or D when the plateau or dip is significant and readily distinguishable.  The utility of the classes is then that the novae in the class will all share an uncommon underlying property that makes the distinctive light curve feature relatively extreme, so the collective study of the novae in the class can help for understanding the relevant physics.  At this time, we do not feel comfortable or knowledgeable enough to place quantitative limits on our class definitions.  Nevertheless, we can still easily place each nova into a unique class based on the visual appearance of its light curve and the verbal definitions in Table 3.
	
	So far, we are able to place all the well-observed nova light curves into just one of seven classes.  This represents a substantial simplification of the complexity of the myriad of light curves.  (One of us remembers with horror a similar task of creating morphological light curve classes for Gamma-Ray Bursts in the Konus catalog, and coming up with over 20 classes yet still having half the bursts remaining as unique and outside the classes.)  The existence of only seven classes covering all nova light curves implies that there are only six basic physical mechanisms that substantially modify the `universal decline law' for the smooth S light curves.
	
	Nevertheless, we should not be complacent enough to think that no exceptions will ever be found.  For example, from 1999-2006, the single known C nova (V1493 Aql) was unique and had a significantly different light curve shape.  We propose that for later cases of unique novae, the classification letter `U' be used for these unique or unusual events.  Later, when more distinct examples are discovered, a separate class can be defined.  But for now, the U class is empty.
	
	Given our new classification scheme, we have typed all 93 novae, as reported in the last column of  Table 1.  Given the class for any nova, we can draw out a rough light curve for that event with the correct shape and time scale.
	
	Out of our unbiased sample of 85 novae, we can count the numbers in each class.  The numbers are 32 for S (38\%), 18 for P (21\%), 15 for D (18\%), 1 for C (1\%), 3 for O (4\%), 2 for F (2\%), and 14 for J (16\%).  The uncertainties on these percentages are given by the usual standard deviation from the binomial distribution.
	
\section{Comparison With Other Classes}

	Novae can be classed by many other properties than just light curve shape.  A prominent way to divide the novae is based on their orbital periods, especially in relation to the `period gap', where the distribution of periods for cataclysmic variables has a narrow range of periods from roughly 2-3 hours with few systems (Patterson 1984).  Another prominent subdivision is for those systems that have a very high magnetic field on the white dwarfs such that the accretion disk is disrupted and the material falls directly onto the white dwarf through an accretion stream, the so-called `polar' systems (Warner 1995).  Or if the magnetic field is only moderate in strength, then a partial disk and an accretion stream form, the so-called `intermediate polars' (Warner 1995).  Novae can be further subdivided based on their spectra (Williams et al. 1991), whether they have shells (Downes \& Duerbeck 2000), the light curve behavior in quiescence (Collazzi et al. 2009; Schaefer \& Collazzi 2010), and recurrence (S2010).  These other categorizations can have connections to the same physical mechanisms that determine the light curve shape, so we should seek connections between our light curve classes and other nova groups.
	
	Table 4 has collected many classifications of novae.  The first column gives the nova identification.  The same 93 novae are included in Table 4 as in Table 1, except that Table 4 is ordered by our light curve classes.  The second column gives our nova light curve class.  The third column gives a spectral classification, pointing to those novae that have the common `FeII' class (Williams 1992; Williams et al. 1991) or that have very high abundance of neon (Starrfield et al. 1986).  The fourth column identifies the RN systems which have had more than one observed nova eruption (S2010).  The fifth column points to those systems that have had expanding nova ejecta observed after their eruptions (Downes \& Duerbeck 2000).  This does not include light echoes.  Certainly, all novae ejecta shells and not all novae have been carefully searched at the right epoch such that a bright shell can be detected, so this list merely notes those novae for which the shell is relatively bright and for which appropriate deep images have been searched.  The sixth column identifies those old novae for which the white dwarf is known to have a high magnetic field, including the polars and intermediate polars.  The seventh column indicates those novae with high-excitation lines (He II, [N V], [Fe X] or higher) around the time of peak or soon thereafter (``High''), as well as those with low excitation emission lines (``Low'').  The eighth column gives the orbital period, with classes given for being below the period gap ($<$2 hours) labeled ``V. Short'', with orbital periods in the period gap from 2 to 3 hours (labeled ``In Gap''), with orbital periods from 8 hours to 6 days (labeled ``Long''), and with orbital periods longer than 200 days (labeled ``V. Long'').  The ninth column collects various properties relating to the quiescent light curve, including whether the nova had an anticipatory pre-eruption rise (``Pre-rise''), an anticipatory pre-eruption dip (``Pre-Dip''), a post-eruption secondary maximum (``Post-Max''), a post-eruption dip (``Post-Dip''), a large change in the minimum magnitude from before eruption to long after eruption (``Hi-$\Delta$m'', see Collazzi et al. 2009; Schaefer \& Collazzi 2010), post-eruption dwarf nova events (``w/DN''), and very long tails on the light curve returning to quiescence (``v. long tails'' with $T_Q \ge 5000$ days).  The tenth column gives the classification from Rosenbush (1999a).  Unfortunately, many of the entries in columns 3-10 are blank, with this implying either that the nova has not been examined for this property or that it has been examined for the property without positive detection.
	
	The most energetic nova events appear to be S and P classes, while the least energetic appear to be the D and J classes.  This is seen for the characterization by the excitation levels of the lines, where the S and P classes have 10 out of 12 that are `High' and only two that are `Low', where the D and J classes have all 4 characterized as `Low'.  This is also seen with the $t_3$, where small values correspond to high mass white dwarfs and large values correspond to relatively low mass white dwarfs (H\&K06).  The $t_3$ value is also directly related to the energetics of the peak, with the fast events being super-Eddington and the slow events being of relatively low luminosity (Downes \& Duerbeck 2000; Shara 1981).  The median $t_3$ values are 23 and 36 days for the S and P classes, and are 49 and 81 days for the D and J classes, respectively.  The RNe must also have white dwarfs near the Chandrasekhar limit, and these are exclusively in the S and P classes.

	All novae have ejected shells, and we can only tabulate the ones whose shells are above some changing and largely unknown threshold.  Nevertheless, those with discovered shells will be those that are the brightest, and the discovery efficiency will be independent of the light curve class.  Therefore the statistics of detected shells as a function of class will provide some measure of the rate at which each class produces conditions conducive to making a bright shell.  Overall, 27 out of our 93 novae have observed shells (30\%).  A high proportion of D novae (9 out of 16, for 56\%) have detected nova shells.  A low portion of S novae (5 out of 39, for 13\%) have detected nova shells.  The binomial probability that 9 or more of the D novae will have shells is 1.4\%, while the probability that 5 or fewer of the S novae will have shells is 10\%.  These probabilities are not formally significant, but they are suggestive.  We can think of various possible physical mechanisms for this correlation (perhaps high metallicity will make for much dust and bright shells, or perhaps low expansion velocities make for dense dust forming conditions and shells that are bright due to high electron densities).  We encourage theorists to investigate possible physical mechanisms.
	
	From Table 4, we note that most of the O class novae have magnetic white dwarfs and preferentially have massive white dwarfs (as shown by their systematically low $t_3$ values).  We think that these two correlations are significant, and strong clues to the physical cause of the oscillations.  Some novae with high magnetic field white dwarfs do not show oscillations while other systems with massive white dwarfs do not show oscillations, so apparently the white dwarfs must be {\it both} highly-magnetic and high-mass so as to produce oscillations.  But this cannot be the full set of requirements for oscillations, as V1500 Cyg (a polar with $t_3$= 4 days) does not show oscillations.  Many theoretical ideas have been briefly sketched as explanations for the nova oscillations.  Pecker (1964) proposed that the oscillations arise within the low-density ejected shell as instabilities in the ionization equilibrium.  Sparks et al. (1976) sketched an idea that pulsations of the white dwarf envelope will have periods of four days (as appropriate for GK Per) when the envelope radius is larger than the binary orbit (however, this is over an order of magnitude larger than they derive for the system).  Bianchini et al. (1992) suggested that the outer layers of the white dwarf have some pulsational instability, much like ordinary stellar pulsations in a polytrope, with changes in the inter-peak intervals caused by variations between the fundamental and first overtone frequencies.  Bianchini et al. (1992) also suggested the possibility of repeated thermonuclear flashes.  Leibowitz (1993) conjectured that the oscillations arise in the disk and are caused by DN-like instabilities, with the only proffered evidence being that the oscillations in GK Per look like the eruptions of SS Cyg, with no calculations or tests of this idea.  These many ideas cover all possible sites for the oscillations, none have been developed past a bare idea, none have been tested with independent observations, and none make any predictions.  For our new result, none of the proposed models connects in any way with highly-magnetic or high-mass white dwarfs.  With this, we judge that there are no realistic models of any useable confidence to explain the oscillations.
	
	The V1500 Cyg stars have a high $\Delta$m (i.e., are over a factor of roughly ten brighter in their post-eruption quiescence than in their pre-eruption quiescence), have magnetic white dwarfs, have very long slow declines after the eruption ends, have long-lasting supersoft sources, and generally have short orbital periods (Schaefer \& Collazzi 2010).  Now we see that five out of six V1500 Cyg stars listed in Table 4 (with `Hi-$\Delta$m') have shells.  For the two V1500 Cyg stars not in Table 4, GQ Mus has no detected shell while RW UMi does have a shell (Downes \& Duerbeck 2000).  Out of the eight known V1500 Cyg stars, six of them have discovered shells.  With a 30\% probability of a nova having a shell discovered, the likelihood of six or more having shells is 1.1\%, which can be taken as suggestive but not convincing.

	V1330 Cyg has an apparent flattening of its light curve from days 440-1235, with the nova light being constant at close to 15.0 mag.  From the AAVSO data, the flattening is a result from 21 independent magnitudes by 3 observers, and we have no reason to doubt the reality of the flattening.  (For example, there is no nearby star of around fifteenth magnitude that could provide any confusion.)  It appears that the nova light curve has gone to quiescence.  But magnitude estimates from much later give V=17.53 (Szkody 1994, with B-V=0.74), and the second epoch sky survey gives B=17.57 (Collazzi et al. 2009), hence V $\approx$ 16.83.  So it looks like V1330 Cyg might have a decades-long secular decline from 15.0 to 17.53.  The pre-eruption magnitude on the first Palomar Sky Survey was B=18.74 (Collazzi et al. 2009) hence V$\approx$18.00, which would make the nova be a  high $\Delta$m star.  With both a secular decline and an apparently significant $\Delta$m, V1330 Cyg is a good candidate for being a V1500 Cyg star (Schaefer \& Collazzi 2010).  With this, the flat light curve from days 440-1235 might be characterized as the light curve going nearly flat at the end of the eruption and also just the start of a long slow decline continuing from at least from 1971 to the 1990s.  It would be of interest to get a recent magnitude of V1330 Cyg to see if the decline is continuing, or whether the system has returned to its pre-eruption level.
	
	The novae with very long tails ($T_Q \ge 5000$ days) might represent an intermediate case between the V1500 Cyg stars and the normal novae that return to their quiescent level within a few years.  That is, the V1500 Cyg stars have slow declines on a time scale of a century or more (Schaefer \& Collazzi 2010) while most novae return to quiescence within years, so it is reasonable to look for novae that are intermediate in returning to quiescence on a time scale of a few decades.  Table 4 has six novae with very long tails in their light curves (V603 Aql, T Aur, V476 Cyg, HR Del, DQ Her, and RR Pic), and these might be the intermediate novae.  If there is a continuum of decline time scales, then the intermediate case might simply represent a less extreme situation than is present in the V1500 Cyg stars.  The root causes of the V1500 Cyg phenomenon are that the white dwarf is highly magnetic and the orbital period is short (Schaefer \& Collazzi 2010), and so maybe the very long tails are caused by intermediate cases of magnetic fields and orbital periods.  Of these novae, five have determinations in the literature that they have highly magnetic white dwarfs, while the sixth has no useful measures reported that would detect such.  This is a distinct pointer that the novae with very long tails could well be intermediate between V1500 Cyg stars and the other novae, but the small number statistics will always force this conclusion to not have high significance.  For the five very long tail novae with known periods, all are just above the period gap (with an average period of 0.18 days), again consistent with an intermediate case, although again this cannot be a significant result due to small number statistics.  We can also test whether the very long tail nova are intermediate in terms of their resultant properties.  Of the six novae with very long tails, {\it all} have nova shells.  For this one property alone, the statistics suggest a likely connection, but again, any one property will not be of high significance.  If we take the two properties of both magnetic white dwarfs and nova shells (as predicted from the idea that they are intermediate cases of the V1500 Cyg phenomenon), the odds are rather long unless there is a causal connection.  With this, we take the novae with very long tails to simply be V1500 Cyg stars for which the timescale for the decline is 10-50 years instead of a century or longer.
	
	Symbiotic stars (featuring a cool giant star plus a hot emitting region usually around a white dwarf) are a prominent class of interacting binary (Kenyon 1986), and some novae systems fit the formal definition for inclusion into this class.  Technically, the four RNe T CrB, RS Oph, V3890 Sgr, and V745 Sco are symbiotic stars.  However, these four are substantially different from other symbiotic stars, in that they have greatly shorter orbital periods, lack the massive dusty winds, and have the mass transfer by Roche lobe overflow (S2010; Miko{\l}ajewska 2009).  The symbiotic stars are represented in Table 4 as those whose orbital periods are in the `V. long' class.  ``Symbiotic novae'' are events that occur on symbiotic stars that are defined by Kenyon (1986) and Iben (2003) to be {\it very slow} eruptions involving systems that do {\it not} fill their Roche lobe.  Symbiotic novae have {\it very} long eruptions (with $t_3$ values from years to decades) and {\it very} low amplitudes (2-6 mag).  As such they are greatly different from all nova eruptions.  So different that any result from symbiotic novae cannot be transfered with any useable confidence to novae.  While the four RNe might be novae that occur on (technically) symbiotic stars, they are {\it not} symbiotic novae because they fail the definition on two counts (being both very fast and with Roche lobe overflow).  In all, the very long period systems in Table 4 might be symbiotic stars, but they are so greatly different from other symbiotic stars and from symbiotic novae that there is no utility in making this connection.

	With Table 4, we recover some known correlations.  The RNe are concentrated in the P class, as predicted and described by Hachisu et al. (2000a; 2002; 2003; 2009a).  Also, all RNe have high excitation spectral lines, with this fact being used as an indicator of which ÔclassicalÕ novae are actually recurrent (Pagnotta et al. 2009).  We see that the V1500 Cyg stars with high $\Delta$m are associated with very short orbital periods and magnetic white dwarfs (Schaefer \& Collazzi 2010).
	
\section{S Class Light Curves}

	The stereotypical nova light curve has a fast rise, a fairly sharp peak, and a smooth decline that is initially steep but which slows with time.  We are taking such novae to be in our S class.  All the other light curve classes are just variants on the S class, so it is as if the S class light curves are the basic phenomenon, lacking in any of the complications that cause the differences that define the other classes.  Indeed, H\&K06 have proposed a `universal decline law' based on the simple radiative transfer in an expanding shell that prescribes a smooth decline that matches our S class.
	
	The universal decline law specifies that the continuum flux should fall off as $(T-T_0)^{-1.75}$ for early times and then break to fall off as $(T-T_0)^{-3.0}$ at late times.  Here, $T-T_0$ is the time after the initial explosion.  When the light curve is plotted on a graph of magnitude versus $\log(T-T_0)$, the predicted early and late slopes should be -4.4 and -7.5.  The break time depends primarily on the mass of the white dwarf, with a small dependency on the composition of the accreted material.  The light curve around the time of the break is expected to have complexities due to a short segment with flux going as $(T-T_0)^{-3.5}$ and due to emission lines changing differently from the continuum.  Nevertheless, we can test the prediction for the early and late slopes.
	
	Our sample has 32 S class light curves.  These are displayed in the six panels of Figure 3.  This figure shows the horizontal axis as linear time, because this is the traditional way of displaying light curves, and covers the first 200 days, so that details around the peak can be well-resolved.  We have also repeated all the light curves in the four panels of Figure 4 with the horizontal axis as $\log(T-T_0)$.  The reason for this is so that the power law slopes and the breaks are easily visible, as well as because this is the only way that the late time data can be usefully displayed.
	
	We have derived various properties of the S light curves (see Table 5), and can give the results from fits to broken power laws.  The first column in Table 5 gives the name of the nova.  The second column gives the number of breaks ($N_b$) that we adopted.  The various light curves have either 0, 1, or 2 breaks (with 1, 2, or 3 power law segments respectively).  Given the normal small variations in the light curves, the number of segments is occasionally not clear, and we can only report the fitted segments based on our best judgment as guided by the usual F-Test analysis for additional parameters.  For each power law segment, the table has three columns, the first giving the slope (in magnitudes per logarithmic time, i.e., dimensionless), the second giving the time of the break at the end of the segment ($T_{b1}$ and $T_{b2}$) with respect to $T_0$ (in days), and the third giving the V-band magnitude of the break ($V_{b1}$ and $V_{b2}$).  For the last fitted segment in each light curve, we only give the slope.  Some of the light curves are well represented by one or two power law segments, while some others require up to seven segments.  The existence of the many segments are highly significant (as based on the F-Test), but it is somewhat daunting to think that each segment has different physical regimes.  Nevertheless, our goal here is to well-represent the data, and attempts to combine the segments only leave large systematic residuals, so we are left with presenting the many segments.
	
	 Our analysis shows that for the 32 novae we categorized as smooth, the average of the first slope is -3.9, and the average for the late slope is -5.3.  These slopes can be translated to a power law index $\alpha$, for the flux going as $(T-T_0)^{\alpha}$, by dividing by 2.5.  So the average S class light curve goes as $(T-T_0)^{-1.6}$ for the early times after the peak and goes as $(T-T_0)^{-2.1}$ at late times.  This should be directly comparable to the theoretical prediction of the universal decline law with $\alpha=-1.75$ and $\alpha=-3.0$ for early and late times.  The observed average early-time slope is close to the predicted value, albeit with substantial variations.  The observed average late-time slope is substantially steeper than the earlier slope, but is still substantially shallower than the prediction.  Indeed, only two of our S class novae are as steep as the prediction.  We speculate that the reason for this difference might be due to the rise of the emission lines at late times, such that the V-band flux will not fall off as fast as continuum flux (for which the universal decline law is applicable).
	
	On average, the universal decline law is successful.  However, the early and late slopes show substantial scatter, much more than could be allowed by the physical basis behind the law.  The range of observed early slopes is from -7.8 to -1.3, with an RMS scatter of 1.3.  The range of observed late slopes is from -10.0 to -2.1, with an RMS scatter of 1.8.  (This scatter is for the S class alone, so there are no effects from plateaus, dust dips, cusps, oscillations, flat tops, or jitters. This scatter is also greatly larger than any possible measurement uncertainties.)  We take this large scatter to imply that there must be some additional mechanism that creates variability on top of the universal decline law.

\section{P Class Light Curves}

	The P class follows the stereotypical nova light curve (i.e., the S class light curve) except that as the light curve approaches 3-6 mag below peak, the smooth decline is interrupted by a long-lasting, nearly-flat interval that abruptly ends and is followed by a steeper decline.  Such a flattened interval is a plateau.  These plateaus range in duration from 15 to 500 days and vary slightly in slopes.  
	
	Our sample has 19 P class light curves.  These are displayed in the four panels of Figure 5.  This figure shows the horizontal axis as linear time for the first 500 days, so that details around the peak can be well-resolved.  These same light curves are also shown in the four panels of Figure 6 with the horizontal axis as $\log(T-T_0)$, such that the power law slopes and the breaks are easily visible and the late-time behavior can also be shown.  In a number of cases the plateaus were too subtle to pick up in linear time, but are readily apparent in a log-plot. 
	
	A substantial difficulty is that we have no formal definition for what constitutes a `plateau'.  Specifically, we have no quantitative limits for the flatness or duration.  Our expectation is that a plateau should end with a sudden steepening of the light curve, but magnitudes might not be available to test this for some novae.  A related difficulty is that a high $\Delta$m nova with a secular decline can have partial coverage of its tail produce the appearance of a plateau.  CP Pup is a V1500 Cyg star (Schaefer \& Collazzi 2010) and our light curve here shows a second apparent plateau starting at day 2000.  BY Cir and V1974 Cyg have more than two sections that are substantially flatter than the adjacent segments, therefore leaving ambiguity as to which segment is the actual plateau.  
	
	We must also attempt to decide whether the plateau is a `true plateau' or a `false plateau'.  H\&K06 distinguish a `true plateau' to occur when the continuum radiation is roughly constant, while a `false plateau' occurs when the continuum is fading but the emission lines in the observer's band are increasing relative to the continuum so that the total flux is roughly constant.  Hachisu et al. (2009a) point to the true plateau as being uniquely from RNe (relating to the extended supersoft source illuminating the accretion disk).  In this paper it is hard for us to distinguish between a `true' or `false' plateau because we are using only V-band data, with this band including emission lines that can brighten at late times.  Observations would need to be done in other band filters to see if the plateaus were in the continuum and not just the V-band.  The plateaus for CI Aql, V838 Her, RS Oph, V2487 Oph, and U Sco have already been examined (Hachisu et al. 2000; 2002; 2003; 2009a; Hachisu \& Kato 2009) and found to be true plateaus.
	
	Another problem with late plateaus is that there might be confusion of the nova with a nearby star.  In the case of V838 Her, the visual observers were reporting it to remain at a constant level nearly equal to that of a nearby star of that brightness while CCD observers reported the nova to fade after the end of a real plateau.  The other problematic case is for the late tail of QU Vul, which apparently goes flat at near the brightness of a close neighbor.  However, in this case, many observers, both visual and with CCDs, report the nova as being on the plateau, and we have no grounds for impeaching the observations, so we choose to present the data as received.  The only other questionable case is for T Pyx, but we think that its late plateau is real as there is no possible confusing neighbor star.
	
	We have derived various properties for the P class light curves, tabulated in Table 6.  For the P class, as with the S class, we can only give the results from fits to broken power laws.  Therefore, Table 6 is set up in the same way as Table 5.
	
	In Table 7, we have picked out the information on the plateaus and also present some derived quantities.  The first column is the nova name.  The second and third column list the start time of the plateau relative to $T_0$ ($\Delta T_{plat}$) in days and the duration of the plateau ($D_{plat}$) in days.  The fourth column gives the V-band magnitude of the start of the plateau relative to the peak magnitude ($\Delta V_{plat}$).  The last column gives the slope of the plateau in units of magnitudes per logarithmic time interval.  The plateaus that start more than 100 days after the explosion are the flattest of all the plateaus.  There is an approximate scaling where $\Delta T_{plat} \propto t_3$, with the fast novae having plateaus that start early and the slow novae having plateaus that start late.
	
	The slopes of the plateaus vary from 0.0 (i.e. flat) to -2.4, with a median slope of -0.5.  This median is certainly flat compared to the normal range of slopes for S class novae and for the other segments of P class novae, and this justifies calling them plateaus even though they are not perfectly flat.  The plateaus of RS Oph and U Sco are well supported (Hachisu et al. 2000; 2009a), so their slopes (-2.0 and -2.4 respectively) set a useful limit on the allowed range of plateau slopes.
	
	The P class should follow the universal decline law as predicted by H\&K06, with the initial slope being -4.4 and the late slope being -7.5 when plotted on a graph of magnitude versus $\log(T-T_0)$, with a plateau in the middle.  The first light curve segments have an average slope of -4.1 while the segments just before the plateau have an average slope of -5.2, both of which are in reasonable agreement with the prediction.  The last observed segments have an average slope of -6.7 while the segments immediately after the plateaus have an average slope of -7.9, again in reasonable agreement with the universal decline law.
	
	For just the classical novae alone (excluding the known RNe) from our unbiased sample of 85 novae, the P class contains 13 of 77 novae.  Of the ten known galactic RNe, 6 certainly have plateaus, while up to 9 might have plateaus with the uncertainty arising from poor late-time light curves (S2010).  There is a stark difference in the plateau frequency between classical novae (17\%) and RNe (60-90\%).  Hachisu et al. (2009a) have already suggested that only RNe can show plateaus.  So the obvious reconciliation is that the `classical novae' with plateaus are really RNe (with multiple nova eruptions in the last century) for which only one event has been discovered.  (This is quite plausible, because the discovery efficiency for novae below the naked eye limit is below 10\%, see S2010; Pagnotta et al. 2009.)  With this, we can suspect that the following novae might really be recurrent: V838 Her (see also Hachisu et al. 2009b; Pagnotta 2009), CP Pup, DD Cir, LZ Mus, V1229 Aql, QU Vul, V4633 Sgr, HS Sge, V1974 Cyg, DN Gem, BY Cir, V351 Pup, and V4021 Sgr, in order of $t_3$.  However, three of these systems (CP Pup, V4633 Sgr, and V1974 Cyg) are identified as V1500 Cyg stars, for which an alternative mechanism is proposed to create a light curve plateau.  Indeed, the white dwarf in V1974 Cyg has a mass of only 0.83 M$_{\odot}$ (Chochol 1999), so it cannot be an RN.  So the existence of a plateau in a light curve does not guarantee the nova to be recurrent.  Nevertheless, given the low frequency of the V1500 Cyg stars and the high frequency of short $t_3$ values (and hence a high white dwarf mass), it appears that the majority of the P classical novae are likely to be RNe.
	
	With this, we can make an estimate of the number of RNe hiding in D\&SCat.  That is, some number of RNe (with more than one eruption in the last century) will have only {\it one} discovered eruption and be mistakenly labeled as a `classical nova'.  With 17\% of the well-observed classical novae in our unbiased sample having plateaus, and likely the majority of these being recurrent, we can approximate the fraction with RN plateaus as 10\%.  With only 60-90\% of RNe showing a plateau, the RN fraction should be increased to between $\sim10/0.9$=11\% and $\sim10/0.6$=17\%.  The D\&SCat has $\sim$250 novae with one known eruption.  So that means that there are roughly 28-42 RNe hiding in the list of classical novae.  Alternative arguments (based on the discovery efficiencies of novae) suggest that the number of hiding RNe is more like 75 (S2010).  In either case, the only convincing proof will be to discover prior undetected eruptions in archival data.

\section{D Class Light Curves}

		The D class novae seem to have their initial decline with substantial jitter and a less-steep decline than the S and P classes.  The defining feature of D class light curves is a dust dip, which starts suddenly after the initial decline.  Dust dips appear as steep drops in the brightness, although the depth of this drop can vary.  The dust dips then reach some minimum brightness, followed by a recovery in brightness to a fainter level than the start of the dip, followed by a slow decline.
		
	The classic D light curve is that of DQ Her.  However, dust dips occur with a very wide range of depths and durations.  The dust dip of V445 Pup is over ten magnitudes in depth, taking the star to over four magnitudes fainter than its normal quiescent level, with the dip remaining nearly this deep at least until 2009 (Woudt et al. 2009).  At the other extreme are shallow dips (e.g., OS And, V476 Cyg, and NQ Vul) for which we can question whether the shape is even produced by a dust dip.
		
 	Our sample has 16 D class light curves.  These are displayed in the four panels of Figure 7.  This figure shows the horizontal axis as linear time, and only covers the first 500 days, so that details around the peak and dip can be well-resolved.  Again, we have repeated all the light curves in Figure 8 with the horizontal axis as $\log(T-T_0)$.  

	We have derived various properties about the D light curves.  In Table 8, we give the results from fits to broken power laws to the pre- and post-dip segments of the light curve, as well as the several parameters relating to the dip.  The first column gives the name of the nova, while the second column gives the initial slope (in units of magnitudes per logarithmic time).  The next six columns give pairs of numbers, each pair consisting of a time with respect to $T_0$ ($\Delta$T in days) and a magnitude ($V$), for the start of the dip, the minimum, and the end of the dip.  The ninth column presents the slope for the first segment of the post-dip decline.  Columns 10-12 report the time of the next break, the magnitude of that break, and the slope after that break.  Columns 13-15 report the next break, magnitude, and slope, as needed.  For three of the novae, the light curve has further significant breaks, which are specified in footnotes.

	Examination of the figures shows the D class light curves that have the deeper dips coming later in time, and with flatter initial declines.  This can be seen in each of the panels of Figure 7, for example in the second panel, where the light curves from bottom to top have deeper dips, later dips, and flatter initial slopes.  The correlation coefficient between $\log (\Delta T_{start})$ and $V_{min}-V_{start}$ is 0.72.  This is a significant correlation, with only a 0.18\% probability (corresponding to over a 3-sigma confidence level) by chance alone.  The correlation coefficient between the initial slope and $V_{min}-V_{start}$ is 0.49.  This corresponds to a probability of 5\% (i.e., the 2-sigma confidence level) which is suggestive but not convincing.  We speculate that a dip might start later in time if the ejecta velocity is slow, and that the ejected gas would then reach the temperature of dust condensation at a time when the gas density (and hence the dust density) is thicker so as to make for a deeper dip.
	
	V445 Pup displays an extreme dust dip.  It is unique as being the only known `helium nova' (Ashok \& Banerjee 2003; Kato \& Hachisu 2003), and also for having a bipolar nova shell (Woudt et al. 2009).  We might expect a particularly deep dust dip both because the metallicity of the ejecta is high and because we are apparently in the equatorial plane of the bipolar ejecta.  In the infrared, the minimum of the dust dip occurs around 600 days after the explosion, while in both the optical and infrared the depth of the minimum is over six magnitudes below the pre-eruption quiescent level.  Despite being extreme and unique, V445 Pup fits well into the pattern that the deeper dust dips come later in time. 
	
	We also see that the D class light curves take the longest time to reach quiescence out of all the classes.  The D class novae have an average $T_Q$ of 9000 days, while the other nova classes have an average $T_Q$ of 3400 days.  All but one of the D novae (V705 Cas is the exception) have a higher $T_Q$ than all but one of the non-D novae (CP Pup being the exception).  This statement can be transformed into a quantitative analysis by the Kolmogorov-Smirnoff test, with the result being that the difference in populations is highly significant.  This is further illustrated by the average $t_9$ values being 1850 and 840 days for the D class and the non-D classes respectively.  However, the two groups do not have significantly different average $t_6$ values, so the difference is entirely in the late tails of the light curve.

\section{C Class Light Curves}

	The C class novae have a fast rise, and a quick initial decline.  The novae of the C class have a characteristic secondary maximum 1-8 months after the primary peak, with this extra light appearing to be added on top of the basic S class light curve.  These cusps in the light curve have a distinctive shape where the additive component starts near zero, then rises at an accelerating rate until it turns off sharply, giving the light curve its ÔcuspÕ shape.    Hachisu \& Kato (2009) have proposed that the cusp is caused by the input of magnetic energy from rotating white dwarfs.  The sharp drop in the cusp is associated with the sudden formation of dust (Lynch et al. 2008).
	
	Our original sample contained only one C class nova (V1493 Aql), which was discovered in 1999.  No other examples were found until 2006 (V2362 Cyg) and 2008 (V2491 Cyg), which are beyond our original cutoff date of early 2006 from D\&SCat.  So now with three examples, we have a distinctive class.  These light curves are displayed in Figure 9 (for linear time on the horizontal axis) and in Figure 10 (for logarithmic time on the horizontal axis).   In these plots, a power law segment is interpolated through the cusp so that the added light can be seen.
	
	We have derived various properties of the C light curves, with these being tabulated in Table 9.  The first column gives the name of the nova, while the second column gives the initial slope (in units of magnitudes per logarithmic time).  The third and fourth columns give the time after $T_0$ ($\Delta T_{min}$) and the magnitude of the minimum ($V_{min}$) as the brightness starts rising to the cusp.  The next two columns give the time and magnitude of the peak of the cusp ($\Delta T_{cusp}$ and $V_{cusp}$).  Columns seven and eight give the time and magnitude of the base of the cusp where the added light goes to zero and the light curve resumes its interrupted decline ($\Delta T_{base}$ and $V_{base}$).  Then in the usual format, columns 9-12 give the slope after the base of the cusp, the next break time ($\Delta T_{b1}$, the V-band magnitude at that break point ($V_{b1}$), and the slope after that break.
	
	It is striking that the amplitude of the cusp appears correlated with the  time after peak of the cusp maximum.  For V2362 Cyg, V1493 Aql, and V2491 Cyg, the cusp amplitudes ($V_{min}-V_{cusp}$) are correlated with the time of the cusp maximum ($\log(\Delta T_{cusp})$) with a correlation coefficient of 0.991.  That is a very good correlation, but with only three novae it is still only a 2-sigma result.  Certainly, we will need further examples to follow this correlation before we can get any strong confidence in it.

\section{O Class Light Curves}

	O class novae follow the basic S class shape but have a time interval after the peak where the brightness varies with moderate amplitude in an apparently periodic and quasi-sinusoidal motion.  Outside of the oscillations, the light curve looks like an S class event.  Oscillations can be distinguished from jitters (or flares) in the light curve by three means.  First, oscillation peak times are quasi-periodic, whereas jitters are apparently randomly timed.  Second, oscillations generally start around three magnitudes below peak, whereas jitters are prominent from before the peak until around the $t_3$ time.  Third, oscillations extend from below to above the smoothed light curve as interpolated from before-to-after the interval, whereas jitters apparently only flare brighter than the base level of the light curve.
	
	The classic examples of the O class light curve are GK Per and V603 Aql.  Both are very well-observed and are nearly the all-time brightest novae from near the start of the previous century.  We had originally been suspicious about the existence of the oscillations, partly because we can imagine problems that could make apparent sinusoidal artifacts, but mainly because no further examples were found after 1918.  However, after examinations of the Harvard plates by A. Pagnotta and the presence of the oscillations on multiple independent data sets, we are satisfied that the oscillations are real.  In addition, we have found three more examples in the AAVSO data base (V1494 Aql, V888 Cen, and V2467 Cyg), even though these new cases are not as well-observed nor are the oscillations as regularly placed.
	
	Our sample has five O class light curves.  These are displayed in Figure 11 (for the linear plot) and in Figure 12 (for the logarithmic plot).  Close-ups of the oscillation sections of each light curve, with the overall power law decline subtracted out, are displayed in Figure 13.
	
	We have derived various properties of the O light curves, with these being tabulated in Table 10.  The first nine columns have the same format and content as in Table 5 for the S class novae. The last two columns then tell when the oscillations start ($\Delta T_{start}$) and end ($\Delta T_{end}$)  with respect to $T_0$ in days.  We have also identified the time and magnitude for the maxima and minima of each oscillation, as given in Table 11.  The first column gives the nova, the second and third columns list the time after $T_0$ (in days) and the V-band magnitude for each maximum, while the last two columns list the same for each minimum.
	
	The O light curves appear to be the same as S light curves, except that there are oscillations superposed.  With this, we expect that the universal decline law will be applicable.  The average slope over the first segments is -3.9, while the average slope over the last segments is -7.6.  This is close (given the few O light curves and their scatter) to the averages for the S class and to the theoretical predictions (-4.4 and -7.5 respectively).  The scatter is these slopes is substantial, and this implies that there must be other physical mechanisms operating that superpose variations on top of the basic universal decline law.
	
	At first glance, the oscillations appear to be regularly spaced in time.  Figure 14 plots the time between successive peaks ($T_{i+1}-T_i$) as a function of the time of the {\it ith} peak ($T_i$).  The use of time differences (as opposed to cumulative time measures) limits the problems that might arise from missing or extra peaks.  For example, if a single peak is missed, then one point will be roughly twice as high on the plot and perhaps recognized as an outlier.  If the peaks are linearly spaced in time, then $T_{i+1}-T_i$ will be nearly a constant.  If the peaks are uniformly spaced in logarithmic-time, then $T_{i+1}-T_i$ will be rising as an exponential.  A linearly rising $T_{i+1}-T_i$ curve would imply that the {\it ith} peaks are spaced quadratically in time.

	The most famous O class light curve is for GK Per, and the oscillations have a typical large amplitude of 1.5 mag.  These oscillations are indeed quite regular, with the average interval being 4.1 days, with an RMS scatter of 1.2 days (a 30\% typical variation).  From Figure 14, we see that the overall curve for GK Per has no trend or slope, and this says that the oscillations are linear in time.  Pejcha (2009) ignored all the oscillations except the first four and concluded that the interval between peaks was increasing exponentially with time, but this is refuted in his Figure 3 (and our Figure 14). The first two intervals are fairly short, but not as short as the last interval.
	
	The only other well-known case of an O class nova is V603 Aql.  This case has a different waveform (more sinusoidal, yet with relatively narrow minima) than for GK Per (fairly narrow peaks, with most of the time spent near minimum).  V603 Aql also has fairly regular oscillations, with an average inter-peak interval of 10.7 days and an RMS scatter of 1.5 days (a 14\% typical variation).  Again, the $T_{i+1}-T_i$ curve is consistent with being flat, implying that the oscillations are linear in time.
	
	V1494 Aql displays substantial variations in both the shape of the waveform (see Fig. 13) and in the time intervals between peaks (see Fig. 14).  The inter-peak intervals vary from 4 to 17 days, with the average being 10 days and an RMS scatter of 4.6 days (a 47\% typical variation).  With this, we have to consider whether V1494 Aql is O class or J class.  We have placed the nova in the O class because the peaks are substantially more regular than the jitters of J class, because the variations extend below the interpolated light curve as characteristic of the O class (see Figure 13), and because the time interval around peak is free of the variations (see Fig. 12).  Even with the substantial scatter, the $T_{i+1}-T_i$ curve displays no significant trend, and so we take the underlying process to approximately produce peaks at linearly spaced times.

	V888 Cen follows the GK Per example of having fairly narrow peaks for its waveform.  The average inter-peak interval is 14 days, with an RMS scatter of 4.7 days (a 34\% typical variation).  Again, the lack of trends in Figure 14 implies a linear timing mechanism.
	
	V2467 Cyg has oscillations, as previously noted by Poggiani (2009) as based on amateur data.  The average inter-peak interval is 19 days, with an RMS scatter of 4 days (a 21\% typical variation.  Again, the inter-peak times show no continuing trend with time (Fig. 14), so that it looks like the inter-peak time is a constant with noise added.
	
	We have to wonder whether there are more O class light curves that remain unrecognized due to incomplete data.  The best possibility that we know of is T Pyx, which is known to have displayed large amplitude variations on a one-day time scale (Landolt 1970).  We do not think that any of the J class light curves are really O class, because all the ones in our catalog do not have the three distinctions identified at the start of this section.  In all, we do not think that the fraction of novae in the O class can be much larger than the 4\% that we estimated at the end of Section 3. 

\section{F Class Light Curves}

	F class novae seem to be ordinary S class novae except that they have an extended interval of nearly constant magnitude at peak.  BT Mon and V849 Oph have a flat peak for about two months, and DO Aql has one of about 7 months.  V2295 Oph has a flat top that last only 11 days, or perhaps much longer before discovery.  These flat-topped light curves are distinct from all others.  We know of no theoretical prediction or explanation for the long-lasting flat tops.
	
	In our basic AAVSO sample, V849 Oph and V2295 Oph are the only well-observed examples of novae with a flat top.  The light curves of BT Mon and DO Aql are sufficiently similar to each other and they are distinct from all the other light curves, so we have added B-band light curves from the literature for these two novae.  Now, our sample has 4 F class light curves.  These are displayed in Figure 15 (for linear time) and Figure 16 (for logarithmic-time).    The F light curves can be represented by simple power law segments.  We have presented our fitted segments in Table 12, with the same format as for Table 5.
	
	The only commonly known example of a flat topped light curve is the symbiotic nova PU Vul.  PU Vul was discovered at close to ninth magnitude and stayed constant at V=9.0 for eleven months.  Then a spectacular dip dimmed the star to V=13.6, with a total duration of 16 months and a full recovery to V=8.6.  The star remained constant at V=8.6 from middle 1981 until late 1987.  This well-observed flat topped light curve lasted either 78 months or 105 months, depending on whether the interval of the dip and before is included.  From 1988 to 1994 PU Vul slowly and steadily faded to V=11.5, from 1995 to 2001 it was constant at V=11.5, and from 2002 to present the system has been slowly fading to V=12.7.  This light curve is completely unique and greatly different from all other known novae (including all symbiotic novae).  As a symbiotic nova, PU Vul is greatly different from the classical novae in that it has a many-year orbital period.  The deep dip in 1980 appears like a dust dip, but spectra show this to likely be an eclipse, for which later small dips in the light curve suggest an orbital period of 13.42 year (Garnavich \& Trammell 1994; Nussbaumer \& Vogel 1996), although this eclipse hypothesis has great difficulties in explaining the `eclipse' depths and shapes.  Since the early 1990s, PU Vul shows quasi-periodic oscillations with an amplitude of half a magnitude and periodicities including $\sim$211 days (Yoon \& Honeycutt 2000).  The quiescent brightness before the eruption was V$\sim$13, with the eruption amplitude then $<$5 magnitudes.

\section{J Class Light Curves}

	J light curves have the same basic power law declines as the S class, except that there is substantial jittering above the base level. These jitters are flare-ups of substantial amplitude usually greater than half a magnitude.  The J class jitters extend above a smoothed light curve and are often isolated flickers.  Generally, the jitter shape is a symmetrical and sharp-topped flare, with the majority of the light curve at some base level (defining a shape similar to the S class light curves) with occasional superposed jitters.  These jitters do {\it not} occur in the late tail of nova light curves, and indeed, almost all jitters are when the base light curve is within three magnitudes below peak.
	
	When sorting through the J class novae, we think that we can distinguish two subclasses based on whether the jitters are visible only near the peak.  One subclass has jittering only up near the peak (e.g., V723 Cas, HR Del, and V2540 Oph), while the second subgroup has jittering sporadically throughout the light curve up until roughly $t_3$ after the peak (e.g., V1039 Cen, DK Lac, and V4745 Sgr).  We do not know how to quantify the significance of the division into these two subclasses, so this possibility must remain a suggestion.
	
	Our sample has 14 J class light curves.  These are displayed in the three panels of Figure 17 (for linear time) and Figure 18 (for logarithmic time).  Figure 19 gives close-ups for three J light curves with the base level subtracted out so that the individual jitters can be readily examined.  We have fitted the base levels of the J light curves to broken power laws, with the results in Table 13.  The content and format of the first nine columns is the same as in Table 5.  The last three columns give the times (with respect to $T_0$ in days) of the start of the interval with jitters ($\Delta T_{start}$), the end of that interval ($\Delta T_{end}$), and the number of flares with amplitude $>0.5$ mag ($N_{jitter}$).  We have also derived various properties of the jitters and flares themselves in Table 14.  The first column gives the nova.  The next five columns give the time after $T_0$ ($\Delta T_{max}$), the time interval between successive maxima ($\Delta T_j$), the V-band peak magnitude ($V_{max}$), the difference in magnitude between the peak and the base level ($\Delta V_{max}$), and the full-width at half-maximum ($FWHM$ in days) for each jitter.
	
	The jitters appear to be scattered at random within the intervals over which they occur.  That is, they are neither periodic nor clustered, and they have occasional large inter-arrival times and occasional small inter-arrival times.  The inter-arrival times are the times between successive jitters ($\Delta T_j$).  For jitter times that are random with a Poisson distribution, the inter-arrival times should have an exponential distribution.  That is, the frequency of inter-arrival times should fall off as an exponential function, with short inter-arrival times being the most common and long inter-arrival times being uncommon.  This can be tested by constructing a distribution function for the inter-arrival times, plotting this on log-frequency versus $\Delta T_j$ axes, and seeing whether it is well fit by a straight line.  Each nova has so few jitters that this test must necessarily be crude.  DK Lac, which, with 14 jitters is the best case, does have an apparently linear plot.  We can get better statistics by making a distribution for all the nova inter-arrival times lumped together (at the cost of smearing together distributions with different averages), and the plot again appears linear to within the error bars, so, it appears that the timing of the jitters is random within the interval over which they occur.
	
	Pejcha (2009) claims that the jitters have intervals between successive peaks that increase greatly with time since the maxima.  In particular, his plot of $\log (\Delta T_j)$ versus $\log (\Delta T_{max})$ shows a nearly linear trend with a slope near unity, so that $\Delta T_j \propto \Delta T_{max}$.  With this, jitters ten days after peak should have a ten times smaller interval than jitters a hundred days after peak.  However, scanning down the values in the second and third columns of Table 14, we do not see this effect.  The only two exceptions are that V4745 Sgr has a short interval at the first and a long interval at the end, while DK Lac has a vague upward trend.  This is in contrast to the plot of Pejcha (2009) which shows both V4745 Sgr and DK Lac having tight linear correlations with a slope of near unity.  As we are operating off the same data, we can only think that the conclusion is very sensitive to the selection of which peaks to include.  With these two cases being questionable, and all other J class light curves showing that $\Delta T_j$ does not depend on position within the burst, we can only conclude that the jitters appear randomly in time with no trends.

\section{Fundamental Open Questions}

	In this paper, we have collected a very large set of well-observed nova light curves, with the number and coverage being greatly better than all previous collections.  We have then used our nova light curve catalog to identify distinct classes and to provide quantitative measures of many light curve properties.  In this process, we have identified many open questions.  Many of these questions address fundamental properties of nova light curves for which we are not aware of any theoretical speculation, and so these open questions can provide a list of key challenges for theorists.
	
	The universal decline law of H\&K06 provides a good and general explanation of the overall shape of light curves.  Nevertheless, even ignoring the effects of the various plateaus, dust dips, cusps, oscillations, flat tops, and jitters, the {\it scatter} in the early and late slopes is much larger than can be explained by either measurement error or emission line fluxes.  We take this to mean that there must be some other mechanism that is operating, and this extra process makes for substantial variations.  What is this extra process?
	
	Oscillations have been well known ever since GK Per in 1901.  No plausible model has been proposed (see discussion in Section 4).  We have found that the oscillations are quasi-periodic (not logarithmically spaced), they vary up to a factor of three in brightness, and they are associated with high magnetic field white dwarfs.  What is the physical mechanism for the oscillations?
	
	What causes the flat topped light curves?  We guess that the the constant brightness implies steady hydrogen burning, but this goes against the idea that nova peaks are caused by explosive burning.  Even if a thermonuclear runaway can be avoided, we do not see how fresh hydrogen can be made available at a steady rate, nor how partial burning can either be steady or produce novae with average peak luminosities.
	
	What causes the jitter around the peak of nova light curves?  A significant fraction of novae display sudden and random large-amplitude flares that carry a lot of energy, so some sort of an explanation is needed.  
	
	At a higher level, what is the root physical cause for why a nova light curve produces one or another of the seven classes?  In principle, a nova can be completely specified by only a few parameters (the white dwarf mass, the white dwarf magnetic field strength, the accretion rate, and the composition of the accreted material), so what is needed is to map the light curve classes into this parameter space.
	
	What fraction of the `classical novae' (with one discovered eruption) are actually RNe (with more than one eruption within the last century), and which nova systems are recurrent?  This question is vital for understanding whether RNe are progenitors of Type Ia supernovae, because the fraction is needed to derive the number and death rate of RNe in our galaxy.
	
	Our work has raised open questions about many smaller issues:  Why do some novae have visible shells while others do not?  What causes the anticipatory phenomenon (pre-eruption rises and dips) where the accretion rate changes in the months-years {\it before} the thermonuclear runaway is triggered?  What causes the unique secondary maximum shown by T CrB?  Why do the big amplitude cusps apparently come later in the eruption?  Why do the deeper dust dips come later during the eruption, and have a shallower initial slope?
	
	This section has summarized the many fundamental questions that provide a challenge and opportunity for theorists.  But this paper only addresses the morphology of the eruption light curves, and much work is needed from observers in the nova community for a variety of overviews and syntheses of data.  The large body of modern spectra allows advances on the overviews of Payne-Gaposhkin (1964) and Williams (1992), for example paying attention to line profiles and abundances, as well as providing an expansion of nova classification to include the spectra.  Another observational task is to provide an overview, synthesis, and classification of nova light curve {\it in quiescence}, where a large amount of data is available from the AAVSO and from various widely scattered sources.  With the wonderful advances in theory and data over the last decade, we are living in a `golden age' for nova studies, or at least until the next decade.

~

The many AAVSO observers over the last century have provided a wonderful data set, with their work providing by far the best collection of nova light curves.  We are grateful and thankful for their many sleepless nights.  We thank Mariko Kato, Izumi Hachisu, Fred Walter, and Ashley Pagnotta for their help in improving this paper.    This work is supported under a grant from the National Science Foundation (AST 0708079).

{}

\begin{deluxetable}{llllllllllll}
\tabletypesize{\scriptsize}
\tablewidth{0pc}
\tablecaption{Cataloged Novae and Their Primary Light Curve Properties}
\tablehead{\colhead{Nova} & \colhead{Year} & \colhead{$T_0$} & \colhead{$T_{peak}$} & \colhead{$V_{peak}$} & \colhead{$t_2$} & \colhead{$t_3$} & \colhead{$t_6$} & \colhead{$t_9$} & \colhead{$V_Q$}	&	\colhead{$T_Q$}&	\colhead{LC Class}}
\startdata
OS And	&	1986	&	2446767	&	2446773	&	6.5	&	11	&	23	&	199	&	770	&	17.5	&	\ldots	&	D(23)	\\
CI Aql      	&	2000	&	2451660	&	2451672	&	9.0	&	25	&	32	&	417	&	\ldots	&	16.7	&	550	&	P(32)	\\
DO Aql	&	1925	&	2424411	&	2424470	&	8.5	&	295	&	900	&	1280	&	\ldots	&	18.0	&	\ldots	&	F(900)	\\
V356 Aql	&	1936	&	2428399	&	2428445	&	7.0	&	127	&	140	&	580	&	\ldots	&	18.3	&	\ldots	&	J(140)	\\
V528 Aql	&	1945	&	2431693	&	2431696	&	6.9	&	16	&	38	&	119	&	\ldots	&	18.5	&	\ldots	&	S(38)	\\
V603 Aql	&	1918	&	2421749	&	2421755	&	-0.5	&	5	&	12	&	149	&	743	&	11.7	&	6800	&	O(12)	\\
V1229 Aql	&	1970	&	2440689	&	2440690	&	6.6	&	18	&	32	&	56	&	\ldots	&	18.1	&	\ldots	&	P(32)	\\
V1370 Aql	&	1982	&	2444998	&	2444999	&	7.7	&	15	&	28	&	199	&	\ldots	&	18.0	&	\ldots	&	D(28)	\\
V1419 Aql	&	1993	&	2449124	&	2449130	&	7.6	&	25	&	32	&	486	&	2000	&	$\sim$21	&	\ldots	&	D(32)	\\
V1425 Aql	&	1995	&	2449755	&	2449757	&	8.0	&	27	&	79	&	564	&	1276	&	$\sim$20	&	\ldots	&	S(79)	\\
V1493 Aql	&	1999	&	2451372	&	2451374	&	10.1	&	9	&	50	&	170	&	\ldots	&	$\sim$21	&	\ldots	&	C(50)	\\
V1494 Aql	&	1999	&	2451514	&	2451516	&	4.1	&	8	&	16	&	170	&	478	&	17.1	&	2500	&	O(16)	\\
T Aur	&	1891	&	2412077	&	2412083	&	4.5	&	80	&	84	&	96	&	6300	&	14.9	&	\ldots	&	D(84)	\\
V705 Cas	&	1993	&	2449325	&	2449340	&	5.7	&	33	&	67	&	75	&	1529	&	16.4	&	2500	&	D(67)	\\
V723 Cas	&	1995	&	2449956	&	2450072	&	7.1	&	263	&	299	&	1215	&	\ldots	&	15.7	&	3000\tablenotemark{a}	&	J(299)	\\
V842 Cen	&	1986	&	2446756	&	2446760	&	4.9	&	43	&	48	&	60	&	1213	&	15.8	&	\ldots	&	D(48)	\\
V868 Cen	&	1991	&	2448347	&	2448349	&	8.7	&	31	&	82	&	450	&	\ldots	&	19.9	&	\ldots	&	J(82)	\\
V888 Cen	&	1995	&	2449771	&	2449774	&	8.0	&	38	&	90	&	302	&	\ldots	&	15.2	&	400	&	O(90)	\\
V1039 Cen	&	2001	&	2452185	&	2452185	&	9.3	&	25	&	174	&	850	&	\ldots	&	$\sim$21	&	\ldots	&	J(174)	\\
BY Cir      	&	1995	&	2449745	&	2449746	&	7.4	&	35	&	124	&	628	&	\ldots	&	17.9	&	\ldots	&	P(124)	\\
DD Cir     	&	1999	&	2451410	&	2451415	&	7.6	&	5	&	16	&	\ldots	&	\ldots	&	20.2	&	\ldots	&	P(16)	\\
V693 CrA	&	1981	&	2444696	&	2444698	&	7.0	&	10	&	18	&	\ldots	&	\ldots	&	$>$21	&	\ldots	&	S(18)	\\
T CrB	&	1946	&	2431860	&	2431860	&	2.5	&	4	&	6	&	15	&	\ldots	&	9.8	&	93	&	S(6)	\\
V476 Cyg	&	1920	&	2422559	&	2422560	&	1.9	&	6	&	16	&	43	&	1076	&	16.2	&	14000	&	D(16)	\\
V1330 Cyg	&	1970	&	2440761	&	2440767	&	9.9	&	161	&	217	&	\ldots	&	\ldots	&	17.5	&	\ldots	&	S(217)	\\
V1500 Cyg	&	1975	&	2442654	&	2442655	&	1.9	&	2	&	4	&	32	&	263	&	17.9	&	3400\tablenotemark{a}	&	S(4)	\\
V1668 Cyg	&	1978	&	2443761	&	2443766	&	6.2	&	11	&	26	&	219	&	948	&	19.7	&	\ldots	&	S(26)	\\
V1819 Cyg	&	1986	&	2446645	&	2446648	&	9.3	&	95	&	181	&	1073	&	\ldots	&	17.0	&	\ldots	&	J(181)	\\
V1974 Cyg	&	1992	&	2448671	&	2448675	&	4.3	&	19	&	43	&	321	&	847	&	16.9	&	4000\tablenotemark{a}	&	P(43)	\\
V2274 Cyg	&	2001	&	2452103	&	2452115	&	11.5	&	22	&	33	&	200	&	\ldots	&	$>$20	&	\ldots	&	D(33)	\\
V2275 Cyg	&	2001	&	2452139	&	2452142	&	6.9	&	3	&	8	&	75	&	472	&	18.4	&	\ldots	&	S(8)	\\
V2362 Cyg	&	2006	&	2453829	&	2453832	&	8.1	&	9	&	246	&	312	&	884	&	$>$21	&	\ldots	&	C(246)	\\
V2467 Cyg	&	2007	&	2454174	&	2454175	&	7.4	&	8	&	20	&	207	&	917	&	$\sim$19	&	\ldots	&	O(20)	\\
V2491 Cyg	&	2008	&	2454567	&	2454568	&	7.5	&	4	&	16	&	38	&	196	&	$\sim$20	&	\ldots	&	C(16)	\\
HR Del	&	1967	&	2439645	&	2439838	&	3.6	&	167	&	231	&	1387	&	\ldots	&	12.1	&	6000	&	J(231)	\\
DN Gem	&	1912	&	2419472	&	2419477	&	3.6	&	16	&	35	&	447	&	1395	&	15.6	&	\ldots	&	P(35)	\\
DQ Her	&	1934	&	2427786	&	2427794	&	1.6	&	76	&	100	&	645	&	2010	&	14.3	&	12000	&	D(100)	\\
V446 Her	&	1960	&	2437003	&	2437004	&	4.8	&	20	&	42	&	197	&	993	&	16.1	&	\ldots	&	S(42)	\\
V533 Her	&	1963	&	2438056	&	2438061	&	3.0	&	30	&	43	&	238	&	688	&	15.0	&	1800	&	S(43)	\\
V827 Her	&	1987	&	2446820	&	2446831	&	7.5	&	21	&	53	&	394	&	\ldots	&	18.1	&	\ldots	&	S(53)	\\
V838 Her	&	1991	&	2448339	&	2448341	&	5.3	&	1	&	4	&	13	&	40	&	19.1	&	\ldots	&	P(4)	\\
CP Lac	&	1936	&	2428337	&	2428340	&	2.0	&	5	&	9	&	78	&	436	&	15.0	&	4300	&	S(9)	\\
DK Lac	&	1950	&	2433303	&	2433309	&	5.9	&	55	&	202	&	488	&	\ldots	&	13.8	&	2100	&	J(202)	\\
LZ Mus	&	1998	&	2451175	&	2451177	&	8.5	&	4	&	12	&	129	&	\ldots	&	$>$18	&	\ldots	&	P(12)	\\
BT Mon	&	1939	&	2429516	&	2429518	&	8.1	&	118	&	182	&	610	&	\ldots	&	15.7	&	5000	&	F(182)	\\
IM Nor   	&	2002	&	2452264	&	2452289	&	8.5	&	50	&	80	&	253	&	471	&	18.3	&	761	&	P(80)	\\
RS Oph	&	2006	&	2453778	&	2453779	&	4.8	&	7	&	14	&	88	&	\ldots	&	11.0	&	93	&	P(14)	\\
V849 Oph	&	1919	&	2422179	&	2422216	&	7.6	&	140	&	270	&	2010	&	\ldots	&	18.8	&	\ldots	&	F(270)	\\
V2214 Oph	&	1988	&	2447248	&	2447262	&	8.5	&	60	&	89	&	777	&	\ldots	&	20.5	&	\ldots	&	S(89)	\\
V2264 Oph	&	1991	&	2448357	&	2448361	&	10.0	&	22	&	30	&	\ldots	&	\ldots	&	$>$21	&	\ldots	&	S(30)	\\
V2295 Oph	&	1993	&	2449091	&	2449096	&	9.3	&	9	&	16	&	129	&	\ldots	&	$>$21	&	\ldots	&	F(16)	\\
V2313 Oph	&	1994	&	2449504	&	2449506	&	7.5	&	8	&	17	&	281	&	\ldots	&	$>$20	&	\ldots	&	S(17)	\\
V2487 Oph	&	1998	&	2450979	&	2450980	&	9.5	&	6	&	8	&	52	&	\ldots	&	17.7	&	\ldots	&	P(8)	\\
V2540 Oph	&	2002	&	2452299	&	2452373	&	8.1	&	66	&	115	&	\ldots	&	\ldots	&	$>$21	&	\ldots	&	J(115)	\\
GK Per	&	1901	&	2415437	&	2415440	&	0.2	&	6	&	13	&	179	&	444	&	13.0	&	\ldots	&	O(13)	\\
RR Pic	&	1925	&	2424298	&	2424309	&	1.0	&	73	&	122	&	1008	&	5805	&	12.2	&	16000	&	J(122)	\\
CP Pup	&	1942	&	2430672	&	2430676	&	0.7	&	4	&	8	&	114	&	748	&	$>$19.5	&	5600\tablenotemark{a}	&	P(8)	\\
V351 Pup	&	1991	&	2448617	&	2448618	&	6.4	&	9	&	26	&	436	&	\ldots	&	19.6	&	\ldots	&	P(26)	\\
V445 Pup	&	2000	&	2451872	&	2451877	&	8.6	&	215	&	240	&	\ldots	&	\ldots	&	14.6	&	\ldots	&	D(240)	\\
V574 Pup	&	2004	&	2453330	&	2453331	&	7.0	&	12	&	27	&	199	&	\ldots	&	17.2	&	840	&	S(27)	\\
T Pyx	&	1967	&	2439466	&	2439506	&	6.4	&	32	&	62	&	175	&	\ldots	&	15.5	&	830	&	P(62)	\\
HS Sge	&	1977	&	2443149	&	2443151	&	7.2	&	15	&	21	&	537	&	\ldots	&	20.7	&	\ldots	&	P(21)	\\
V732 Sgr	&	1936	&	2428275	&	2428286	&	6.4	&	65	&	75	&	481	&	\ldots	&	$\sim$16	&	\ldots	&	D(75)	\\
V3890 Sgr	&	1990	&	2448007	&	2448009	&	8.1	&	6	&	14	&	28	&	\ldots	&	15.5	&	\ldots	&	S(14)	\\
V4021 Sgr	&	1977	&	2443229	&	2443234	&	8.9	&	56	&	215	&	570	&	\ldots	&	18.0	&	\ldots	&	P(215)	\\
V4160 Sgr	&	1991	&	2448457	&	2448467	&	7.0	&	2	&	3	&	22	&	\ldots	&	$>$19	&	\ldots	&	S(3)	\\
V4169 Sgr	&	1992	&	2448806	&	2448814	&	7.9	&	24	&	36	&	375	&	\ldots	&	$>$17	&	\ldots	&	S(36)	\\
V4444 Sgr	&	1999	&	2451293	&	2451295	&	7.6	&	5	&	9	&	61	&	\ldots	&	$>$21	&	\ldots	&	S(9)	\\
V4633 Sgr	&	1998	&	2450893	&	2450896	&	7.4	&	17	&	44	&	601	&	\ldots	&	18.7	&	3000\tablenotemark{a}	&	P(44)	\\
V4643 Sgr	&	2001	&	2451963	&	2451965	&	7.7	&	3	&	6	&	43	&	\ldots	&	$>$16	&	\ldots	&	S(6)	\\
V4739 Sgr	&	2001	&	2452147	&	2452149	&	7.2	&	2	&	3	&	17	&	\ldots	&	$>$18	&	\ldots	&	S(3)	\\
V4740 Sgr	&	2001	&	2452157	&	2452158	&	6.7	&	18	&	33	&	150	&	\ldots	&	$>$18	&	\ldots	&	S(33)	\\
V4742 Sgr	&	2002	&	2452531	&	2452534	&	7.9	&	9	&	23	&	250	&	\ldots	&	$>$18	&	\ldots	&	S(23)	\\
V4743 Sgr	&	2002	&	2452535	&	2452538	&	5.0	&	6	&	12	&	170	&	680	&	16.8	&	\ldots	&	S(12)	\\
V4745 Sgr	&	2003	&	2452738	&	2452743	&	7.3	&	79	&	190	&	\ldots	&	\ldots	&	$>$17	&	\ldots	&	J(190)	\\
V5114 Sgr	&	2004	&	2453079	&	2453082	&	8.1	&	9	&	21	&	103	&	\ldots	&	$>$21	&	\ldots	&	S(21)	\\
V5115 Sgr	&	2005	&	2453456	&	2453460	&	7.9	&	7	&	13	&	134	&	\ldots	&	$>$18	&	\ldots	&	S(13)	\\
V5116 Sgr	&	2005	&	2453556	&	2453557	&	7.6	&	12	&	26	&	\ldots	&	\ldots	&	$>$16	&	\ldots	&	S(26)	\\
U Sco	&	1999	&	2451235	&	2451235	&	7.5	&	1	&	3	&	11	&	\ldots	&	17.6	&	\ldots	&	P(3)	\\
V992 Sco	&	1992	&	2448763	&	2448769	&	7.7	&	100	&	120	&	765	&	\ldots	&	17.2	&	\ldots	&	D(120)	\\
V1186 Sco	&	2004	&	2453188	&	2453205	&	9.7	&	12	&	62	&	\ldots	&	\ldots	&	$>$18	&	\ldots	&	J(62)	\\
V1187 Sco	&	2004	&	2453218	&	2453223	&	9.8	&	10	&	17	&	\ldots	&	\ldots	&	18.0	&	\ldots	&	S(17)	\\
V1188 Sco	&	2005	&	2453575	&	2453578	&	8.9	&	11	&	23	&	65	&	\ldots	&	$>$19	&	\ldots	&	S(23)	\\
V373 Sct	&	1975	&	2442529	&	2442543	&	6.1	&	47	&	79	&	237	&	700	&	$>$18.3	&	\ldots	&	J(79)	\\
V443 Sct	&	1989	&	2447777	&	2447787	&	8.5	&	33	&	60	&	\ldots	&	\ldots	&	$>$20	&	\ldots	&	J(60)	\\
FH Ser	&	1970	&	2440630	&	2440634	&	4.5	&	49	&	62	&	410	&	1666	&	16.8	&	\ldots	&	D(62)	\\
LW Ser	&	1978	&	2443567	&	2443573	&	8.3	&	32	&	52	&	211	&	\ldots	&	19.4	&	\ldots	&	D(52)	\\
V382 Vel	&	1999	&	2451319	&	2451322	&	2.8	&	6	&	13	&	154	&	542	&	16.6	&	\ldots	&	S(13)	\\
LV Vul	&	1968	&	2439960	&	2439964	&	4.5	&	20	&	38	&	280	&	784	&	15.3	&	1500	&	S(38)	\\
NQ Vul	&	1976	&	2443072	&	2443085	&	6.2	&	21	&	50	&	365	&	1269	&	17.2	&	\ldots	&	D(50)	\\
PW Vul	&	1984	&	2445908	&	2445918	&	6.4	&	44	&	116	&	474	&	1366	&	16.9	&	\ldots	&	J(116)	\\
QU Vul	&	1984	&	2446056	&	2446061	&	5.3	&	20	&	36	&	814	&	1445	&	17.9	&	\ldots	&	P(36)	\\
QV Vul	&	1987	&	2447116	&	2447127	&	7.1	&	37	&	47	&	388	&	\ldots	&	18.0	&	\ldots	&	D(47)	\\
\enddata
\tablenotetext{a}{This time of quiescence is when the light curve becomes essentially flat, although the brightness after the end of the eruption is greatly brighter than is known for the time before eruption.  See Schaefer \& Collazzi (2010) for full details.}
\end{deluxetable}

\clearpage
\begin{deluxetable}{llllll}
\tabletypesize{\scriptsize}
\tablecaption{Binned Light Curves for 92 Novae
\label{tbl2}}
\tablewidth{0pt}
\tablehead{
\colhead{Nova}   &
\colhead{JD}   &
\colhead{$T-T_0$ (days)}   &
\colhead{V (mag)}  &
\colhead{$\sigma_V$ (mag)}  &
\colhead{N}
}
\startdata

OS And	&	2446768.75	&	0.75	&	8.00	&	0.15	&	1	\\
OS And	&	2446769.75	&	1.75	&	7.50	&	0.15	&	1	\\
OS And	&	2446772.25	&	4.25	&	6.55	&	0.11	&	2	\\
OS And	&	2446773.25	&	5.25	&	6.55	&	0.05	&	8	\\
OS And	&	2446774.25	&	6.25	&	6.72	&	0.06	&	13	\\
OS And	&	2446774.75	&	6.75	&	6.87	&	0.17	&	3	\\
OS And	&	2446775.25	&	7.25	&	6.96	&	0.06	&	11	\\
OS And	&	2446775.75	&	7.75	&	7.30	&	0.11	&	2	\\
OS And	&	2446776.25	&	8.25	&	7.46	&	0.07	&	5	\\
OS And	&	2446776.75	&	8.75	&	7.43	&	0.09	&	3	\\
OS And	&	2446777.25	&	9.25	&	7.65	&	0.11	&	2	\\
OS And	&	2446777.75	&	9.75	&	7.65	&	0.07	&	6	\\
OS And	&	2446778.25	&	10.25	&	7.68	&	0.09	&	5	\\
OS And	&	2446778.75	&	10.75	&	7.85	&	0.10	&	8	\\
OS And	&	2446779.25	&	11.25	&	7.95	&	0.09	&	13	\\
\enddata
\end{deluxetable}

\clearpage
\begin{deluxetable}{lll}
\tabletypesize{\scriptsize}
\tablecaption{Definitions and Examples of Light Curve Classes
\label{tbl3}}
\tablewidth{0pt}
\tablehead{
\colhead{Class}   &
\colhead{Definition}   &
\colhead{Examples}
}
\startdata

S	&	Smooth; Power law decline with no major fluctuations	&	CP Lac,	\\
	&										&	V1668 Cyg, V2275 Cyg	\\
P	&	Plateau; Smooth decline interrupted by a long-lasting	&	V4633 Sgr, \\
	&	nearly-flat interval followed by steeper decline		&	CP Pup, RS Oph	\\
D	&	Dust dip; Decline interrupted by fast decline, minimum,	&	DQ Her, 	\\
	&	and recovery to just below original decline		&	FH Ser, V705 Cas	\\
C	&	Cusp; Power law decline plus secondary maximum with	& 	V2362 Cyg, 	\\
	&	steepening rise then steep decline			&	V1493 Aql, V2491 Cyg	\\
O	&	Oscillations; Smooth decline with interval showing	&	V603 Aql		\\
	&	quasi-sinusoidal variations				&	GK Per, V1494 Aql	\\	
F	&	Flat-topped; Smooth light curve with an extended interval		&	DO Aql,	\\ 
	&	at peak with near constant brightness			&	V849 Oph, BT Mon	\\
J	&	Jitter; Decline displays substantial variability often as 	&	DK Lac	\\
	&	short-duration brightenings				&	HR Del, V723 Cas	\\

\enddata
\end{deluxetable}

\clearpage
\begin{deluxetable}{llllllllll}
\tabletypesize{\scriptsize}
\tablecaption{Nova Classes of Various Types
\label{tbl4}}
\tablewidth{0pt}
\tablehead{
\colhead{Nova}   &
\colhead{LC Class}   &
\colhead{Spectrum}   &
\colhead{RN}   &
\colhead{Shell}   &
\colhead{Magn. WD}   &
\colhead{Excitation}   &
\colhead{$P_{orb}$}   &
\colhead{Min. LC}   &
\colhead{Rosenbush}
}
\startdata

V4160 Sgr	&	S(3)	&	Neon	&	\ldots	&	\ldots	&	\ldots	&	\ldots	&	\ldots	&	\ldots	&	\ldots	\\
V4739 Sgr	&	S(3)	&	\ldots	&	\ldots	&	\ldots	&	\ldots	&	\ldots	&	\ldots	&	\ldots	&	\ldots	\\
V1500 Cyg	&	S(4)	&	Neon	&	\ldots	&	Shell	&	Yes	&	\ldots	&	\ldots	&	Pre-Rise, Hi-$\Delta$m	&	CP Lac	\\
V4643 Sgr	&	S(6)	&	\ldots	&	\ldots	&	\ldots	&	\ldots	&	\ldots	&	\ldots	&	\ldots	&	\ldots	\\
T CrB	&	S(6)	&	\ldots	&	RN	&	\ldots	&	\ldots	&	High	&	V. Long	&	Pre-Dip, Post-Max	&	\ldots	\\
V2275 Cyg	&	S(8)	&	\ldots	&	\ldots	&	\ldots	&	Yes	&	\ldots	&	\ldots	&	\ldots	&	\ldots	\\
CP Lac	&	S(9)	&	Hybrid	&	\ldots	&	Shell	&	\ldots	&	High	&	\ldots	&	\ldots	&	CP Lac	\\
V4444 Sgr	&	S(9)	&	\ldots	&	\ldots	&	\ldots	&	\ldots	&	\ldots	&	\ldots	&	\ldots	&	\ldots	\\
V4743 Sgr	&	S(12)	&	Neon	&	\ldots	&	\ldots	&	\ldots	&	\ldots	&	\ldots	&	\ldots	&	\ldots	\\
V5115 Sgr	&	S(13)	&	\ldots	&	\ldots	&	\ldots	&	\ldots	&	\ldots	&	\ldots	&	\ldots	&	\ldots	\\
V382 Vel	&	S(13)	&	Neon	&	\ldots	&	\ldots	&	\ldots	&	\ldots	&	\ldots	&	\ldots	&	\ldots	\\
V3890 Sgr	&	S(14)	&	\ldots	&	RN	&	\ldots	&	\ldots	&	High	&	V. Long	&	\ldots	&	\ldots	\\
V2313 Oph	&	S(17)	&	\ldots	&	\ldots	&	\ldots	&	\ldots	&	\ldots	&	\ldots	&	\ldots	&	\ldots	\\
V1187 Sco	&	S(17)	&	Neon	&	\ldots	&	\ldots	&	\ldots	&	\ldots	&	\ldots	&	\ldots	&	\ldots	\\
V693 CrA	&	S(18)	&	Neon	&	\ldots	&	\ldots	&	\ldots	&	\ldots	&	\ldots	&	\ldots	&	V630 Sgr	\\
V5114 Sgr	&	S(21)	&	\ldots	&	\ldots	&	\ldots	&	\ldots	&	\ldots	&	\ldots	&	\ldots	&	\ldots	\\
V4742 Sgr	&	S(23)	&	\ldots	&	\ldots	&	\ldots	&	\ldots	&	\ldots	&	\ldots	&	\ldots	&	\ldots	\\
V1188 Sco	&	S(23)	&	\ldots	&	\ldots	&	\ldots	&	\ldots	&	\ldots	&	\ldots	&	\ldots	&	\ldots	\\
V1668 Cyg	&	S(26)	&	Neon	&	\ldots	&	\ldots	&	\ldots	&	\ldots	&	\ldots	&	\ldots	&	DQ Her	\\
V5116 Sgr	&	S(26)	&	\ldots	&	\ldots	&	\ldots	&	\ldots	&	\ldots	&	\ldots	&	\ldots	&	\ldots	\\
V574 Pup	&	S(27)	&	Neon	&	\ldots	&	\ldots	&	\ldots	&	\ldots	&	\ldots	&	\ldots	&	\ldots	\\
V2264 Oph	&	S(30)	&	\ldots	&	\ldots	&	\ldots	&	\ldots	&	\ldots	&	\ldots	&	\ldots	&	GQ Mus	\\
V4740 Sgr	&	S(33)	&	\ldots	&	\ldots	&	\ldots	&	\ldots	&	\ldots	&	\ldots	&	\ldots	&	\ldots	\\
V4169 Sgr	&	S(36)	&	\ldots	&	\ldots	&	\ldots	&	\ldots	&	\ldots	&	\ldots	&	\ldots	&	CP Lac	\\
V528 Aql	&	S(38)	&	Neon	&	\ldots	&	\ldots	&	\ldots	&	Low	&	\ldots	&	\ldots	&	CP Pup	\\
LV Vul	&	S(38)	&	FeII	&	\ldots	&	Shell	&	\ldots	&	\ldots	&	\ldots	&	\ldots	&	CP Lac	\\
V446 Her	&	S(42)	&	He/N	&	\ldots	&	Shell	&	\ldots	&	\ldots	&	\ldots	&	w/DN	&	CP Lac	\\
V533 Her	&	S(43)	&	FeII	&	\ldots	&	Shell	&	Yes	&	\ldots	&	\ldots	&	Pre-Rise	&	\ldots	\\
V827 Her	&	S(53)	&	\ldots	&	\ldots	&	\ldots	&	\ldots	&	\ldots	&	\ldots	&	\ldots	&	DQ Her	\\
V1425 Aql	&	S(79)	&	FeII	&	\ldots	&	\ldots	&	Yes	&	\ldots	&	\ldots	&	\ldots	&	V1974 Cyg	\\
V2214 Oph	&	S(89)	&	Neon	&	\ldots	&	\ldots	&	Yes	&	\ldots	&	\ldots	&	\ldots	&	\ldots	\\
V1330 Cyg	&	S(217)	&	\ldots	&	\ldots	&	\ldots	&	\ldots	&	\ldots	&	\ldots	&	Hi-$\Delta$m?	&	\ldots	\\
U Sco	&	P(3)	&	\ldots	&	RN	&	\ldots	&	\ldots	&	High	&	Long	&	\ldots	&	\ldots	\\
V838 Her	&	P(4)	&	Neon	&	\ldots	&	\ldots	&	\ldots	&	\ldots	&	\ldots	&	\ldots	&	DQ Her	\\
CP Pup	&	P(8)	&	He/N	&	\ldots	&	Shell	&	Yes	&	High	&	V. Short	&	Hi-$\Delta$m	&	CP Pup	\\
V2487 Oph	&	P(8)	&	\ldots	&	RN	&	\ldots	&	?	&	High	&	Long	&	\ldots	&	\ldots	\\
LZ Mus	&	P(12)	&	\ldots	&	\ldots	&	\ldots	&	\ldots	&	\ldots	&	\ldots	&	\ldots	&	\ldots	\\
RS Oph	&	P(14)	&	\ldots	&	RN	&	\ldots	&	\ldots	&	High	&	V. Long	&	Post-Dip	&	\ldots	\\
DD Cir     	&	P(16)	&	\ldots	&	\ldots	&	\ldots	&	\ldots	&	\ldots	&	\ldots	&	\ldots	&	\ldots	\\
HS Sge	&	P(21)	&	\ldots	&	\ldots	&	\ldots	&	\ldots	&	\ldots	&	\ldots	&	\ldots	&	CP Pup	\\
V351 Pup	&	P(26)	&	Neon	&	\ldots	&	Shell	&	\ldots	&	\ldots	&	\ldots	&	\ldots	&	GQ Mus	\\
CI Aql      	&	P(32)	&	\ldots	&	RN	&	\ldots	&	\ldots	&	High	&	Long	&	\ldots	&	\ldots	\\
V1229 Aql	&	P(32)	&	FeII	&	\ldots	&	Shell	&	\ldots	&	\ldots	&	\ldots	&	\ldots	&	DQ Her	\\
QU Vul	&	P(36)	&	Neon	&	\ldots	&	Shell	&	\ldots	&	\ldots	&	\ldots	&	\ldots	&	V1974 Cyg	\\
V1974 Cyg	&	P(43)	&	Neon	&	\ldots	&	Shell	&	Yes	&	\ldots	&	\ldots	&	Hi-$\Delta$m	&	V1974 Cyg	\\
V4633 Sgr	&	P(44)	&	\ldots	&	\ldots	&	\ldots	&	Yes	&	\ldots	&	\ldots	&	Hi-$\Delta$m	&	\ldots	\\
DN Gem	&	P(50)	&	FeII	&	\ldots	&	\ldots	&	\ldots	&	Low	&	Long	&	\ldots	&	GQ Mus	\\
IM Nor   	&	P(80)	&	\ldots	&	RN	&	\ldots	&	\ldots	&	High	&	In Gap	&	\ldots	&	\ldots	\\
T Pyx	&	P(62)	&	\ldots	&	RN	&	Shell	&	?	&	High	&	V. Short	&	Hi-$\Delta$m	&	\ldots	\\
BY Cir      	&	P(124)	&	\ldots	&	\ldots	&	\ldots	&	\ldots	&	\ldots	&	\ldots	&	\ldots	&	\ldots	\\
V4021 Sgr	&	P(215)	&	\ldots	&	\ldots	&	\ldots	&	\ldots	&	\ldots	&	\ldots	&	\ldots	&	\ldots	\\
V476 Cyg	&	D(16)	&	\ldots	&	\ldots	&	Shell	&	\ldots	&	\ldots	&	\ldots	&	v. long tail	&	CP Pup	\\
OS And	&	D(23)	&	\ldots	&	\ldots	&	\ldots	&	\ldots	&	\ldots	&	\ldots	&	\ldots	&	GQ Mus	\\
V1370 Aql	&	D(28)	&	Neon	&	\ldots	&	\ldots	&	\ldots	&	\ldots	&	\ldots	&	\ldots	&	CP Pup	\\
V1419 Aql	&	D(32)	&	\ldots	&	\ldots	&	\ldots	&	\ldots	&	\ldots	&	\ldots	&	\ldots	&	DQ Her	\\
V2274 Cyg	&	D(33)	&	\ldots	&	\ldots	&	\ldots	&	\ldots	&	\ldots	&	\ldots	&	\ldots	&	\ldots	\\
QV Vul	&	D(47)	&	FeII	&	\ldots	&	Shell	&	\ldots	&	\ldots	&	\ldots	&	\ldots	&	DQ Her	\\
V842 Cen	&	D(48)	&	FeII	&	\ldots	&	Shell	&	\ldots	&	\ldots	&	\ldots	&	\ldots	&	DQ Her	\\
NQ Vul	&	D(50)	&	FeII	&	\ldots	&	Shell	&	\ldots	&	\ldots	&	\ldots	&	\ldots	&	DQ Her	\\
LW Ser	&	D(52)	&	\ldots	&	\ldots	&	\ldots	&	\ldots	&	\ldots	&	\ldots	&	\ldots	&	DQ Her	\\
FH Ser	&	D(62)	&	FeII	&	\ldots	&	Shell	&	\ldots	&	\ldots	&	\ldots	&	\ldots	&	DQ Her	\\
V705 Cas	&	D(67)	&	\ldots	&	\ldots	&	Shell	&	\ldots	&	\ldots	&	\ldots	&	\ldots	&	DQ Her	\\
V732 Sgr	&	D(75)	&	\ldots	&	\ldots	&	\ldots	&	Yes	&	Low	&	Long	&	\ldots	&	\ldots	\\
T Aur	&	D(84)	&	FeII	&	\ldots	&	Shell	&	Yes	&	Low	&	\ldots	&	v. long tail	&	\ldots	\\
DQ Her	&	D(100)	&	Neon / FeII	&	\ldots	&	Shell	&	Yes	&	Low	&	\ldots	&	v. long tail	&	DQ Her	\\
V992 Sco	&	D(120)	&	\ldots	&	\ldots	&	\ldots	&	\ldots	&	\ldots	&	\ldots	&	\ldots	&	\ldots	\\
V445 Pup	&	D(240)	&	Helium	&	\ldots	&	Shell	&	\ldots	&	\ldots	&	\ldots	&	\ldots	&	\ldots	\\
V2491 Cyg	&	C(16)	&	\ldots	&	\ldots	&	\ldots	&	\ldots	&	\ldots	&	\ldots	&	\ldots	&	\ldots	\\
V1493 Aql	&	C(50)	&	\ldots	&	\ldots	&	\ldots	&	\ldots	&	\ldots	&	\ldots	&	\ldots	&	\ldots	\\
V2362 Cyg	&	C(246)	&	\ldots	&	\ldots	&	\ldots	&	\ldots	&	\ldots	&	\ldots	&	\ldots	&	\ldots	\\
V603 Aql	&	O(12)	&	Hybrid	&	\ldots	&	Shell	&	Yes	&	High	&	\ldots	&	v. long tail	&	CP Pup	\\
GK Per	&	O(13)	&	Neon/FeII	&	\ldots	&	Shell	&	Yes	&	Low	&	Long	&	w/DN	&	\ldots	\\
V1494 Aql	&	O(16)	&	Neon	&	\ldots	&	\ldots	&	Yes	&	\ldots	&	\ldots	&	\ldots	&	\ldots	\\
V2467 Cyg	&	O(20)	&	FeII	&	\ldots	&	\ldots	&	Yes	&	Low	&	\ldots	&	\ldots	&	\ldots	\\
V888 Cen	&	O(90)	&	FeII	&	\ldots	&	\ldots	&	\ldots	&	\ldots	&	\ldots	&	\ldots	&	\ldots	\\
V2295 Oph	&	F(16)	&	\ldots	&	\ldots	&	\ldots	&	\ldots	&	\ldots	&	\ldots	&	\ldots	&	\ldots	\\
BT Mon	&	F(182)	&	\ldots	&	\ldots	&	Shell	&	Yes	&	High	&	Long	&	\ldots	&	\ldots	\\
V849 Oph	&	F(270)	&	Neon / FeII	&	\ldots	&	\ldots	&	\ldots	&	Low	&	\ldots	&	\ldots	&	RR Pic	\\
DO Aql	&	F(900)	&	\ldots	&	\ldots	&	\ldots	&	\ldots	&	Low	&	\ldots	&	\ldots	&	\ldots	\\
V443 Sct	&	J(60)	&	\ldots	&	\ldots	&	\ldots	&	\ldots	&	\ldots	&	\ldots	&	\ldots	&	\ldots	\\
V1186 Sco	&	J(62)	&	\ldots	&	\ldots	&	\ldots	&	\ldots	&	\ldots	&	\ldots	&	\ldots	&	\ldots	\\
V373 Sct	&	J(79)	&	\ldots	&	\ldots	&	\ldots	&	\ldots	&	\ldots	&	\ldots	&	\ldots	&	\ldots	\\
V868 Cen	&	J(82)	&	\ldots	&	\ldots	&	\ldots	&	\ldots	&	\ldots	&	\ldots	&	\ldots	&	\ldots	\\
V2540 Oph	&	J(115)	&	\ldots	&	\ldots	&	\ldots	&	?	&	\ldots	&	\ldots	&	\ldots	&	\ldots	\\
PW Vul	&	J(116)	&	FeII	&	\ldots	&	Shell	&	\ldots	&	\ldots	&	\ldots	&	\ldots	&	RR Pic	\\
RR Pic	&	J(122)	&	\ldots	&	\ldots	&	Shell	&	Yes	&	\ldots	&	\ldots	&	v. long tail	&	RR Pic	\\
V356 Aql	&	J(140)	&	FeII	&	\ldots	&	\ldots	&	\ldots	&	Low	&	\ldots	&	\ldots	&	RR Pic	\\
V1039 Cen	&	J(174)	&	\ldots	&	\ldots	&	\ldots	&	\ldots	&	\ldots	&	\ldots	&	\ldots	&	\ldots	\\
V1819 Cyg	&	J(181)	&	FeII	&	\ldots	&	Shell	&	\ldots	&	\ldots	&	\ldots	&	\ldots	&	RR Pic	\\
V4745 Sgr	&	J(190)	&	\ldots	&	\ldots	&	\ldots	&	\ldots	&	\ldots	&	\ldots	&	\ldots	&	\ldots	\\
DK Lac	&	J(202)	&	FeII	&	\ldots	&	Shell	&	\ldots	&	\ldots	&	\ldots	&	\ldots	&	GQ Mus	\\
HR Del	&	J(231)	&	Neon	&	\ldots	&	Shell	&	Yes	&	\ldots	&	\ldots	&	v. long tail	&	RR Pic	\\
V723 Cas	&	J(299)	&	Neon	&	\ldots	&	Shell	&	Yes	&	\ldots	&	Long	&	Hi-$\Delta$m	&	RR Pic	\\
\enddata
\end{deluxetable}

\clearpage
\begin{deluxetable}{lllllllll}
\tabletypesize{\scriptsize}
\tablecaption{S Class Light Curves
\label{tbl5}}
\tablewidth{0pt}
\tablehead{
\colhead{Nova}   &
\colhead{$N_{b}$}   &
\colhead{Slope 1}   &
\colhead{$T_{b1}-T_0$}   &
\colhead{$V_{b1}$}   &
\colhead{Slope 2}   &
\colhead{$T_{b2}-T_0$}   &
\colhead{$V_{b2}$}   &
\colhead{Slope 3}
}
\startdata

V528 Aql	&	1	&	-2.7	&	25	&	9.2	&	-4.5	&	\ldots	&	\ldots	&	\ldots	\\
V1425 Aql	&	1	&	-2.7	&	410	&	13.0	&	-7.1	&	\ldots	&	\ldots	&	\ldots	\\
T CrB	&	2	&	-7.8	&	16	&	8.6	&	-4.0	&	\ldots	&	\ldots	&	\ldots	\\
V693 CrA	&	0	&	-3.8	&	\ldots	&	\ldots	&	\ldots	&	\ldots	&	\ldots	&	\ldots	\\
V1330 Cyg	&	3\tablenotemark{a}	&	-0.4	&	60	&	10.9	&	-2.9	&	210	&	12.5	&	-7.8	\\
V1500 Cyg	&	2	&	-5.9	&	10	&	6.2	&	-3.3	&	400	&	11.5	&	-6.0	\\
V1668 Cyg	&	2	&	-5.0	&	45	&	10.3	&	-1.7	&	100	&	10.9	&	-4.3	\\
V2275 Cyg	&	1	&	-3.3	&	62	&	12.6	&	-4.7	&	\ldots	&	\ldots	&	\ldots	\\
V446 Her	&	2	&	-3.3	&	140	&	9.5	&	-9.1	&	190	&	10.7	&	-4.3	\\
V533 Her	&	2	&	-3.4	&	100	&	7.1	&	-4.7	&	200	&	8.5	&	-6.8	\\
V827 Her	&	1	&	-3.6	&	400	&	13.5	&	-10.0	&	\ldots	&	\ldots	&	\ldots	\\
CP Lac	&	2	&	-3.9	&	300	&	10.3	&	-5.4	&	1000	&	13.1	&	-3.0	\\
V2214 Oph	&	2	&	-5.1	&	130	&	11.9	&	-3.3	&	350	&	13.3	&	-6.3	\\
V2264 Oph	&	2	&	-2.2	&	30	&	12.1	&	-20.2	&	34	&	13.2	&	-2.1	\\
V2313 Oph	&	1	&	-3.2	&	20	&	10.7	&	-2.5	&	\ldots	&	\ldots	&	\ldots	\\
V574 Pup	&	0	&	-3.2	&	\ldots	&	\ldots	&	\ldots	&	\ldots	&	\ldots	&	\ldots	\\
V3890 Sgr	&	2	&	-4.0	&	10	&	10.5	&	-5.9	&	33	&	13.5	&	-5.0	\\
V4160 Sgr	&	1	&	-9.0	&	20	&	11.8	&	-4.9	&	\ldots	&	\ldots	&	\ldots	\\
V4169 Sgr	&	2	&	-3.7	&	57	&	10.9	&	-6.0	&	150	&	13.4	&	-2.8	\\
V4444 Sgr	&	1	&	-3.5	&	20	&	11.4	&	-3.9	&	\ldots	&	\ldots	&	\ldots	\\
V4643 Sgr	&	1	&	-2.8	&	19	&	11.7	&	-5.3	&	\ldots	&	\ldots	&	\ldots	\\
V4739 Sgr	&	0	&	-4.7	&	\ldots	&	\ldots	&	\ldots	&	\ldots	&	\ldots	&	\ldots	\\
V4740 Sgr	&	1	&	-3.8	&	45	&	10.2	&	-4.5	&	\ldots	&	\ldots	&	\ldots	\\
V4742 Sgr	&	2	&	-3.4	&	65	&	12.5	&	-2.0	&	185	&	13.4	&	-3.7	\\
V4743 Sgr	&	2	&	-3.5	&	22	&	8.7	&	-2.4	&	270	&	11.3	&	-6.5	\\
V5114 Sgr	&	1	&	-3.3	&	44	&	12.1	&	-5.4	&	\ldots	&	\ldots	&	\ldots	\\
V5115 Sgr	&	0	&	-3.5	&	\ldots	&	\ldots	&	\ldots	&	\ldots	&	\ldots	&	\ldots	\\
V5116 Sgr	&	0	&	-3.0	&	\ldots	&	\ldots	&	\ldots	&	\ldots	&	\ldots	&	\ldots	\\
V1187 Sco	&	0	&	-4.2	&	\ldots	&	\ldots	&	\ldots	&	\ldots	&	\ldots	&	\ldots	\\
V1188 Sco	&	1	&	-3.7	&	37	&	12.6	&	-9.1	&	\ldots	&	\ldots	&	\ldots	\\
V382 Vel	&	2	&	-4.2	&	30	&	7.1	&	-1.7	&	130	&	8.2	&	-5.8	\\
LV Vul	&	2	&	-4.1	&	65	&	8.4	&	-3.0	&	220	&	10.0	&	-6.0	\\

\enddata
\tablenotetext{a}{The third break is at 440 days, V=15.0 mag, with a slope of 0.0.  See Section 5.}
\end{deluxetable}

\clearpage
\begin{deluxetable}{lllllllllllllll}
\rotate
\tabletypesize{\scriptsize}
\tablecaption{P Class Light Curves
\label{tbl6}}
\tablewidth{0pt}
\tablehead{
\colhead{Nova}   &
\colhead{$N_{b}$}   &
\colhead{Slope 1}   &
\colhead{$T_{b1}-T_0$}   &
\colhead{$V_{b1}$}   &
\colhead{Slope 2}   &
\colhead{$T_{b2}-T_0$}   &
\colhead{$V_{b2}$}   &
\colhead{Slope 3}   &
\colhead{$T_{b3}-T_0$}   &
\colhead{$V_{b3}$}   &
\colhead{Slope 4}   &
\colhead{$T_{b4}-T_0$}   &
\colhead{$V_{b4}$}   &
\colhead{Slope 5}
}
\startdata

CI Aql	&	3	&	-1.6	&	30	&	10.5	&	-11.4	&	50	&	13.1	&	-1.6	&	220	&	14.1	&	-3.0	&	\ldots	&	\ldots	&	\ldots	\\
V1229 Aql	&	1	&	-12.5	&	70	&	13.5	&	-0.9	&	\ldots	&	\ldots	&	\ldots	&	\ldots	&	\ldots	&	\ldots	&	\ldots	&	\ldots	&	\ldots	\\
BY Cir	&	4	&	-2.4	&	190	&	11.0	&	0.0	&	280	&	11.0	&	-8.5	&	390	&	12.4	&	0.0	&	580	&	12.4	&	-11.4	\\
DD Cir	&	3	&	-3.2	&	17	&	10.4	&	-5.7	&	30	&	11.8	&	-0.5	&	52	&	11.9	&	-6.0	&	\ldots	&	\ldots	&	\ldots	\\
V1974 Cyg 	&	4	&	-3.7	&	65	&	8.0	&	-2.2	&	250	&	9.3	&	-9.5	&	550	&	12.7	&	0.0	&	700	&	12.7	&	-7.2	\\
DN Gem	&	3	&	-3.4	&	70	&	7.7	&	-0.3	&	290	&	7.9	&	-7.5	&	1000	&	12.0	&	-3.4	&	\ldots	&	\ldots	&	\ldots	\\
V838 Her	&	2	&	-6.1	&	52	&	14.8	&	0.0	&	80	&	14.8	&	-5.5	&	\ldots	&	\ldots	&	\ldots	&	\ldots	&	\ldots	&	\ldots	\\
LZ Mus	&	2	&	-3.3	&	53	&	13.9	&	-1.0	&	125	&	14.3	&	-8.0	&	\ldots	&	\ldots	&	\ldots	&	\ldots	&	\ldots	&	\ldots	\\
IM Nor	&	4	&	-1.5	&	40	&	10.0	&	-5.4	&	180	&	13.5	&	0.0	&	230	&	13.5	&	-21.3	&	295	&	15.8	&	-5.0	\\
RS Oph	&	2	&	-3.7	&	50	&	9.8	&	-2.0	&	80	&	10.2	&	-10.0	&	\ldots	&	\ldots	&	\ldots	&	\ldots	&	\ldots	&	\ldots	\\
V2487 Oph	&	2	&	-3.9	&	12	&	13.7	&	-1.7	&	28	&	14.3	&	-5.2	&	\ldots	&	\ldots	&	\ldots	&	\ldots	&	\ldots	&	\ldots	\\
CP Pup\tablenotemark{a}	&	5	&	-7	&	35	&	6.5	&	-0.3	&	110	&	6.7	&	-3.2	&	504	&	8.8	&	-6.2	&	2000	&	12.5	&	-1.0	\\
V351 Pup	&	2	&	-3.0	&	75	&	10.8	&	-0.9	&	290	&	11.3	&	-7.0	&	\ldots	&	\ldots	&	\ldots	&	\ldots	&	\ldots	&	\ldots	\\
T Pyx	&	4	&	-5.2	&	120	&	9.6	&	-26.8	&	145	&	11.8	&	-3.6	&	220	&	12.5	&	-9.7	&	300	&	13.8	&	0.0	\\
HS Sge	&	2	&	-3.7	&	48	&	11.3	&	-1.3	&	430	&	12.5	&	-8.7	&	\ldots	&	\ldots	&	\ldots	&	\ldots	&	\ldots	&	\ldots	\\
V4021 Sgr	&	2	&	-3.5	&	70	&	11.3	&	-0.9	&	260	&	11.8	&	-6.5	&	\ldots	&	\ldots	&	\ldots	&	\ldots	&	\ldots	&	\ldots	\\
V4633 Sgr	&	2	&	-2.7	&	120	&	11.8	&	-0.2	&	320	&	11.9	&	-6.0	&	\ldots	&	\ldots	&	\ldots	&	\ldots	&	\ldots	&	\ldots	\\
U Sco	&	2	&	-5.9	&	15	&	13.8	&	-2.4	&	30	&	14.5	&	-7.4	&	\ldots	&	\ldots	&	\ldots	&	\ldots	&	\ldots	&	\ldots	\\
QU Vul	&	2	&	-4.2	&	80	&	9.4	&	-1.5	&	700	&	10.8	&	-10.5	&	\ldots	&	\ldots	&	\ldots	&	\ldots	&	\ldots	&	\ldots	\\

\enddata
\tablenotetext{a}{CP Pup has one more break that did not fit on this table, with a break around 4400 days at V=12.8 and with a slope of -3.6 after this break.}
\end{deluxetable}

\clearpage
\begin{deluxetable}{lllll}
\tabletypesize{\scriptsize}
\tablecaption{Plateau Properties
\label{tbl7}}
\tablewidth{0pt}
\tablehead{
\colhead{Nova}   &
\colhead{$\Delta T_{plat}$ (days)}   &
\colhead{$D_{plat}$ (days)}   &
\colhead{$\Delta V_{plat}$ (mag)}  &
\colhead{Slope}
}
\startdata

CI Aql	&	50	&	170	&	6.5	&	-1.6	\\
V1229 Aql	&	70	&	$>$3000	&	6.4	&	-0.9	\\
BY Cir (1)	&	190	&	90	&	3.6	&	0.0	\\
BY Cir (2)	&	390	&	190	&	5.0	&	0.0	\\
DD Cir	&	30	&	22	&	4.2	&	-0.5	\\
V1974 Cyg (1)	&	65	&	185	&	3.7	&	-2.2	\\
V1974 Cyg (2)	&	550	&	150	&	8.4	&	0.0	\\
DN Gem	&	70	&	220	&	3.3	&	-0.3	\\
V838 Her	&	52	&	28	&	9.5	&	0.0	\\
LZ Mus	&	53	&	72	&	5.4	&	-1.0	\\
IM Nor	&	180	&	50	&	5.0	&	0.0	\\
RS Oph	&	50	&	30	&	5.0	&	-2.0	\\
V2487 Oph	&	12	&	16	&	4.2	&	-1.7	\\
CP Pup	&	35	&	75	&	5.8	&	-0.3	\\
V351 Pup	&	75	&	215	&	4.4	&	-0.9	\\
T Pyx	&	300	&	$>$200	&	7.4	&	0.0	\\
HS Sge	&	48	&	382	&	3.5	&	-1.2	\\
V4021 Sgr	&	70	&	180	&	2.4	&	-0.5	\\
V4633 Sgr	&	120	&	200	&	4.4	&	-0.2	\\
U Sco	&	15	&	15	&	6.3	&	-2.4	\\
QU Vul	&	80	&	620	&	4.1	&	-1.5	\\
\enddata
\end{deluxetable}

\clearpage
\begin{deluxetable}{lllllllllllllll}
\rotate
\tabletypesize{\scriptsize}
\tablecaption{D Class Light Curves
\label{tbl8}}
\tablewidth{0pt}
\tablehead{
\colhead{Nova}   &
\colhead{Slope 1}   &
\colhead{$\Delta T_{start}$}   &
\colhead{$V_{start}$}  &
\colhead{$\Delta T_{min}$}   &
\colhead{$V_{min}$}  &
\colhead{$\Delta T_{end}$}   &
\colhead{$V_{end}$}  &
\colhead{Slope 2}   &
\colhead{$T_{b2}-T_0$}   &
\colhead{$V_{b2}$}   &
\colhead{Slope 3}   &
\colhead{$T_{b3}-T_0$}   &
\colhead{$V_{b3}$}   &
\colhead{Slope 4}
}
\startdata

OS And	&	-4.0	&	22	&	9.2	&	43	&	12.0	&	65	&	11.6	&	-1.8	&	190	&	12.4	&	-5.3	&	\ldots	&	\ldots	&	\ldots	\\
V1370 Aql	&	-3.0	&	23	&	9.8	&	53	&	13.5	&	105	&	12.0	&	-5.0	&	190	&	13.3	&	-22.0	&	\ldots	&	\ldots	&	\ldots	\\
V1419 Aql	&	-3.5	&	40	&	10.1	&	68	&	14.1	&	165	&	12.7	&	-1.9	&	450	&	13.5	&	-6.0	&	\ldots	&	\ldots	&	\ldots	\\
T Aur	&	-0.7	&	87	&	5.1	&	118	&	14.2	&	260	&	11.4	&	-0.4	&	700	&	11.5	&	-3.0	&	\ldots	&	\ldots	&	\ldots	\\
V705 Cas	&	-3.0	&	66	&	8.4	&	95	&	16.2	&	162	&	12.3	&	0.0	&	890	&	12.6	&	-29.5	&	940	&	13.3	&	-7.0	\\
V842 Cen	&	-4.0	&	46	&	7.2	&	70	&	14.0	&	133	&	11.3	&	-0.2	&	570	&	11.4	&	-6.5	&	\ldots	&	\ldots	&	\ldots	\\
V476 Cyg\tablenotemark{a}	&	-7.0	&	40	&	7.2	&	86	&	9.2	&	105	&	8.9	&	-1.4	&	512	&	9.9	&	-3.5	&	1350	&	11.4	&	-8.0	\\
V2274 Cyg	&	-2.0	&	40	&	14.1	&	54	&	17.2	&	80	&	16.3	&	-2.3	&	\ldots	&	\ldots	&	\ldots	&	\ldots	&	\ldots	&	\ldots	\\
DQ Her	&	-3.0	&	105	&	4.5	&	140	&	13.0	&	232	&	6.9	&	-2.9	&	1200	&	8.5	&	-8.6	&	4200	&	13.2	&	-3.3	\\
V445 Pup	&	-2.9	&	218	&	10.6	&	570	&	$\sim$20	&	\ldots	&	\ldots	&	\ldots	&	\ldots	&	\ldots	&	\ldots	&	\ldots	&	\ldots	&	\ldots	\\
V732 Sgr	&	-3.0	&	73	&	8.0	&	111	&	15.0	&	185	&	11.8	&	0.0	&	330	&	11.5	&	-5.5	&	\ldots	&	\ldots	&	\ldots	\\
V992 Sco	&	-0.3	&	93	&	9.0	&	155	&	14.3	&	245	&	12.7	&	-0.2	&	500	&	12.8	&	-4.5	&	\ldots	&	\ldots	&	\ldots	\\
FH Ser\tablenotemark{b}	&	-2.9	&	57	&	6.9	&	100	&	11.9	&	175	&	10.0	&	-1.3	&	500	&	10.6	&	-6.6	&	1000	&	12.6	&	-1.7	\\
LW Ser	&	-3.5	&	53	&	10.6	&	96	&	15.3	&	120	&	14.0	&	-0.7	&	\ldots	&	\ldots	&	\ldots	&	\ldots	&	\ldots	&	\ldots	\\
NQ Vul\tablenotemark{c}	&	-4.0	&	64	&	9.1	&	83	&	12.4	&	159	&	11.5	&	-2.1	&	450	&	12.4	&	-4.5	&	940	&	13.6	&	-11.3	\\
QV Vul	&	-0.8	&	35	&	7.8	&	98	&	15.6	&	287	&	12.9	&	-3.0	&	660	&	14.0	&	-5.0	&	\ldots	&	\ldots	&	\ldots	\\
\enddata
\tablenotetext{a}{V476 Cyg has a fifth break at 2310 days after the explosion at V=13.4 with a slope after the break of -2.0.  A sixth break occurs at 6400 days after the explosion at V=14.3 with a slope after the break of -6.5.}
\tablenotetext{b}{FH Ser has a fifth break at 1500 days after the explosion at V=12.9 with a slope after the break of -11.}
\tablenotetext{c}{NQ Vul has a fifth break at 1500 days after the explosion at V=15.9 with a slope after the break of -3.5.}
\end{deluxetable}

\clearpage
\begin{deluxetable}{llllllllllll}
%\rotate
\tabletypesize{\scriptsize}
\tablecaption{C Class Light Curves
\label{tbl9}}
\tablewidth{0pt}
\tablehead{
\colhead{Nova}   &
\colhead{Slope 1}   &
\colhead{$\Delta T_{min}$}   &
\colhead{$V_{min}$}  &
\colhead{$\Delta T_{cusp}$}   &
\colhead{$V_{cusp}$}  &
\colhead{$\Delta T_{base}$}   &
\colhead{$V_{base}$}  &
\colhead{Slope 2}   &
\colhead{$T_{b2}-T_0$}   &
\colhead{$V_{b2}$}   &
\colhead{Slope 3}
}
\startdata

V1493 Aql	&	-4.0	&	24	&	12.8	&	47	&	11.7	&	64	&	14.9	&	-2.0	&	\ldots	&	\ldots	&	\ldots	\\
V2362 Cyg	&	-2.7	&	102	&	12.1	&	239	&	10.1	&	263	&	13.6	&	-2.7	&	550	&	14.7	&	-10.8	\\
V2491 Cyg	&	-2.0	&	10	&	9.9	&	15	&	9.6	&	18	&	11.2	&	-7	&	63	&	14.5	&	-3.8	\\
\enddata
\end{deluxetable}

%\clearpage
\begin{deluxetable}{lllllllllll}
\tabletypesize{\scriptsize}
\tablecaption{O Class Light Curves
\label{tbl10}}
\tablewidth{0pt}
\tablehead{
\colhead{Nova}   &
\colhead{$N_{b}$}   &
\colhead{Slope 1}   &
\colhead{$T_{b1}-T_0$}   &
\colhead{$V_{b1}$}   &
\colhead{Slope 2}   &
\colhead{$T_{b2}-T_0$}   &
\colhead{$V_{b2}$}   &
\colhead{Slope 3}	&
\colhead{$\Delta T_{start}$}   &
\colhead{$\Delta T_{end}$} 
}
\startdata

V603 Aql	&	3	&	-6.2	&	23	&	3.2	&	-2.8	&	400	&	6.6	&	-6.7	&	20	&	100	\\
V1494 Aql	&	2	&	-3.5	&	220	&	10.4	&	-7.2	&	670	&	13.9	&	-4.8	&	15	&	140	\\
V888 Cen\tablenotemark{a}	&	1	&	-2.1	&	190	&	12.1	&	-9.0	&	\ldots	&	\ldots	&	\ldots	&	6	&	100	\\
V2467 Cyg\tablenotemark{b}	&	3	&	-3.6	&	45	&	12.0	&	-2.1	&	210	&	13	&	-3.5	&	30	&	160	\\
GK Per	&	2	&	-4.0	&	23	&	3.9	&	-2.8	&	310	&	7.1	&	-10.1	&	25	&	125	\\
\enddata
\tablenotetext{a}{V888 Cen has an apparent break at 400 days at a magnitude of 15.0 with a slope of 0.0 (i.e. flat) out to at least 746 days.  However, this might well be the return to quiescence, or it might be confusion with a nearby unresolved star of near that magnitude.}
\tablenotetext{b}{V2467 Cyg has a third break at 600 days after the explosion at V=15.0 with a slope after the break of -7.8.}
\end{deluxetable}

\clearpage
\begin{deluxetable}{lllll}
\tabletypesize{\scriptsize}
\tablecaption{Oscillation Times and Magnitudes
\label{tbl11}}
\tablewidth{0pt}
\tablehead{
\colhead{Nova}   &
\colhead{Maximum $T_i$}   &
\colhead{Maximum $V_i$}   &
\colhead{Minimum $T_i$}   &
\colhead{Minimum $V_i$}
}
\startdata

V603 Aql	&	\ldots	&	\ldots	&	24.5	&	3.8	\\
V603 Aql	&	31.5	&	3.2	&	37.5	&	3.9	\\
V603 Aql	&	43.5	&	3.7	&	49.5	&	4.4	\\
V603 Aql	&	53.5	&	3.8	&	61.5	&	4.7	\\
V603 Aql	&	65.5	&	4.0	&	74.5	&	4.7	\\
V603 Aql	&	76.5	&	4.4	&	80.5	&	4.9	\\
V603 Aql	&	84.5	&	4.4	&	90.5	&	4.9	\\
V603 Aql	&	95.5	&	4.6	&	100.5	&	5.1	\\
	&		&		&		&		\\
V1494 Aql	&	14.5	&	6.3	&	21.5	&	7.5	\\
V1494 Aql	&	24.5	&	7.3	&	26.5	&	7.7	\\
V1494 Aql	&	29.5	&	7.4	&	33.5	&	8.1	\\
V1494 Aql	&	37.5	&	7.6	&	46.5	&	8.5	\\
V1494 Aql	&	54.5	&	8.3	&	56.5	&	8.6	\\
V1494 Aql	&	58.5	&	8.4	&	61.5	&	9.5	\\
V1494 Aql	&	69.5	&	8.6	&	70.5	&	8.9	\\
V1494 Aql	&	75.5	&	8.6	&	77.5	&	9.0	\\
V1494 Aql	&	79.5	&	8.8	&	85.5	&	9.8	\\
V1494 Aql	&	93	&	8.9	&	101	&	9.7	\\
V1494 Aql	&	107	&	8.7	&	117	&	9.6	\\
V1494 Aql	&	121	&	9.1	&	125	&	9.8	\\
	&		&		&		&		\\
V888 Cen	&	\ldots	&	\ldots	&	4.25	&	9.1	\\
V888 Cen	&	7.25	&	8.5	&	9.5	&	9.8	\\
V888 Cen	&	12.5	&	9.1	&	18.5	&	10.2	\\
V888 Cen	&	17.5	&	9.5	&	39.5	&	11.0	\\
V888 Cen	&	40.5	&	9.8	&	56.5	&	11.5	\\
V888 Cen	&	61.5	&	10.3	&	70.5	&	12.3	\\
V888 Cen	&	76.5	&	10.6	&	84.5	&	11.6	\\
V888 Cen	&	90.5	&	10.8	&	\ldots	&	\ldots	\\
	&		&		&		&		\\
V2467 Cyg	&	\ldots	&	\ldots	&	34	&	11.8	\\
V2467 Cyg	&	52	&	11.8	&	60	&	12.8	\\
V2467 Cyg	&	70	&	12.1	&	80	&	12.7	\\
V2467 Cyg	&	92	&	12.4	&	100	&	12.9	\\
V2467 Cyg	&	116	&	12.7	&	120	&	13.2	\\
V2467 Cyg	&	134	&	12.9	&	146	&	13.2	\\
V2467 Cyg	&	148	&	12.9	&	\ldots	&	\ldots	\\
	&		&		&		&		\\
GK Per	&	25.5	&	3.8	&	26.5	&	5.0	\\
GK Per	&	27.8	&	3.7	&	29.1	&	5.3	\\
GK Per	&	30.5	&	3.7	&	32.5	&	5.3	\\
GK Per	&	33.8	&	3.9	&	35.8	&	4.8	\\
GK Per	&	37.4	&	4.1	&	38.3	&	4.5	\\
GK Per	&	39.2	&	4.1	&	41.2	&	5.4	\\
GK Per	&	42.7	&	4.0	&	45	&	5.5	\\
GK Per	&	47	&	4.1	&	49	&	5.5	\\
GK Per	&	51	&	4.4	&	53.7	&	5.7	\\
GK Per	&	56	&	4.3	&	59.2	&	5.8	\\
GK Per	&	61.5	&	4.3	&	63.6	&	5.8	\\
GK Per	&	65.6	&	4.2	&	68.1	&	5.8	\\
GK Per	&	70.4	&	4.4	&	73.1	&	5.9	\\
GK Per	&	75.9	&	4.1	&	78.3	&	5.9	\\
GK Per	&	81.2	&	\ldots	&	83.3	&	6.0	\\
GK Per	&	85.8	&	\ldots	&	88.8	&	6.2	\\
GK Per	&	90.5	&	\ldots	&	92.5	&	6.2	\\
GK Per	&	93.9	&	\ldots	&	96.6	&	6.2	\\
GK Per	&	98	&	\ldots	&	100.7	&	6.2	\\
GK Per	&	102.8	&	4.9	&	105.2	&	6.2	\\
GK Per	&	107.3	&	4.9	&	109	&	6.2	\\
GK Per	&	110.2	&	5.2	&	112	&	5.8	\\
GK Per	&	114.7	&	\ldots	&	118.7	&	6.0	\\
GK Per	&	121.5	&	4.7	&	122.6	&	5.9	\\
GK Per	&	123.5	&	5.5	&	\ldots	&	\ldots	\\
\enddata
\end{deluxetable}

\clearpage
\begin{deluxetable}{lllllllll}
\tabletypesize{\scriptsize}
\tablecaption{F Class Light Curves
\label{tbl12}}
\tablewidth{0pt}
\tablehead{
\colhead{Nova}   &
\colhead{$N_{b}$}   &
\colhead{Slope 1}   &
\colhead{$T_{b1}-T_0$}   &
\colhead{$V_{b1}$}   &
\colhead{Slope 2}   &
\colhead{$T_{b2}-T_0$}   &
\colhead{$V_{b2}$}   &
\colhead{Slope 3}
}
\startdata

DO Aql\tablenotemark{a}	&	4	&	0.0	&	244	&	8.7	&	-15.0	&	330	&	10.6	&	-0.6	\\
BT Mon	&	2	&	0.0	&	75	&	8.5	&	-5.9	&	900	&	14.8	&	-1.2	\\
V849 Oph\tablenotemark{b}	&	4	&	0.0	&	160	&	9.0	&	-23.5	&	180	&	10.2	&	-2.3	\\
V2295 Oph	&	2	&	0.0	&	11	&	9.7	&	-20.0	&	13	&	11.2	&	-3.9	\\
\enddata
\tablenotetext{a}{DO Aql has another break at 720 days after the eruption at V=10.8, after which the slope is -5.8, followed by a very sharp drop off at 1160 days from V=12.0 to V=15.2 at 1316 days.}
\tablenotetext{b}{V849 Oph has another break at 950 days after the eruption at V=11.8, afterwhich the slope is -11.0.  A final break occur at 1200 days at V=12.9, afterwhich the slope is -3.5.}
\end{deluxetable}

%\clearpage
\begin{deluxetable}{llllllllllll}
\tabletypesize{\scriptsize}
\tablecaption{J Class Light Curves
\label{tbl13}}
\tablewidth{0pt}
\tablehead{
\colhead{Nova}   &
\colhead{$N_{b}$}   &
\colhead{Slope 1}   &
\colhead{$T_{b1}-T_0$}   &
\colhead{$V_{b1}$}   &
\colhead{Slope 2}   &
\colhead{$T_{b2}-T_0$}   &
\colhead{$V_{b2}$}   &
\colhead{Slope 3}	&
\colhead{$\Delta T_{start}$}   &
\colhead{$\Delta T_{end}$}   &
\colhead{$N_{jitter}$}
}
\startdata
V356 Aql	&	1	&	-0.8	&	120	&	9.0	&	-5.3	&	\ldots	&	\ldots	&	\ldots	&	20	&	125	&	9	\\
V723 Cas	&	2	&	0.0	&	100	&	8.8	&	-2.4	&	425	&	10.3	&	-5.6	&	100	&	400	&	6	\\
V868 Cen	&	2	&	-1.0	&	60	&	12.4	&	-5.4	&	130	&	14.2	&	-0.6	&	13	&	135	&	6	\\
V1039 Cen	&	0	&	-2.5	&	\ldots	&	\ldots	&	\ldots	&	\ldots	&	\ldots	&	\ldots	&	5	&	350	&	8	\\
V1819 Cyg	&	2	&	-2.5	&	200	&	12.7	&	-1.1	&	460	&	13.1	&	-5.4	&	13	&	190	&	8	\\
HR Del\tablenotemark{a}	&	3	&	0.0	&	180	&	5.0	&	-3.4	&	380	&	6.1	&	-5.9	&	190	&	400	&	5	\\
DK Lac	&	2	&	-2.9	&	380	&	11.3	&	-8.1	&	600	&	12.9	&	-1.5	&	8	&	355	&	14	\\
V2540 Oph\tablenotemark{b}	&	3	&	-0.7	&	140	&	10.4	&	-6.0	&	280	&	12.2	&	-0.4	&	14	&	155	&	11	\\
RR Pic\tablenotemark{c}	&	4	&	-0.8	&	110	&	3.6	&	-3.9	&	1600	&	8.1	&	-2.0	&	2	&	100	&	3	\\
V4745 Sgr	&	0	&	-1.9	&	\ldots	&	\ldots	&	\ldots	&	\ldots	&	\ldots	&	\ldots	&	10	&	200	&	7	\\
V1186 Sco	&	2	&	-2.9	&	90	&	13.2	&	-0.1	&	510	&	13.3	&	-2.7	&	7	&	90	&	5	\\
V373 Sct	&	2	&	-4.6	&	290	&	12.4	&	-13.8	&	360	&	13.7	&	-4.0	&	30	&	100	&	4	\\
V443 Sct	&	1	&	-1.5	&	75	&	12.0	&	-2.8	&	\ldots	&	\ldots	&	\ldots	&	5	&	85	&	5	\\
PW Vul	&	2	&	0.0	&	55	&	8.4	&	-2.8	&	160	&	9.7	&	-5.9	&	3	&	48	&	4	\\
\enddata
\tablenotetext{a}{HR Del has one additional break at 3300 days with magnitude V=11.5, after which the slope is -1.3.}
\tablenotetext{a}{V2540 Oph has one additional break at 480 days with magnitude V=12.3, after which the slope is -4.2.}
\tablenotetext{c}{RR Pic has an additional break at 4000 days with magnitude V=8.9, after which the slope is -7.  A final break occurs at 8000 days at magnitude V=10.9, after which the slope is -4.}
\end{deluxetable}

\clearpage
\begin{deluxetable}{llllll}
\tabletypesize{\scriptsize}
\tablecaption{Jitter Details
\label{tbl14}}
\tablewidth{0pt}
\tablehead{
\colhead{Nova}   &
\colhead{$\Delta T_{max}$}   &
\colhead{$\Delta T_j$}   &
\colhead{$V_{max}$}   &
\colhead{$\Delta V_{max}$}   &
\colhead{FWHM}
}
\startdata

V356 Aql	&	27.5	&	19.0	&	7.53	&	0.96	&	10	\\
V356 Aql	&	46.5	&	12.0	&	6.97	&	1.70	&	10	\\
V356 Aql	&	58.5	&	4.0	&	7.90	&	0.85	&	9	\\
V356 Aql	&	62.5	&	9.0	&	8.05	&	0.72	&	9	\\
V356 Aql	&	71.5	&	15.0	&	7.37	&	1.45	&	13	\\
V356 Aql	&	86.5	&	15.0	&	7.80	&	1.09	&	5	\\
V356 Aql	&	101.5	&	10.0	&	7.60	&	1.34	&	11	\\
V356 Aql	&	111.5	&	8.0	&	8.20	&	0.77	&	4	\\
V356 Aql	&	119.5	&	\ldots	&	8.50	&	0.50	&	4	\\
	&		&		&		&		&		\\
V723 Cas	&	115.5	&	59.0	&	7.08	&	1.87	&	9	\\
V723 Cas	&	174.5	&	55.0	&	8.20	&	1.18	&	14	\\
V723 Cas	&	229.5	&	117.5	&	8.38	&	1.28	&	11	\\
V723 Cas	&	347.0	&	6.0	&	9.02	&	1.07	&	6	\\
V723 Cas	&	353.0	&	14.0	&	8.87	&	1.23	&	8	\\
V723 Cas	&	367.0	&	\ldots	&	8.65	&	1.50	&	35	\\
	&		&		&		&		&		\\
V868 Cen	&	1.5	&	26.0	&	8.70	&	3.00	&	4	\\
V868 Cen	&	27.5	&	22.0	&	10.45	&	0.96	&	13	\\
V868 Cen	&	49.5	&	6.0	&	11.10	&	1.33	&	7	\\
V868 Cen	&	55.5	&	12.0	&	11.40	&	1.22	&	7	\\
V868 Cen	&	67.5	&	12.0	&	11.97	&	0.98	&	2	\\
V868 Cen	&	79.5	&	49.5	&	11.25	&	1.97	&	12	\\
V868 Cen	&	129.0	&	\ldots	&	13.30	&	0.73	&	10	\\
	&		&		&		&		&		\\
V1039 Cen	&	5.3	&	17.3	&	9.50	&	0.39	&	1.5	\\
V1039 Cen	&	22.5	&	60.0	&	10.40	&	1.06	&	5	\\
V1039 Cen	&	82.5	&	55.0	&	11.70	&	1.17	&	33	\\
V1039 Cen	&	137.5	&	37.0	&	12.00	&	1.42	&	8	\\
V1039 Cen	&	174.5	&	41.0	&	12.20	&	1.48	&	7	\\
V1039 Cen	&	215.5	&	41.0	&	12.70	&	1.20	&	19	\\
V1039 Cen	&	256.5	&	53.0	&	13.20	&	0.89	&	18	\\
V1039 Cen	&	309.5	&	\ldots	&	12.40	&	1.90	&	360.5	\\
	&		&		&		&		&		\\
V1819 Cyg	&	13.5	&	23.0	&	9.50	&	0.27	&	2	\\
V1819 Cyg	&	36.5	&	27.0	&	10.20	&	0.65	&	4	\\
V1819 Cyg	&	63.5	&	22.0	&	10.20	&	1.25	&	12	\\
V1819 Cyg	&	85.5	&	10.0	&	10.42	&	1.36	&	15	\\
V1819 Cyg	&	95.5	&	44.5	&	11.19	&	0.71	&	19.5	\\
V1819 Cyg	&	140.0	&	25.0	&	11.95	&	0.36	&	6	\\
V1819 Cyg	&	165.0	&	16.0	&	12.03	&	0.46	&	16	\\
V1819 Cyg	&	181.0	&	\ldots	&	11.96	&	0.63	&	16	\\
	&		&		&		&		&		\\
HR Del	&	193.0	&	54.0	&	3.58	&	1.52	&	14	\\
HR Del	&	247.0	&	10.0	&	4.45	&	1.01	&	8	\\
HR Del	&	257.0	&	12.0	&	4.75	&	0.77	&	4	\\
HR Del	&	269.0	&	67.0	&	4.80	&	0.79	&	14	\\
HR Del	&	336.0	&	\ldots	&	4.40	&	1.52	&	48	\\
	&		&		&		&		&		\\
DK Lac	&	18.5	&	11.0	&	6.60	&	0.67	&	2	\\
DK Lac	&	29.5	&	14.0	&	7.37	&	0.58	&	3	\\
DK Lac	&	43.5	&	6.0	&	7.67	&	0.81	&	4	\\
DK Lac	&	49.5	&	10.0	&	7.90	&	0.75	&	3	\\
DK Lac	&	59.5	&	26.0	&	7.50	&	1.40	&	2	\\
DK Lac	&	85.5	&	21.0	&	7.97	&	1.41	&	10	\\
DK Lac	&	106.5	&	20.0	&	8.35	&	1.31	&	4	\\
DK Lac	&	126.5	&	53.0	&	9.20	&	0.68	&	4	\\
DK Lac	&	179.5	&	26.0	&	8.20	&	2.13	&	4	\\
DK Lac	&	205.5	&	49.0	&	8.60	&	1.91	&	4	\\
DK Lac	&	254.5	&	37.0	&	9.67	&	1.11	&	5	\\
DK Lac	&	291.5	&	7.5	&	10.20	&	0.75	&	4	\\
DK Lac	&	299.0	&	46.0	&	9.95	&	1.04	&	6.5	\\
DK Lac	&	345.0	&	\ldots	&	10.23	&	0.94	&	12	\\
	&		&		&		&		&		\\
V2540 Oph	&	7.3	&	12.5	&	8.40	&	1.19	&	7	\\
V2540 Oph	&	19.8	&	9.5	&	8.29	&	1.59	&	4.5	\\
V2540 Oph	&	29.3	&	15.0	&	9.20	&	0.78	&	2.5	\\
V2540 Oph	&	44.3	&	22.5	&	8.55	&	1.54	&	6.5	\\
V2540 Oph	&	66.8	&	7.5	&	8.67	&	1.54	&	4	\\
V2540 Oph	&	74.3	&	25.5	&	8.10	&	2.13	&	12	\\
V2540 Oph	&	99.8	&	2.5	&	9.85	&	0.46	&	2.5	\\
V2540 Oph	&	102.3	&	11.5	&	9.40	&	0.92	&	2	\\
V2540 Oph	&	113.8	&	21.8	&	8.95	&	1.39	&	5.5	\\
V2540 Oph	&	135.5	&	5.0	&	9.93	&	0.46	&	4	\\
V2540 Oph	&	140.5	&	\ldots	&	9.97	&	0.43	&	3	\\
	&		&		&		&		&		\\
RR Pic	&	9.5	&	50.5	&	0.95	&	1.80	&	9	\\
RR Pic	&	60.0	&	12.0	&	1.65	&	1.74	&	12	\\
RR Pic	&	72.0	&	\ldots	&	2.04	&	1.41	&	12	\\
	&		&		&		&		&		\\
V4745 Sgr	&	22.8	&	8.0	&	8.80	&	0.72	&	6	\\
V4745 Sgr	&	30.8	&	21.2	&	8.10	&	2.10	&	8.25	\\
V4745 Sgr	&	52.0	&	26.0	&	8.90	&	1.93	&	10	\\
V4745 Sgr	&	78.0	&	20.0	&	8.95	&	2.19	&	18	\\
V4745 Sgr	&	98.0	&	30.0	&	10.40	&	0.90	&	6	\\
V4745 Sgr	&	128.0	&	60.0	&	9.37	&	2.11	&	12	\\
V4745 Sgr	&	188.0	&	\ldots	&	9.50	&	2.21	&	14	\\
	&		&		&		&		&		\\
V1186 Sco	&	15.8	&	8.0	&	9.65	&	1.38	&	5	\\
V1186 Sco	&	23.8	&	22.3	&	10.31	&	1.24	&	7.5	\\
V1186 Sco	&	46.0	&	22.0	&	12.03	&	0.36	&	6	\\
V1186 Sco	&	68.0	&	10.0	&	12.15	&	0.73	&	16	\\
V1186 Sco	&	78.0	&	\ldots	&	11.89	&	1.16	&	4	\\
	&		&		&		&		&		\\
V373 Sct	&	45.5	&	14.0	&	7.20	&	1.50	&	18	\\
V373 Sct	&	59.5	&	11.0	&	7.85	&	1.07	&	4	\\
V373 Sct	&	70.5	&	20.0	&	8.74	&	0.57	&	2	\\
V373 Sct	&	90.5	&	\ldots	&	8.96	&	0.91	&	5	\\
	&		&		&		&		&		\\
V443 Sct	&	25.8	&	6.0	&	9.66	&	1.65	&	10.5	\\
V443 Sct	&	31.8	&	9.8	&	10.60	&	0.84	&	1.75	\\
V443 Sct	&	41.5	&	19.5	&	9.84	&	1.77	&	8	\\
V443 Sct	&	61.0	&	\ldots	&	11.03	&	0.84	&	17.5	\\
	&		&		&		&		&		\\
PW Vul	&	8.8	&	7.8	&	6.41	&	1.99	&	5.5	\\
PW Vul	&	16.5	&	6.0	&	8.16	&	0.24	&	3	\\
PW Vul	&	22.5	&	14.0	&	7.49	&	0.91	&	5	\\
PW Vul	&	36.5	&	\ldots	&	7.57	&	0.83	&	8	\\
\enddata
\end{deluxetable}

\clearpage
\begin{figure}
\epsscale{0.7}
\plotone{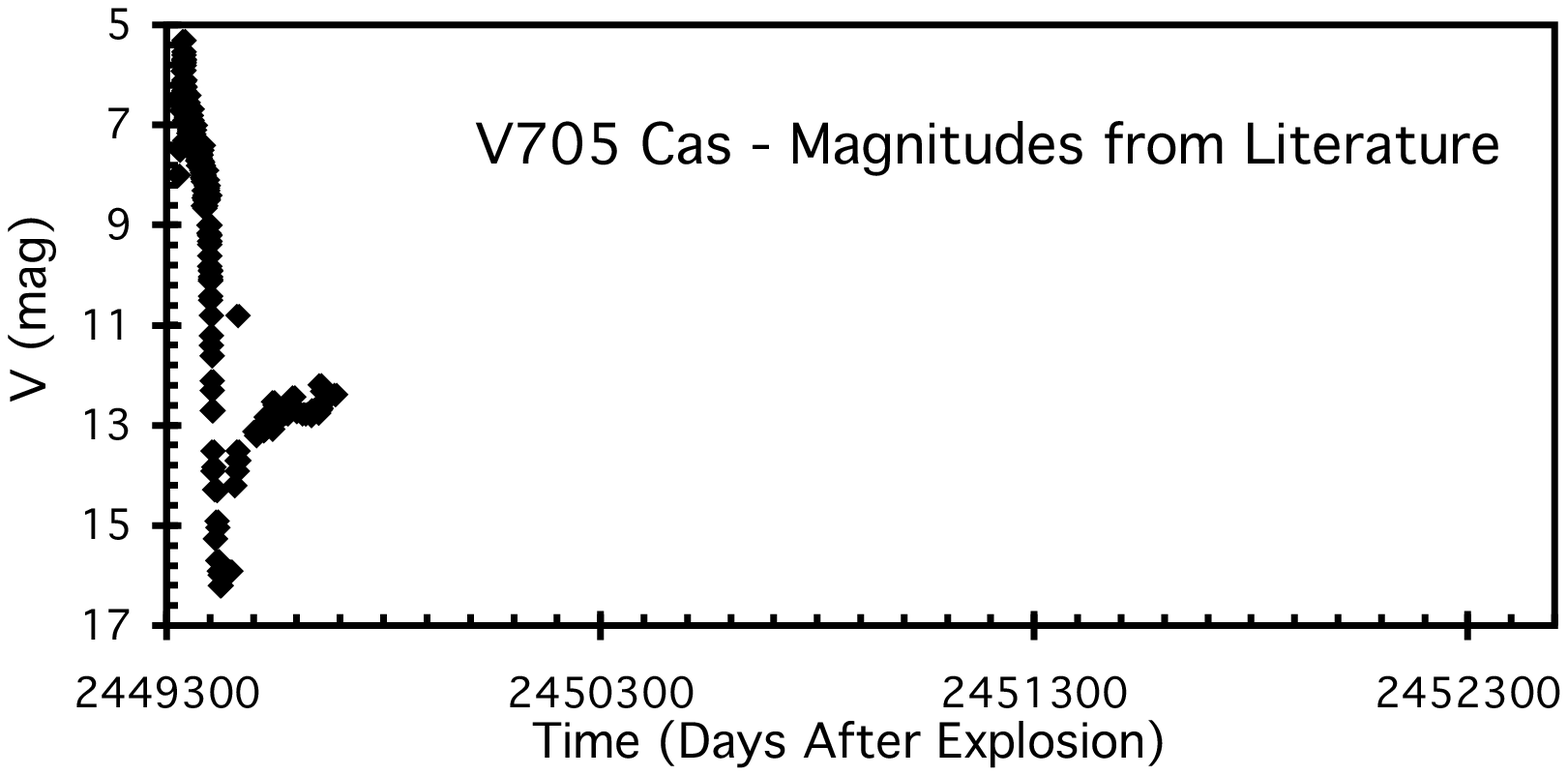}
\plotone{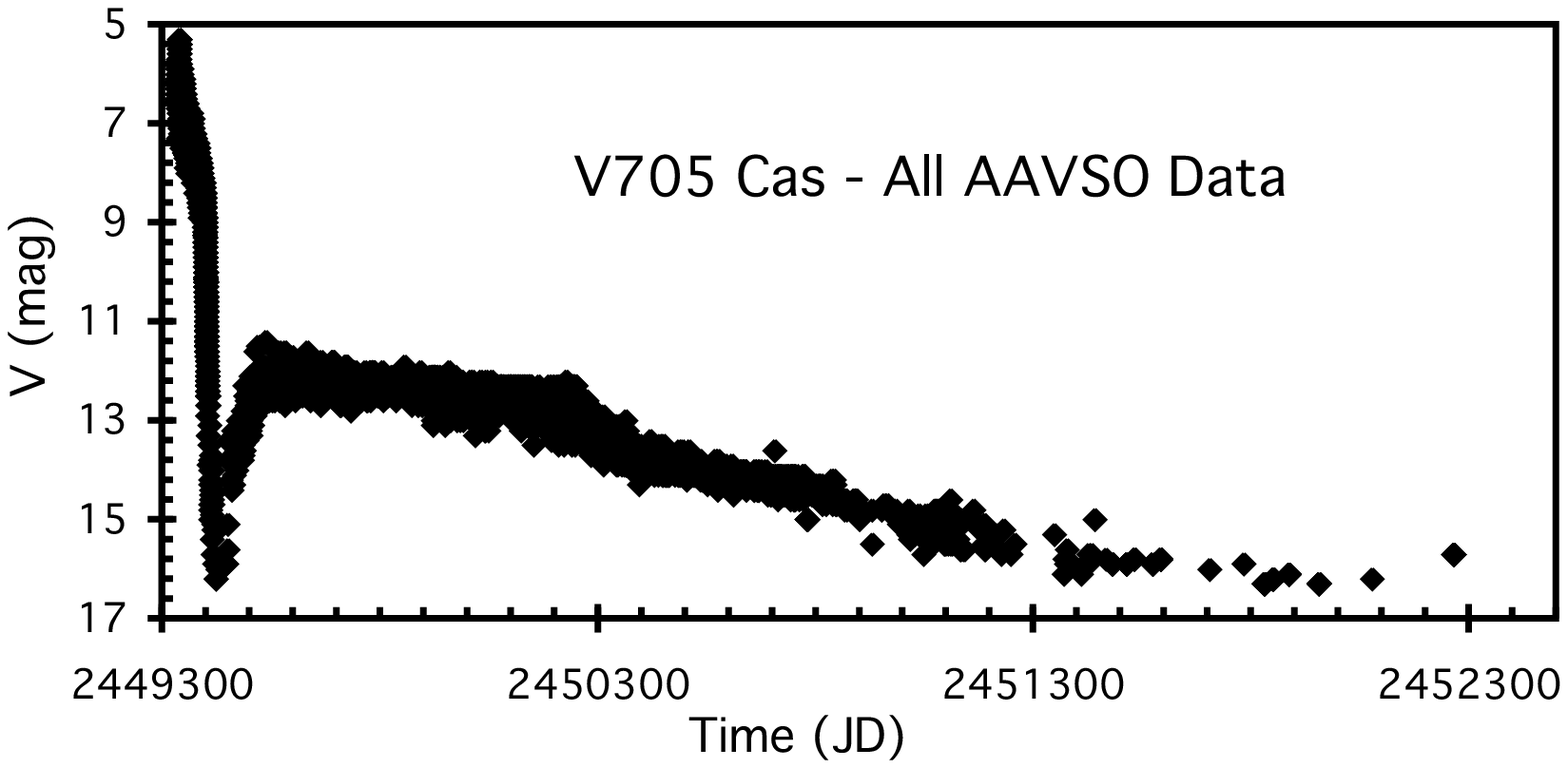}
\epsscale{0.73}
\plotone{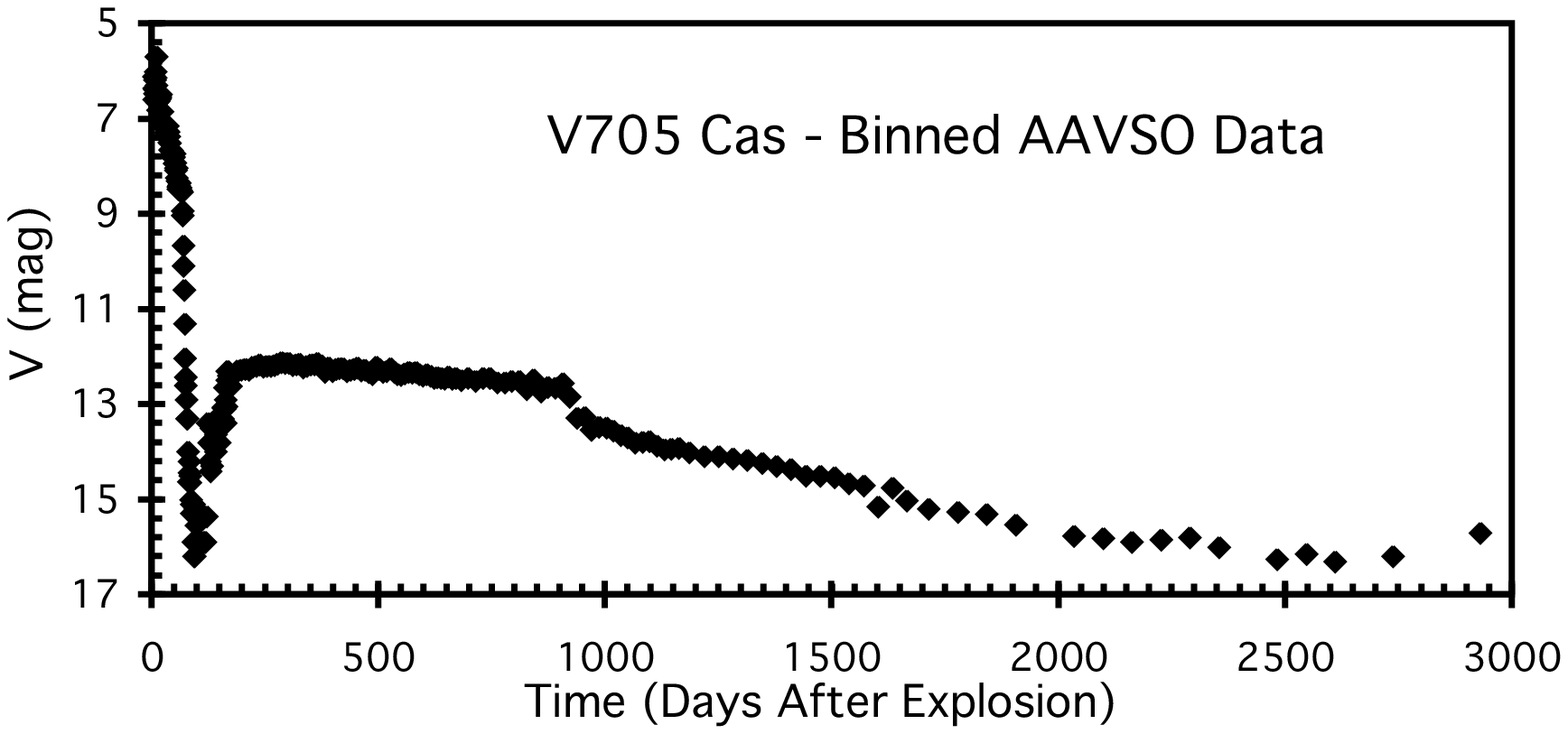}
\caption{
V705 Cas light curves.  The three panels display the light curves for V705 Cas for (a) all V-band magnitudes published in the professional literature (Hric et al. 1998; Kijewski et al. 1995; Munari et al. 1994; Skiff 1994; and from many observers in IAU Circulars 5902, 5904, 5905, 5910, 5912, 5914, 5920, 5925, 5928, 5929, 5934, 5939, 5945, 5954, 5957, 5978, and 5980), (b) all V-band magnitudes in the AAVSO database, and (c) the binned and averaged AAVSO light curve.  V705 Cas is one of the best observed novae as reported in the professional literature.  A comparison between the first two panels demonstrates the strength of the AAVSO program in that it provides wonderful coverage that continues until late times.  A comparison between the last two panels demonstrates the binning process by which we created our final light curve for analysis.}
\end{figure}

\clearpage
\begin{figure}
\epsscale{0.9}
\plotone{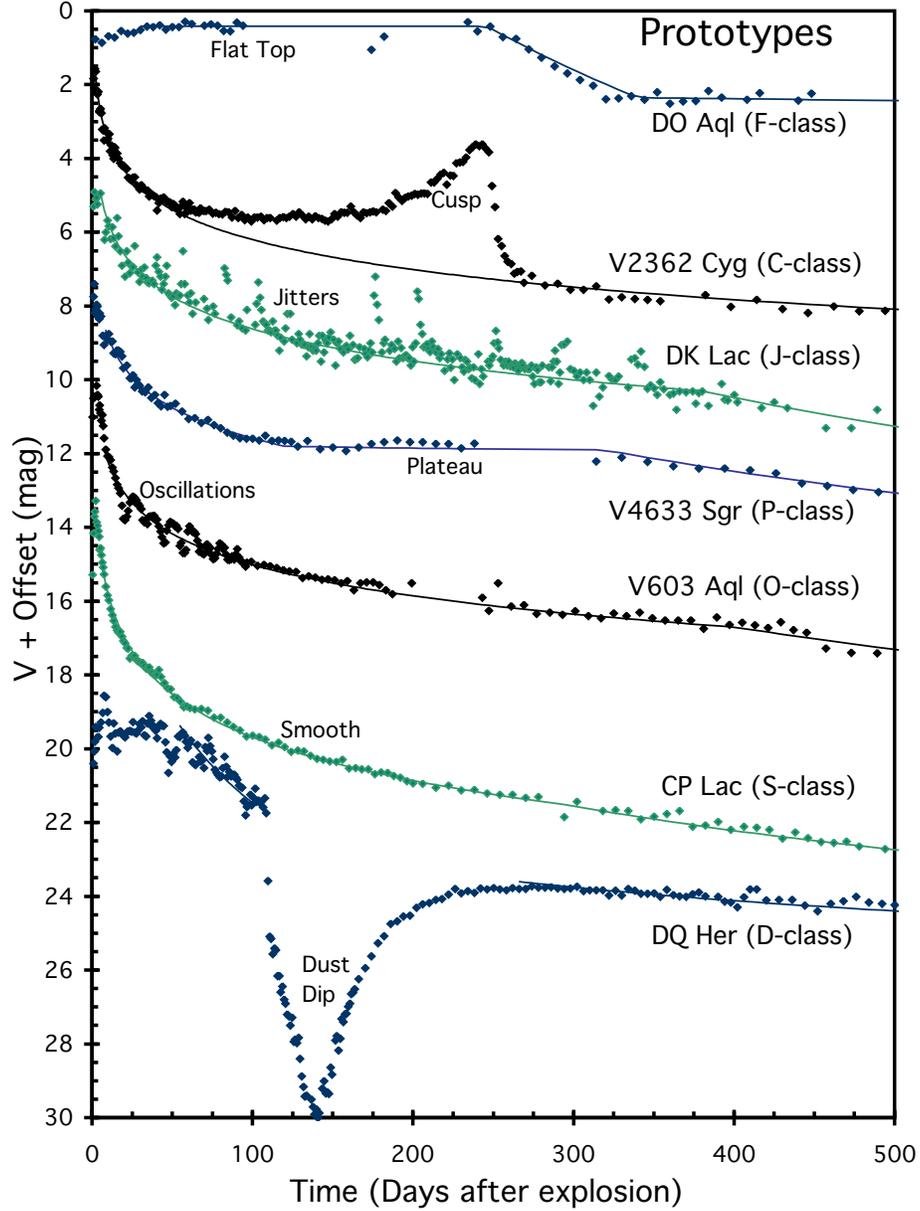}
\caption{
Prototypes of light curves classes.  The seven binned light curves show the distinct features of each class (see Table 3).}
\end{figure}

\clearpage
\begin{figure}
\epsscale{1.0}
\plottwo{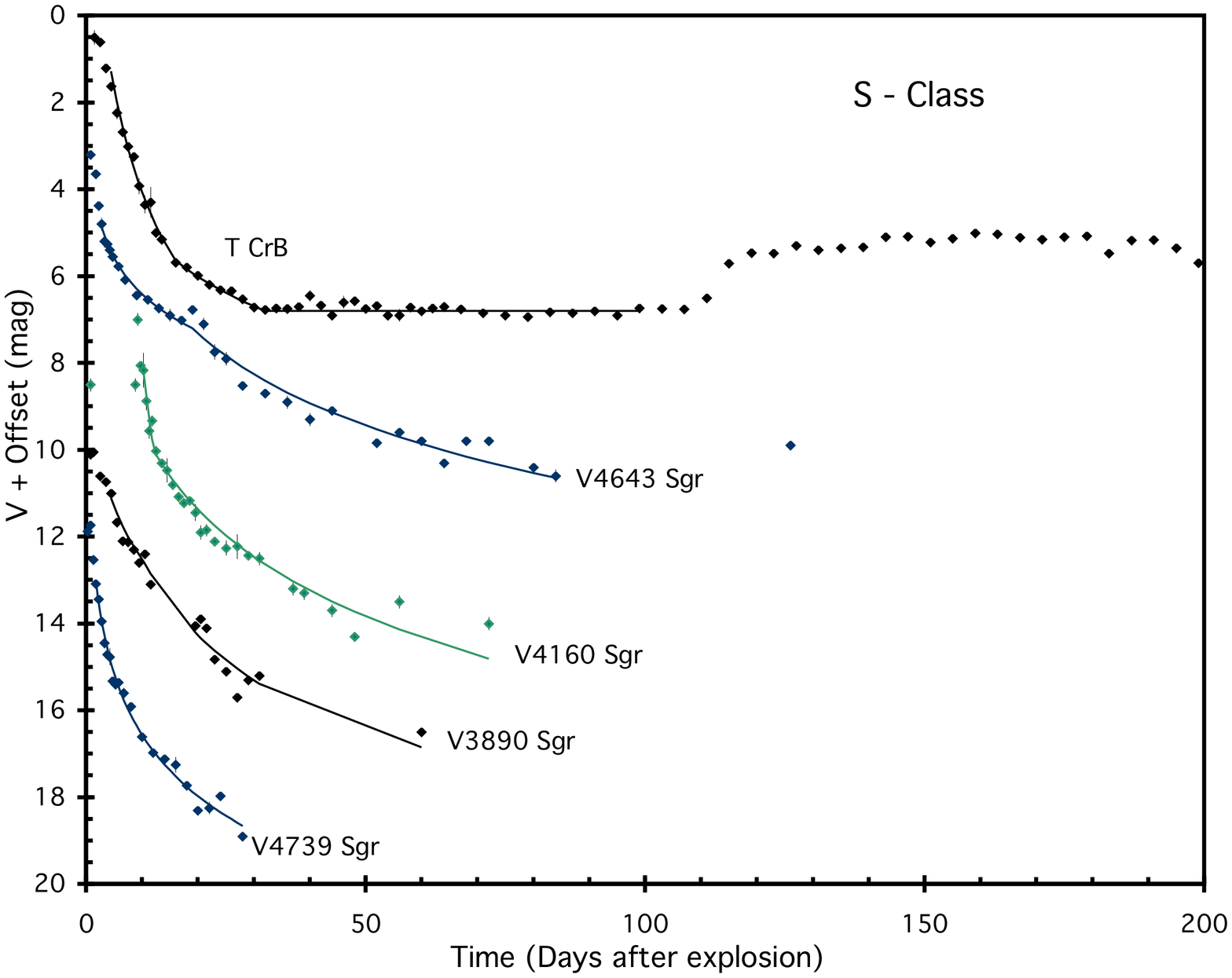}{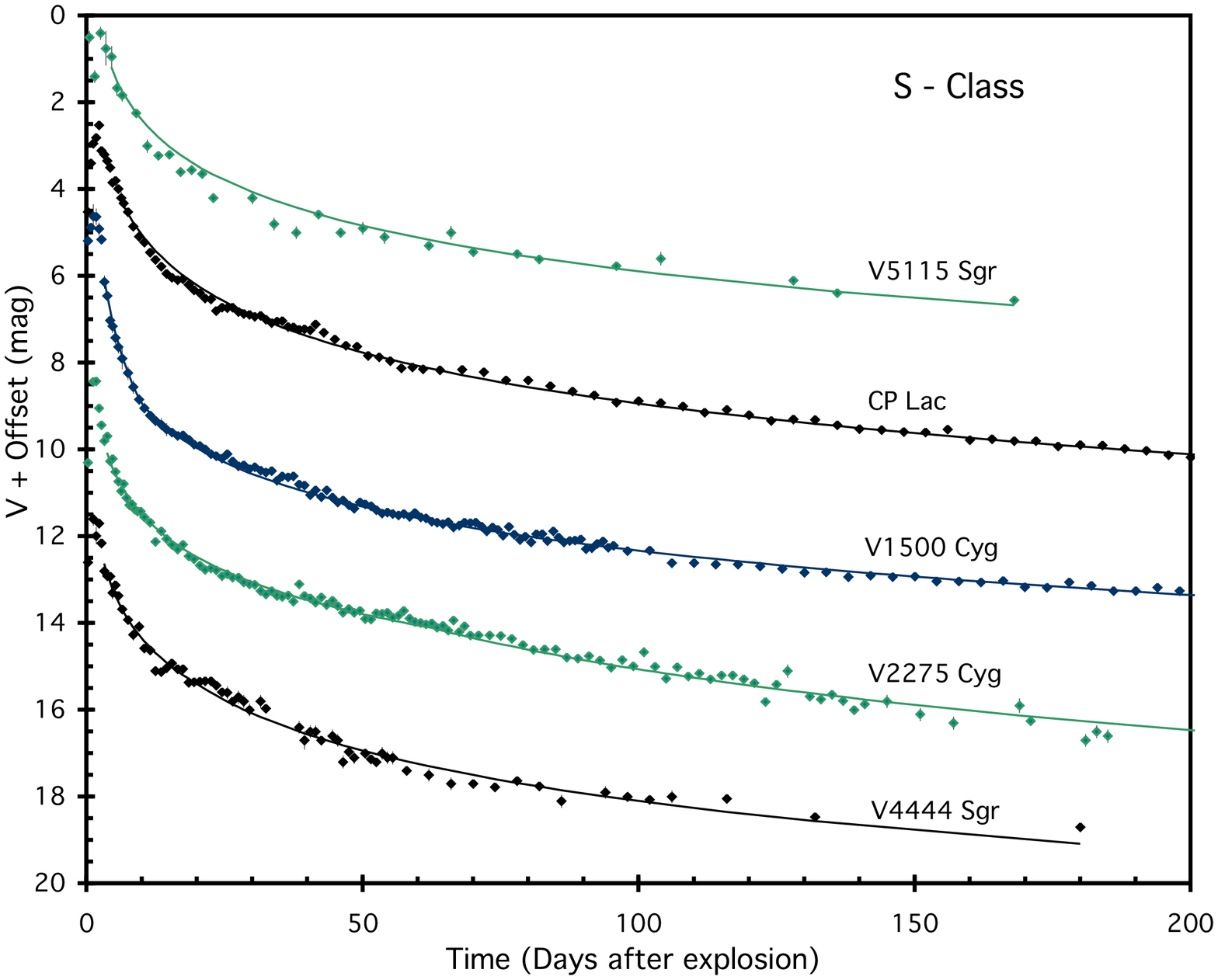}
\plottwo{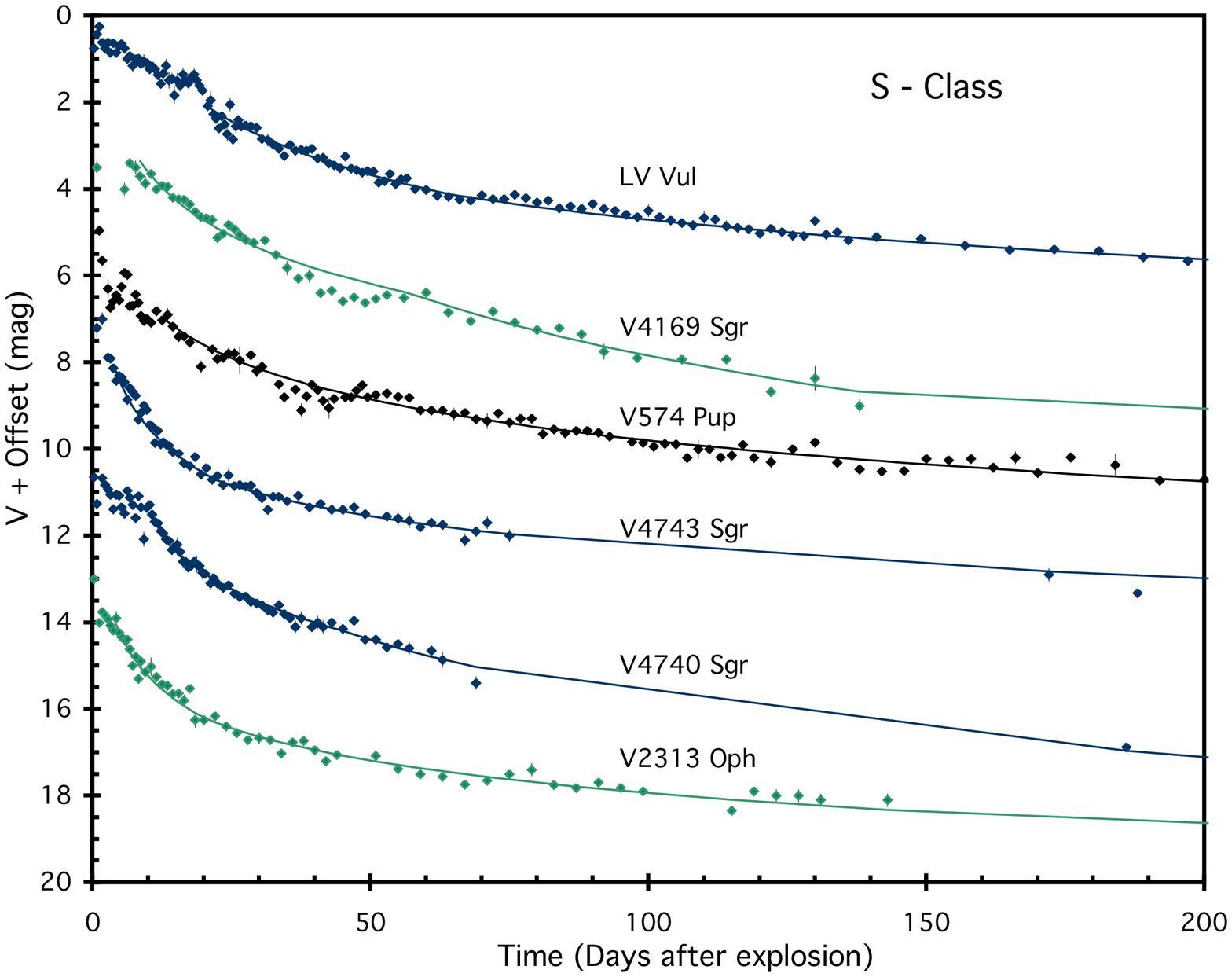}{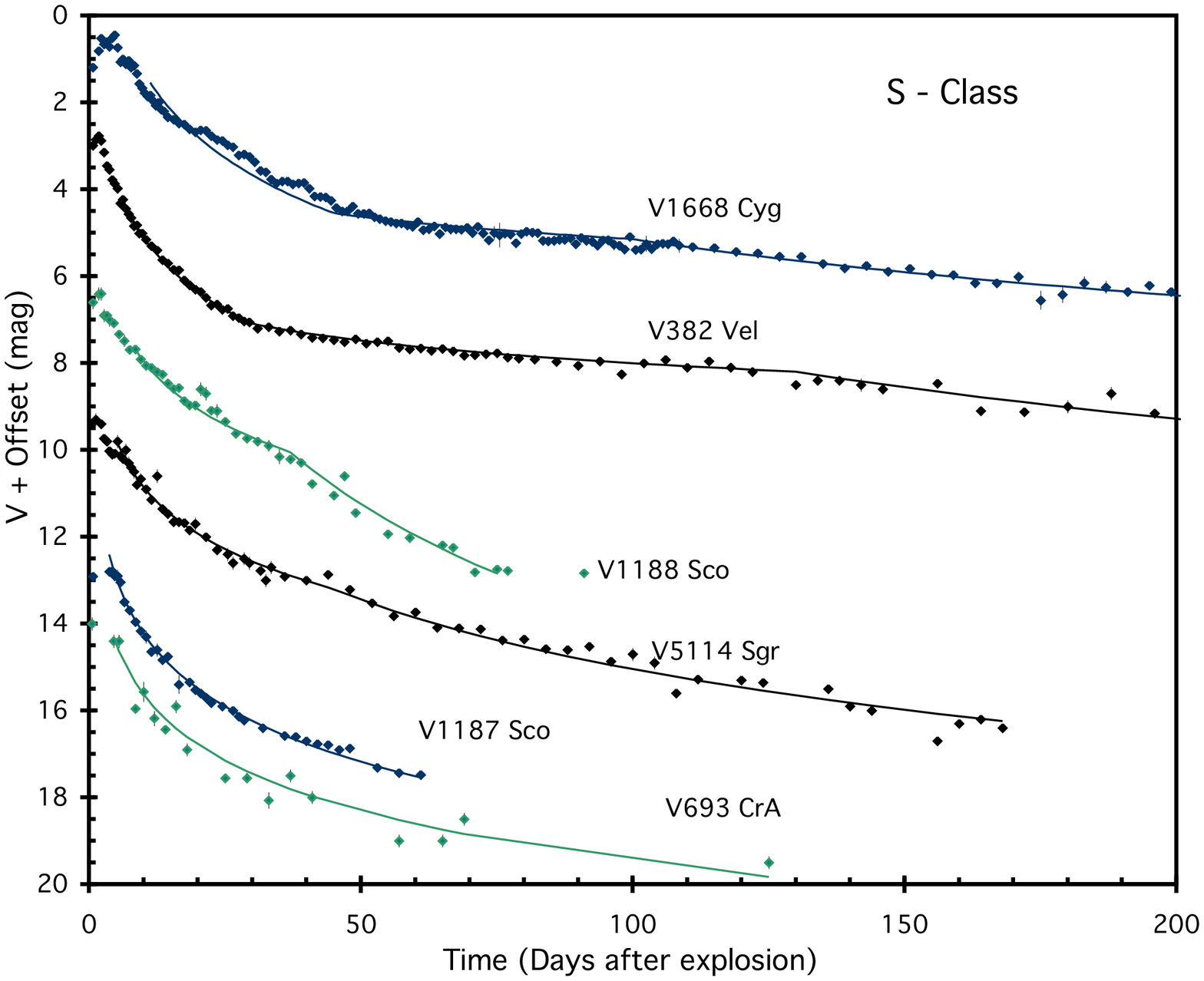}
\plottwo{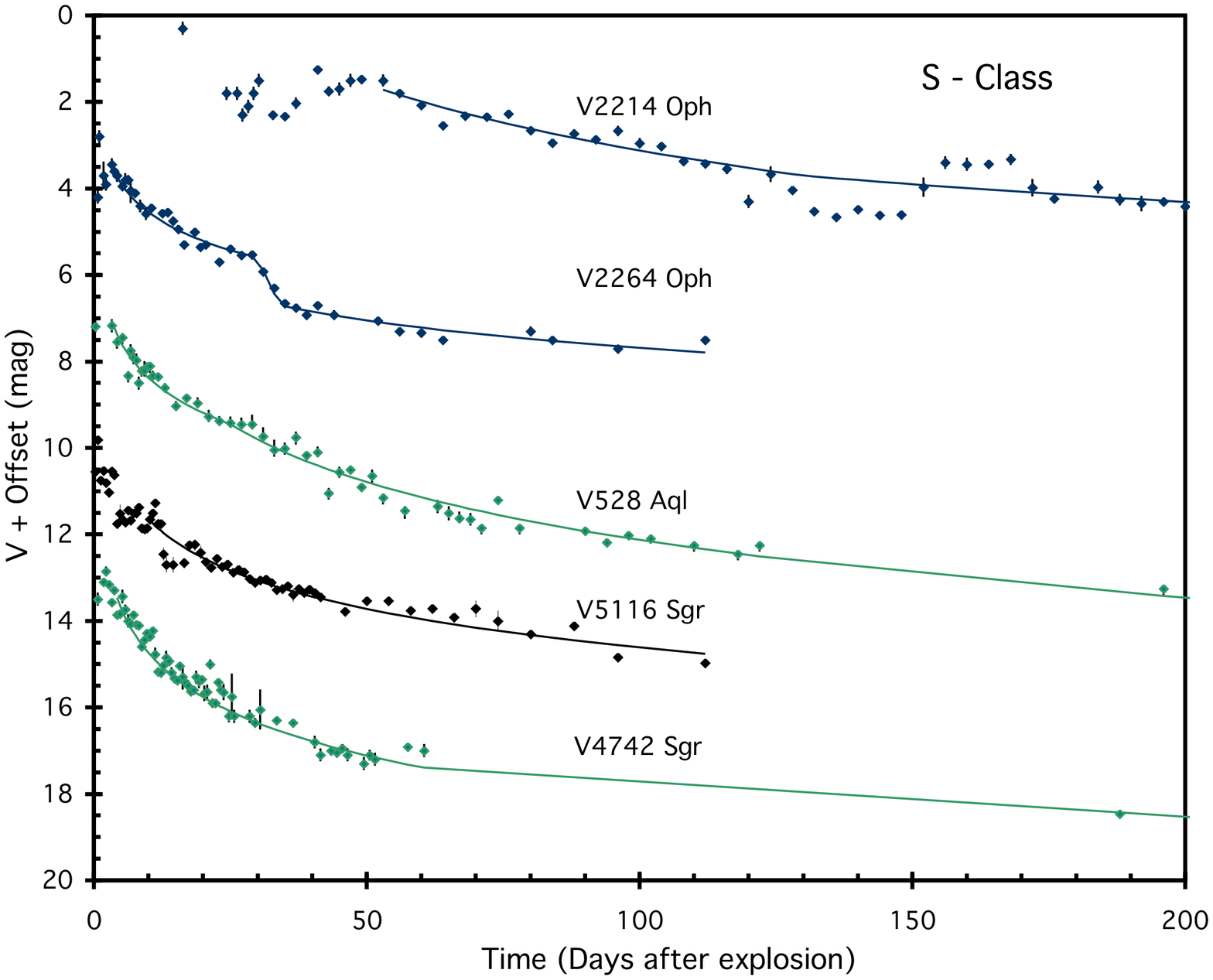}{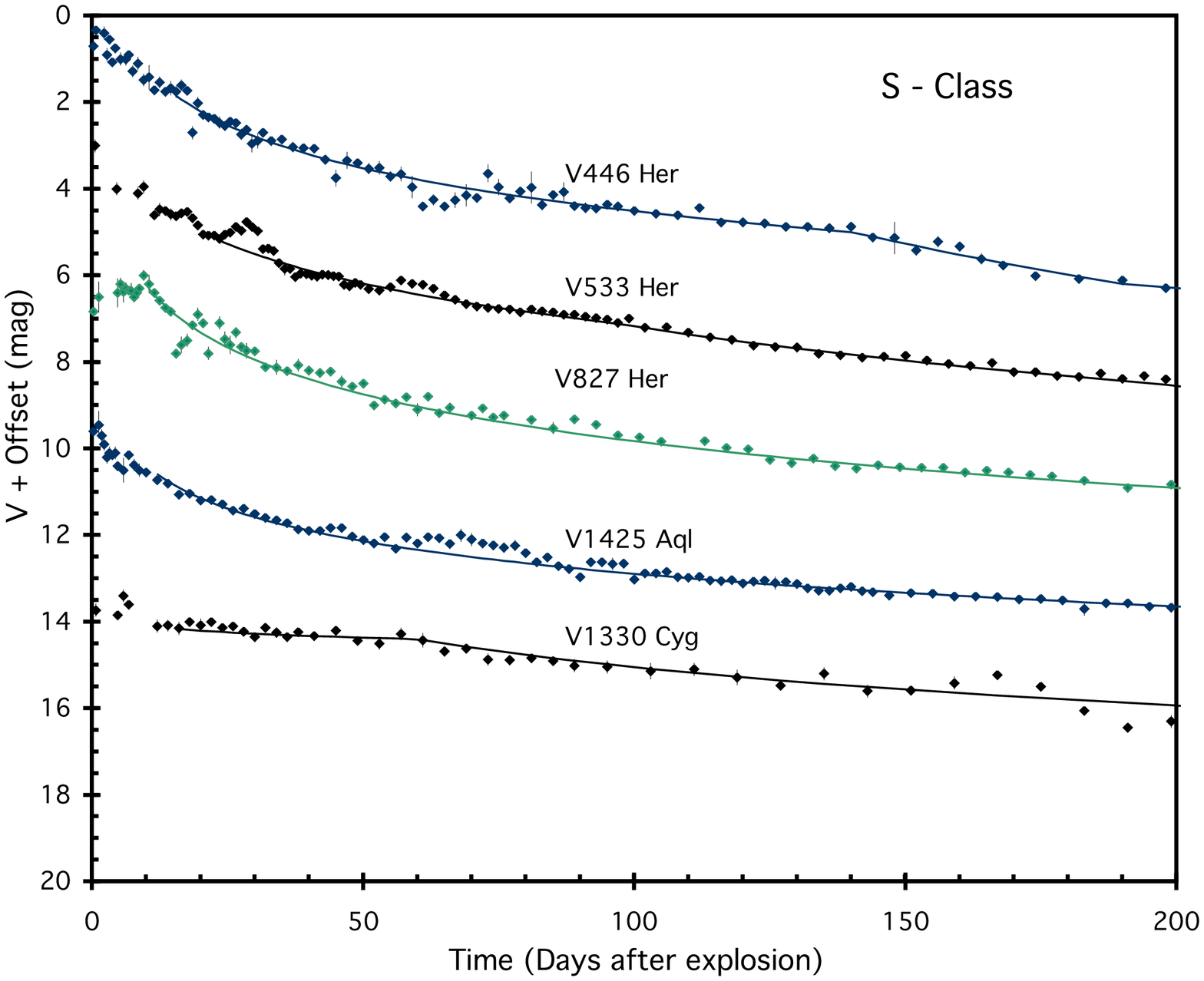}
\caption{
S class light curves.  These six panels show the 32 S class novae with smoothly declining light curves.  The horizontal axis is linear, the traditional display format, and shows only the first 200 days so as to get details around the peak.  Each nova has its V-band light curve with an arbitrary offset added so that the light curves are well separated, with the shape being retained.  Curved line segments are superposed on the binned and averaged magnitudes (diamonds) to represent the fitted broken power law models.}
\end{figure}

\clearpage
\begin{figure}
\epsscale{1.0}
\plottwo{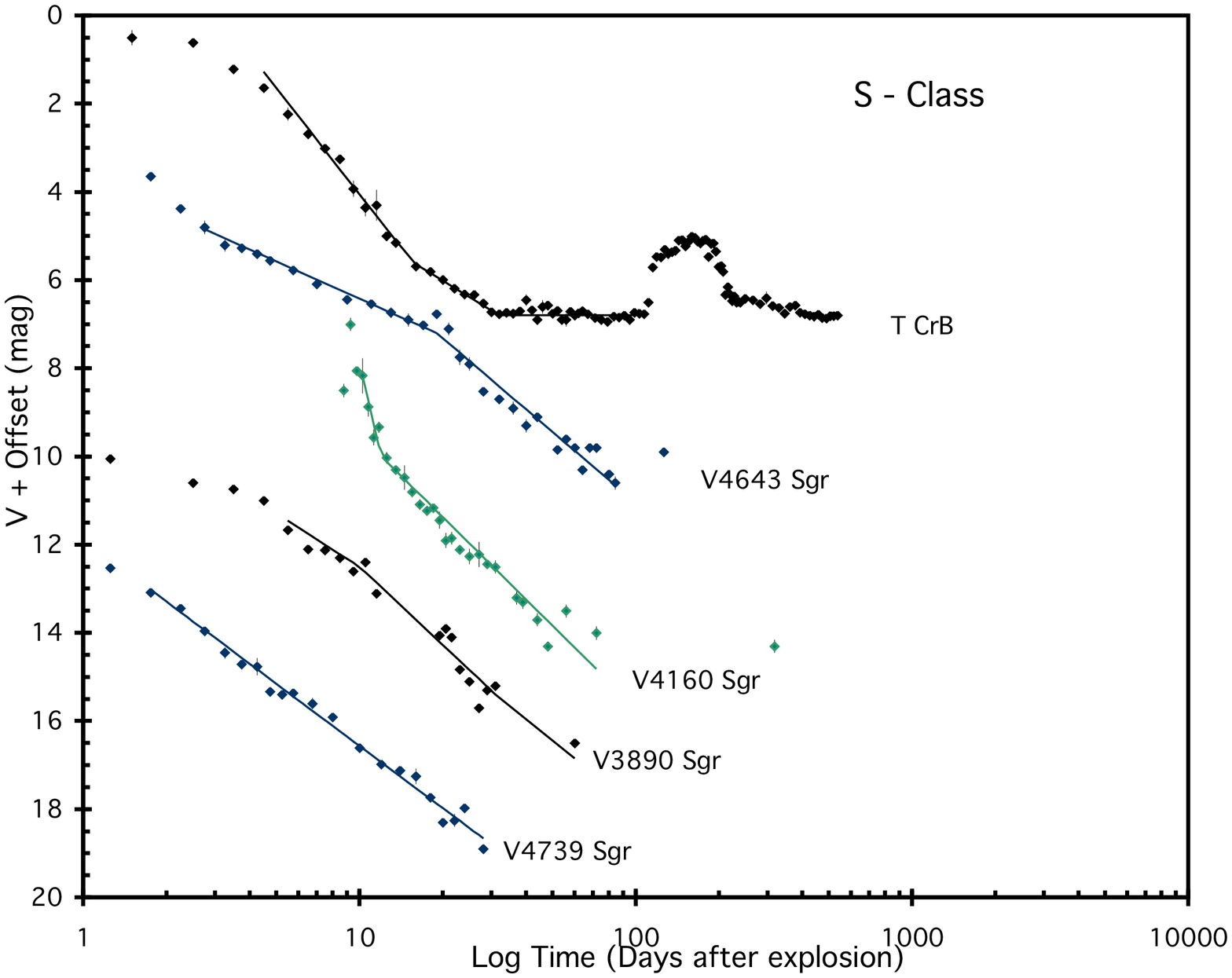}{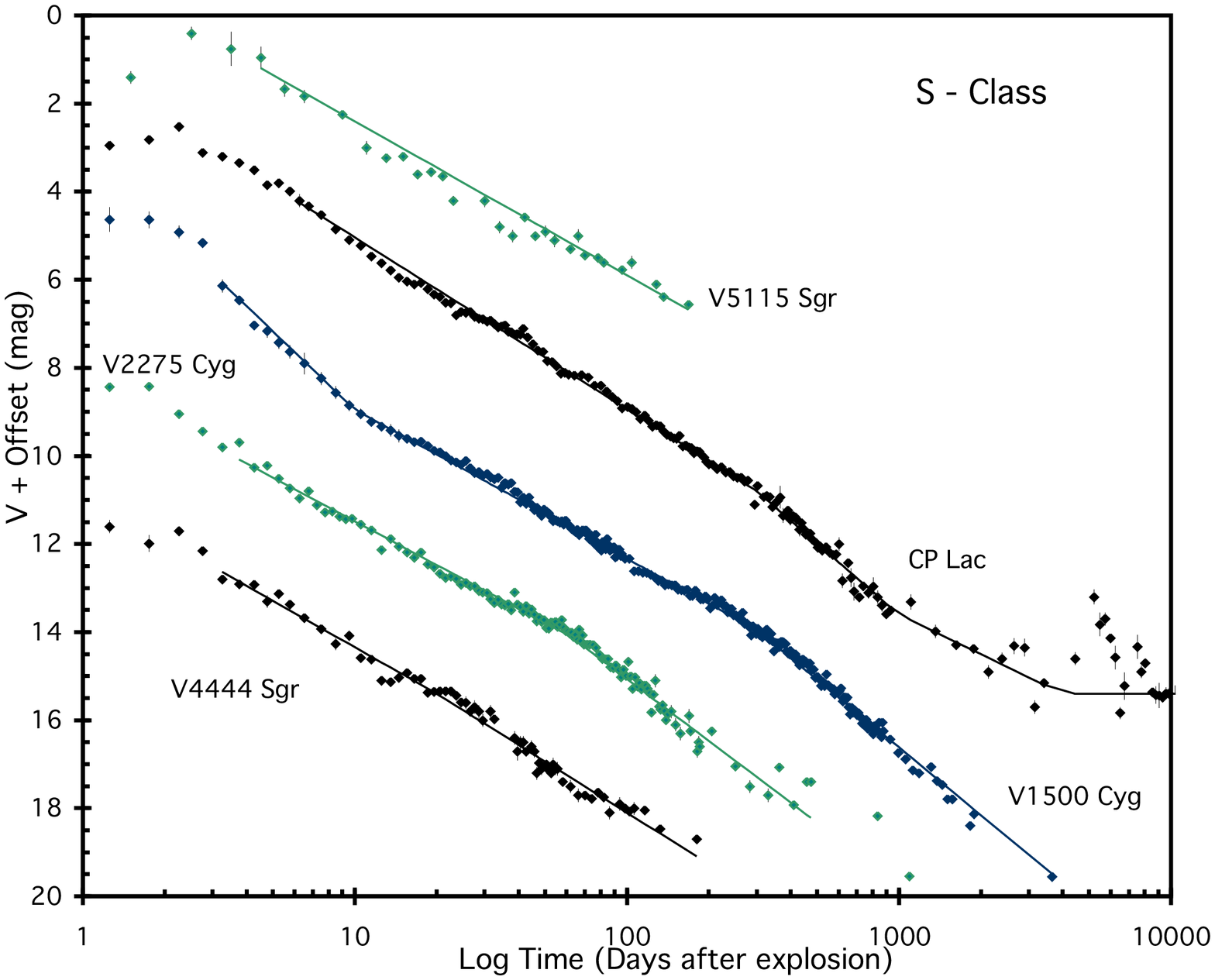}
\plottwo{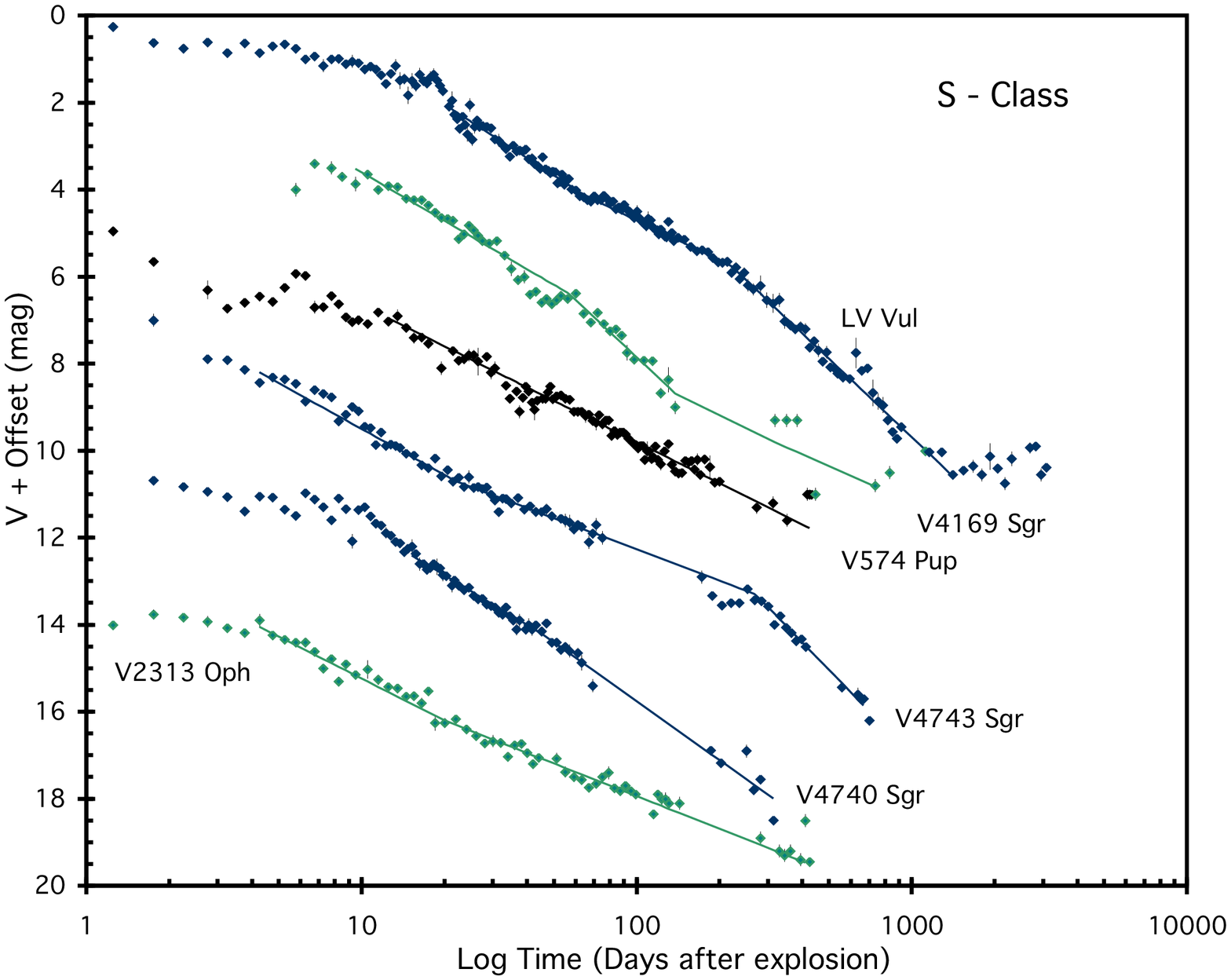}{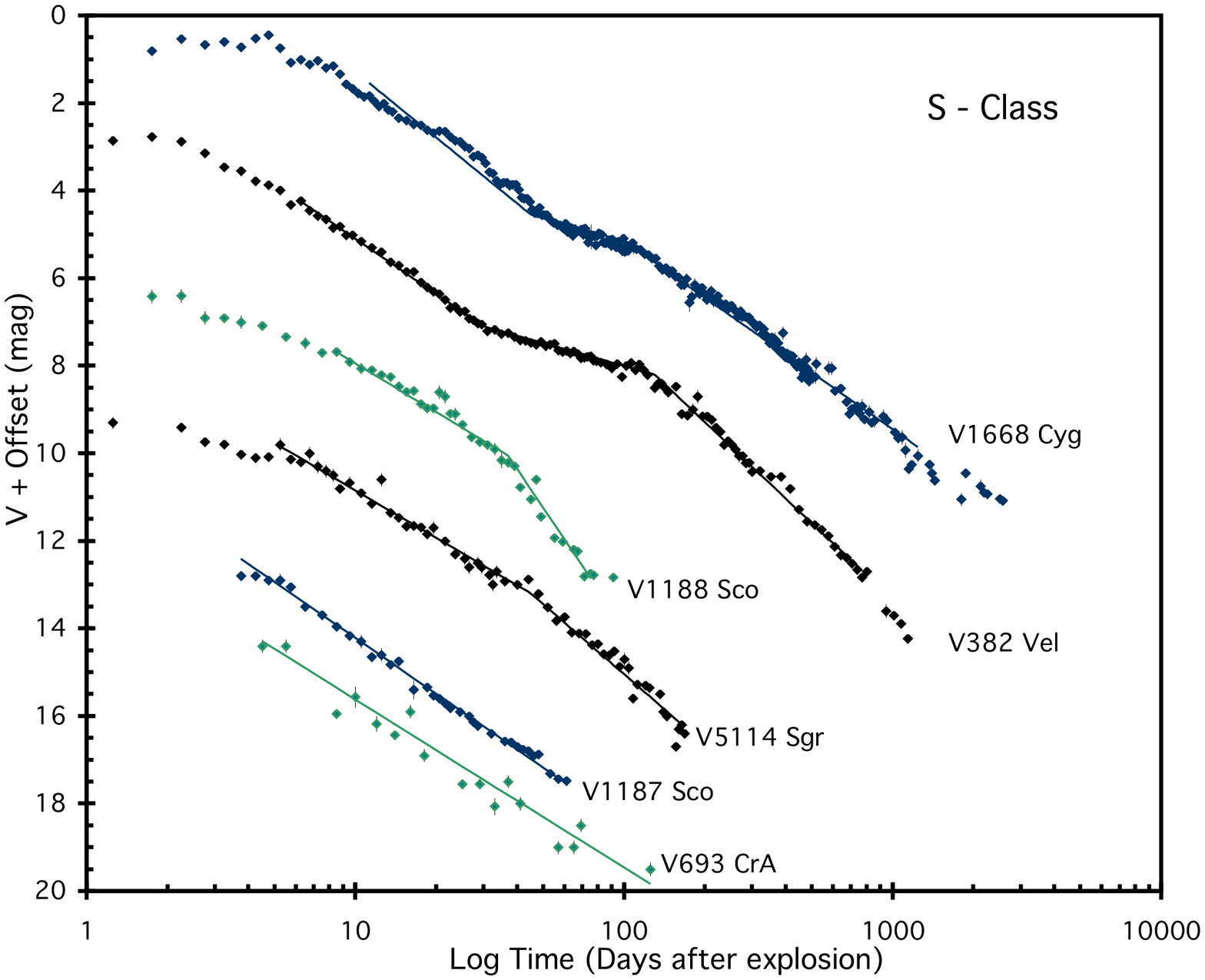}
\plottwo{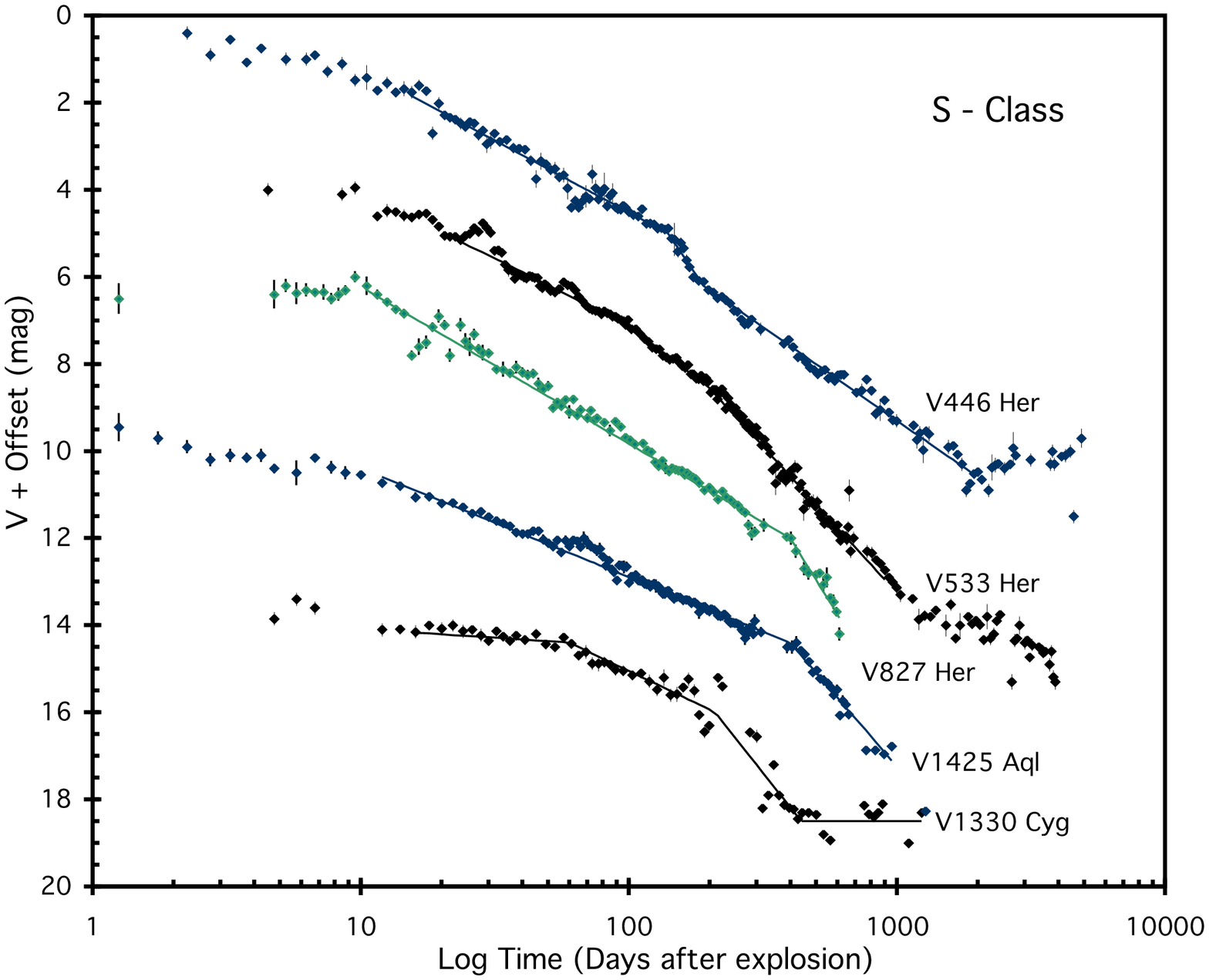}{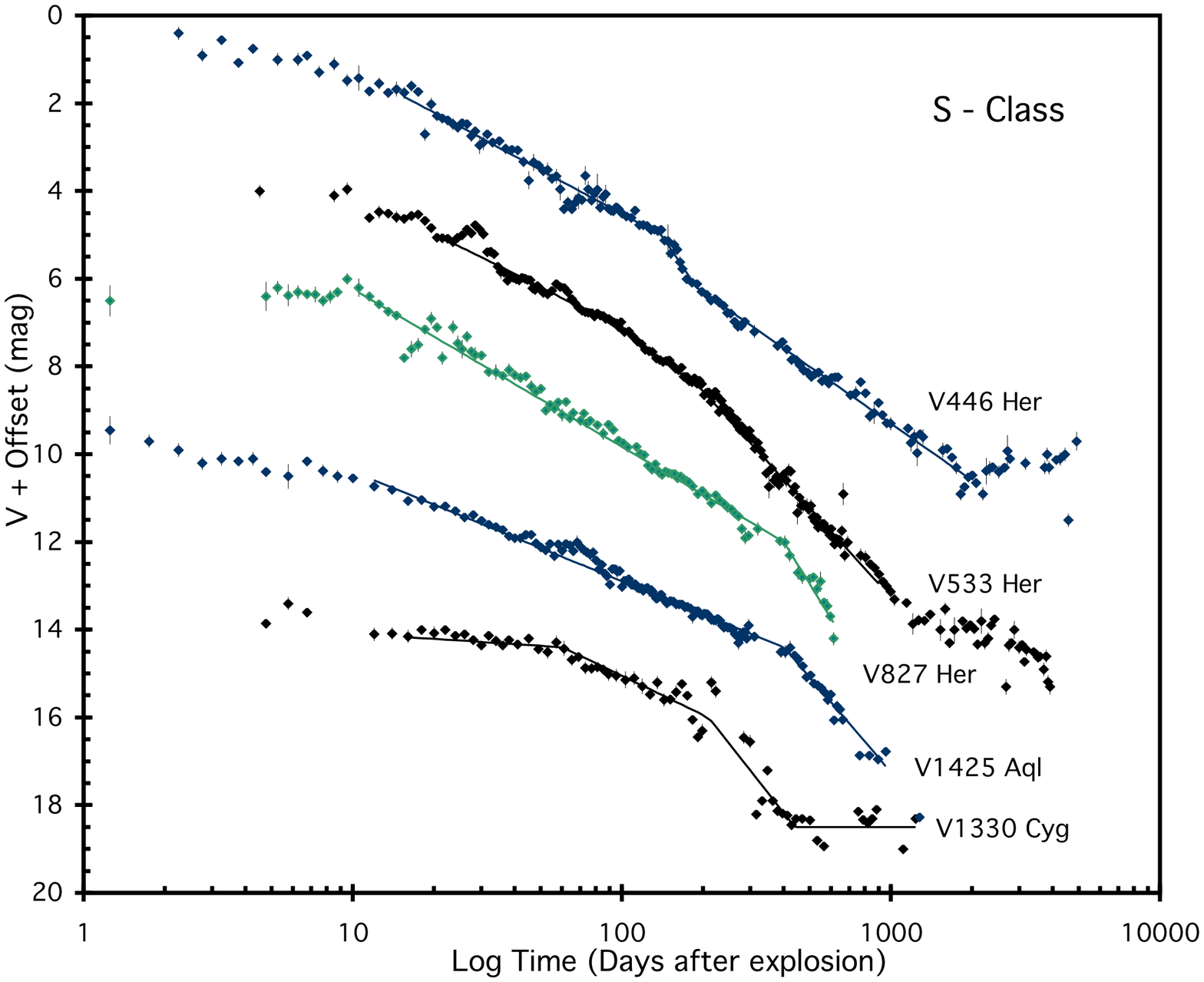}
\caption{
S class light curves on a logarithmic scale.  These six panels show the 32 S class novae with a log scale for the time axis.  The data and symbols are the same as for Figure 3.  However, the log scale allows data all the way out to very late times to all be usefully displayed.  Also, the log scale allows for the power law segments to be readily visible as straight lines.}
\end{figure}

\clearpage
\begin{figure}
\epsscale{1.1}
\plottwo{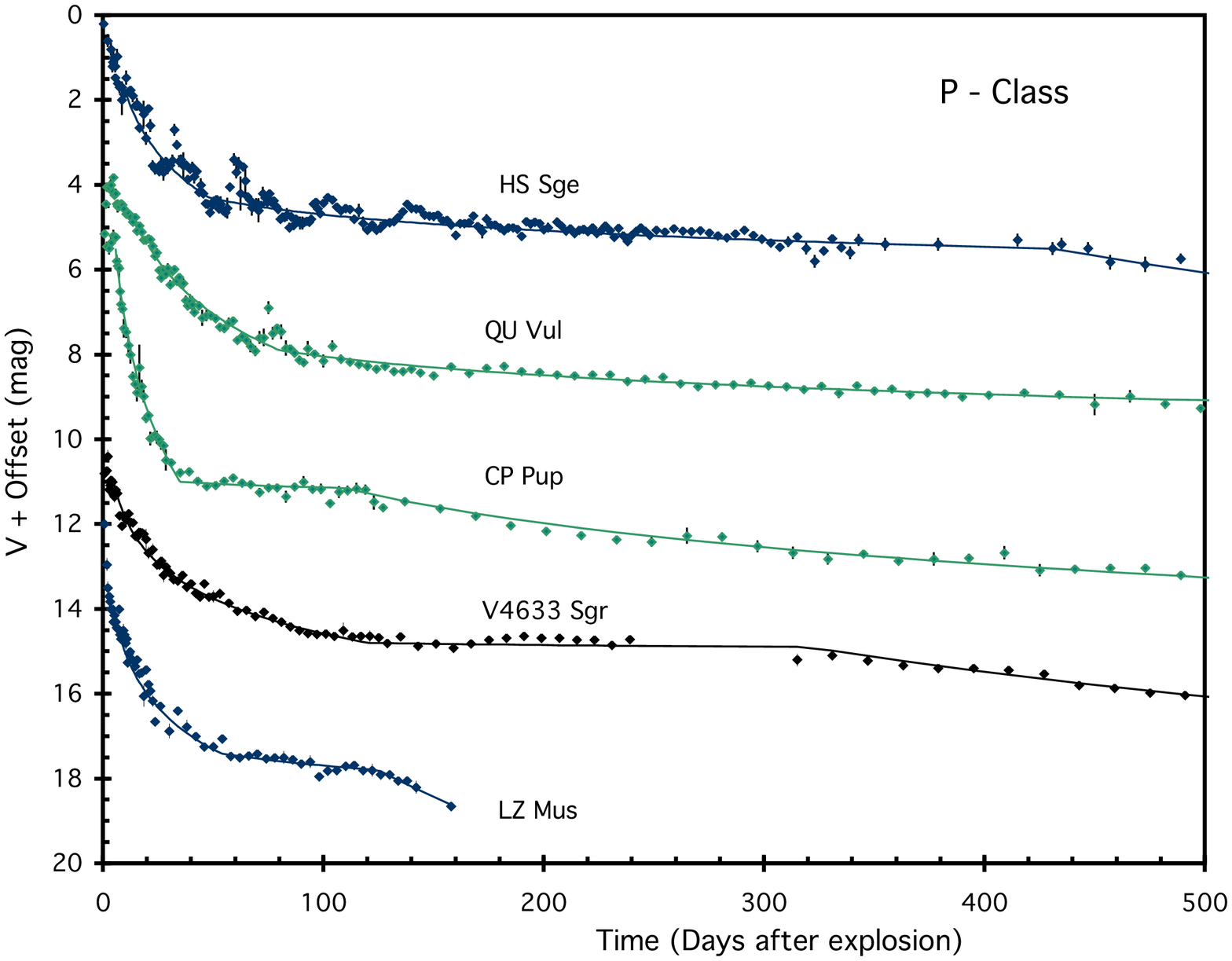}{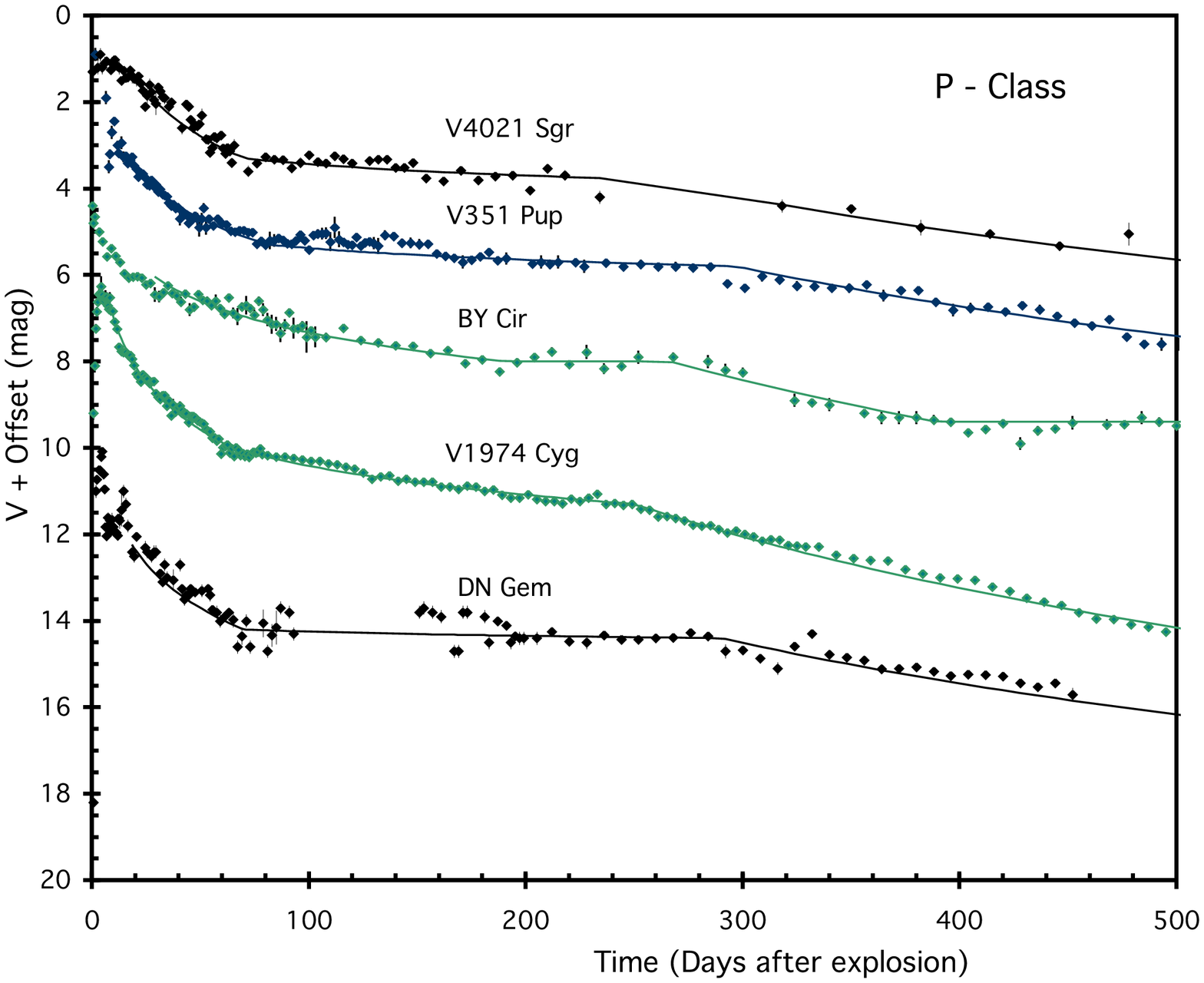}
\plottwo{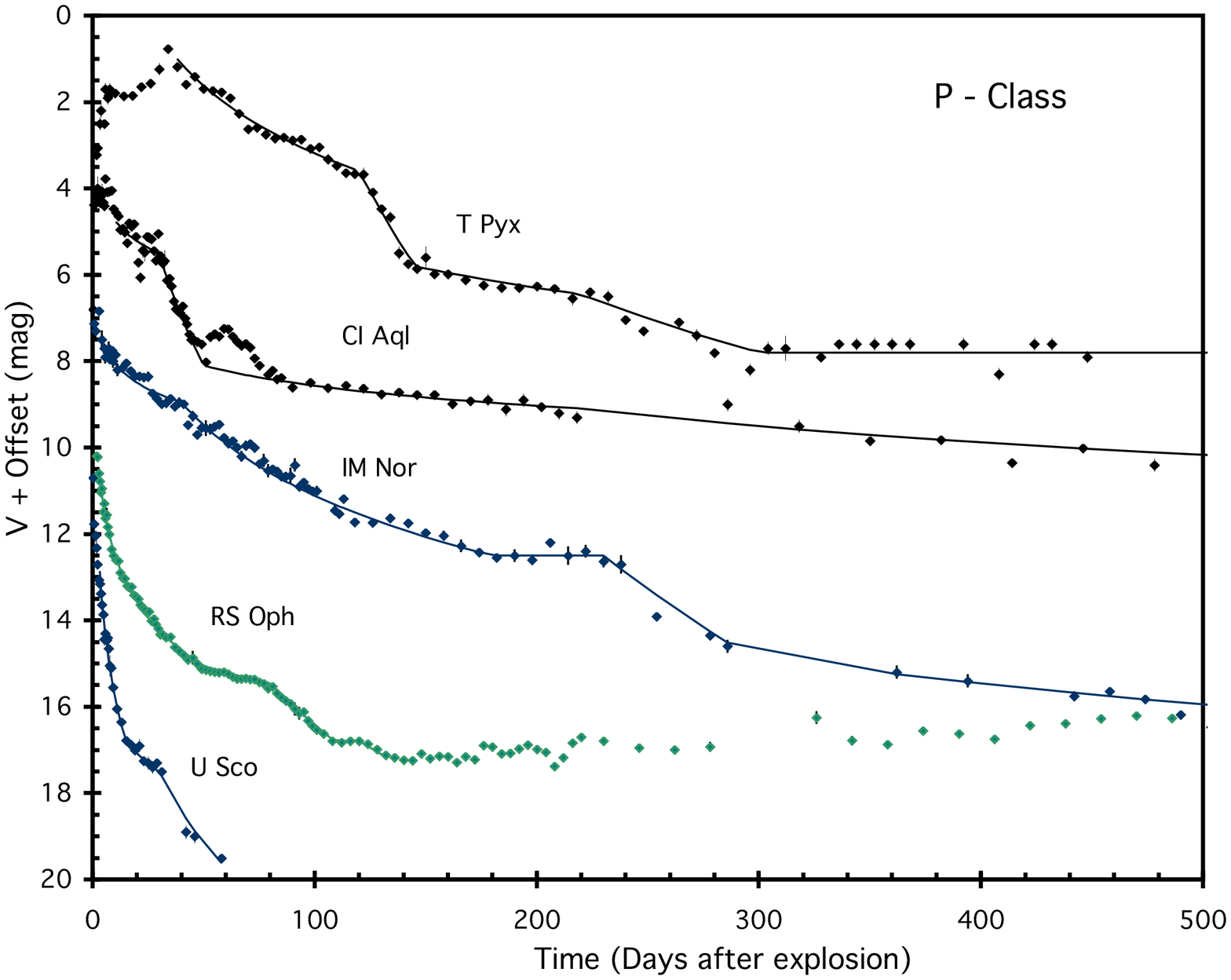}{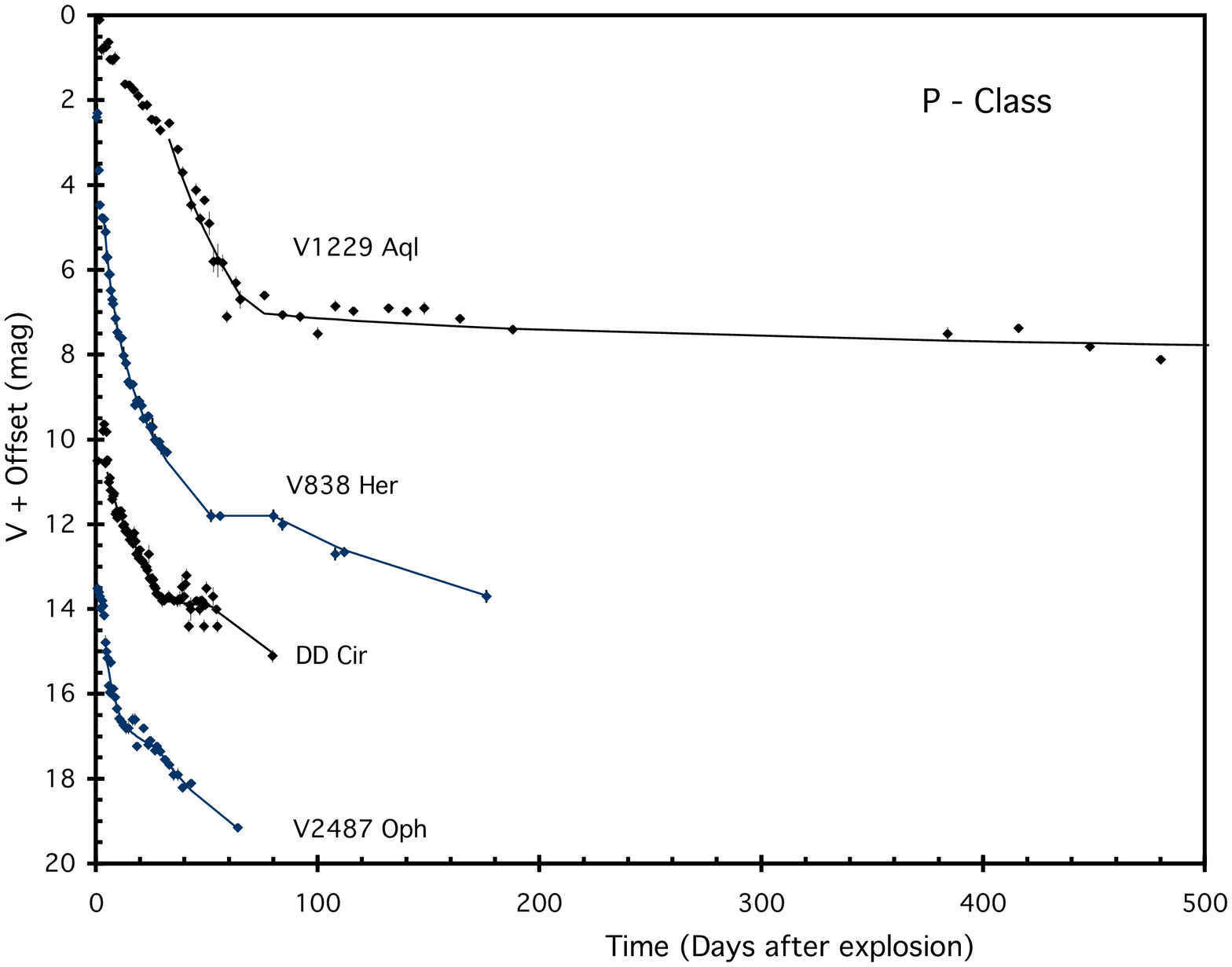}
\caption{
P class light curves.  These four panels show the 19 P class novae with plateaus.  As in all the light curves in this paper, the 1-sigma error bars on magnitudes are displayed as vertical lines, but these are smaller than the plotting symbol in almost all cases.}
\end{figure}

\clearpage
\begin{figure}
\epsscale{1.1}
\plottwo{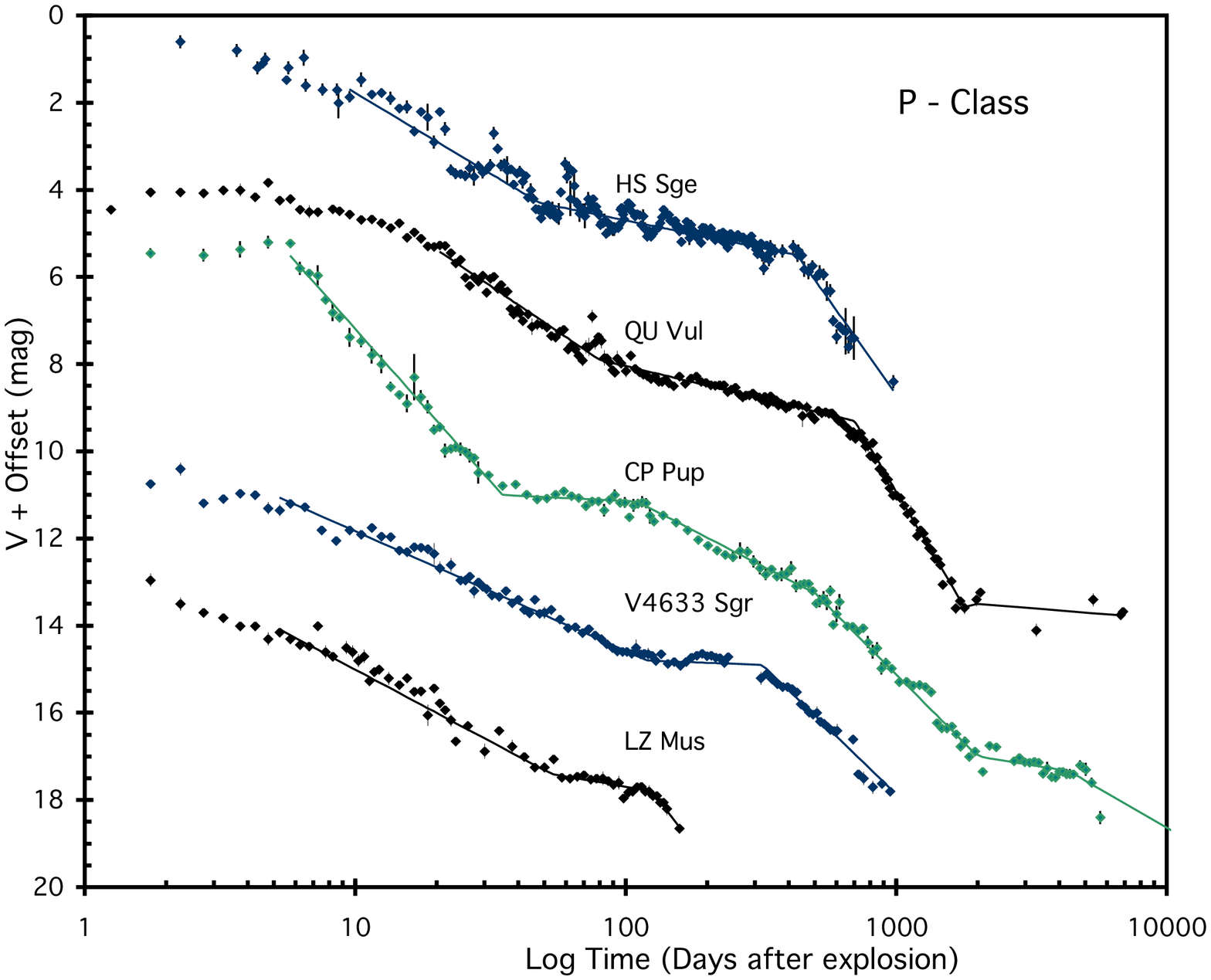}{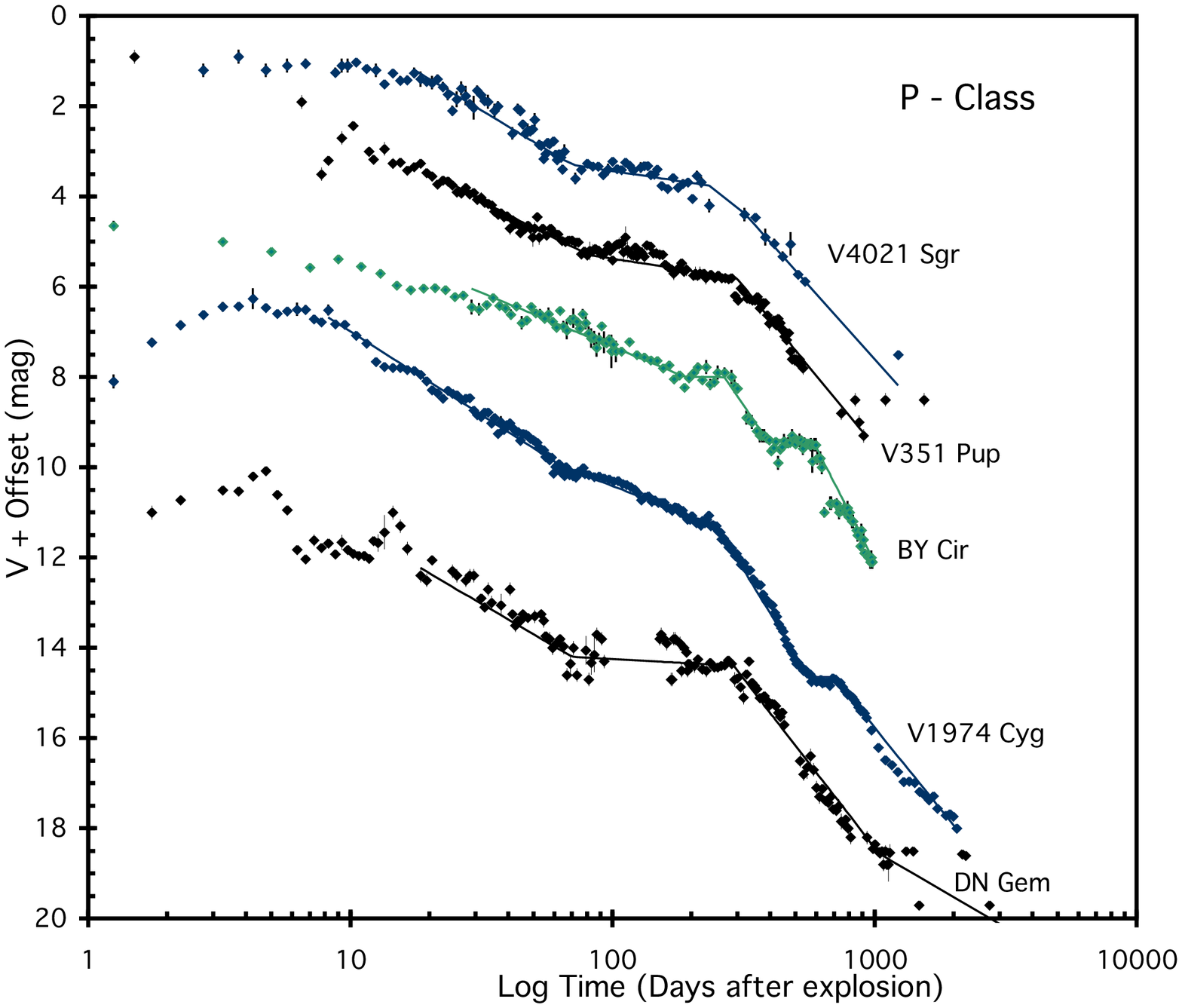}
\plottwo{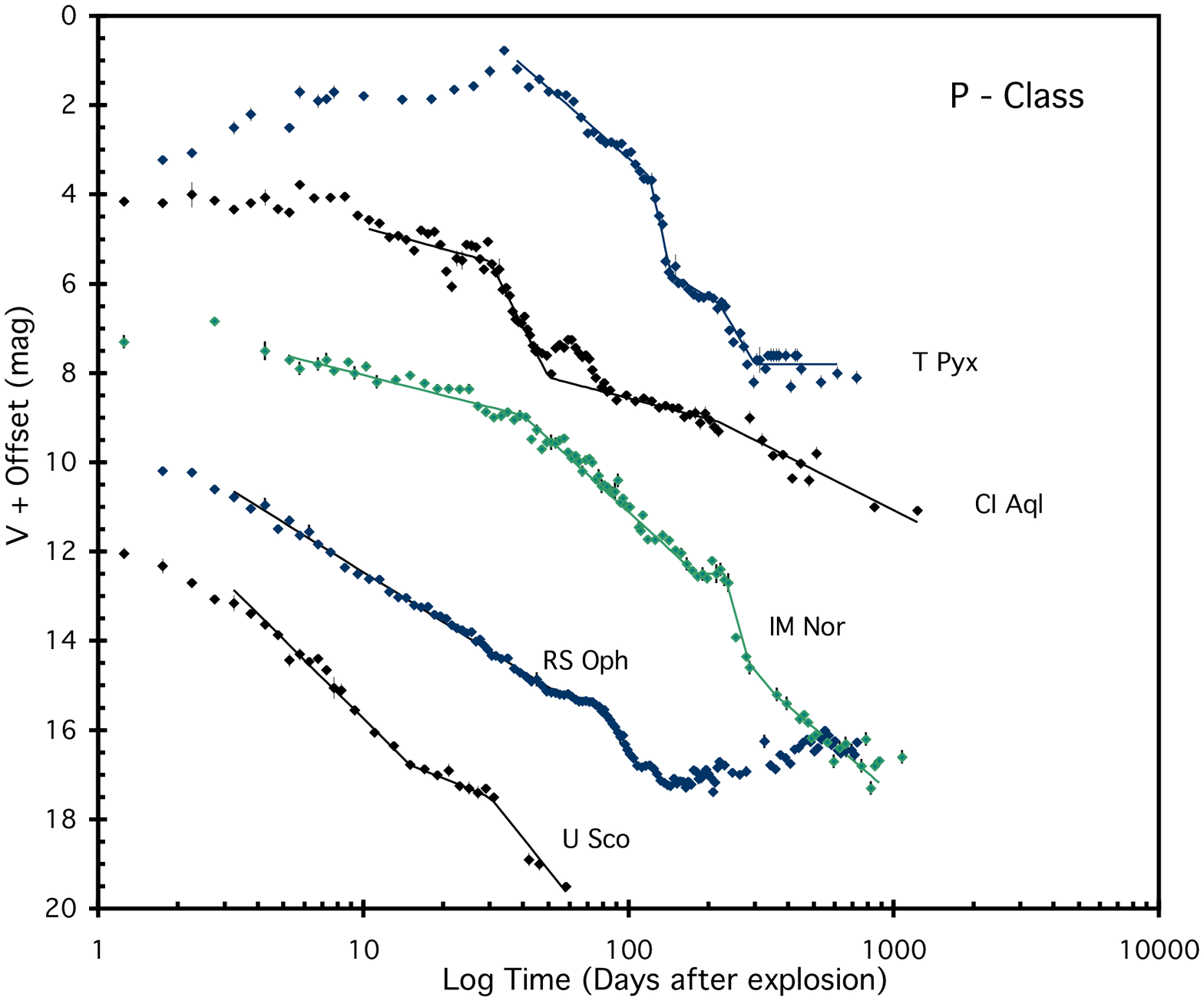}{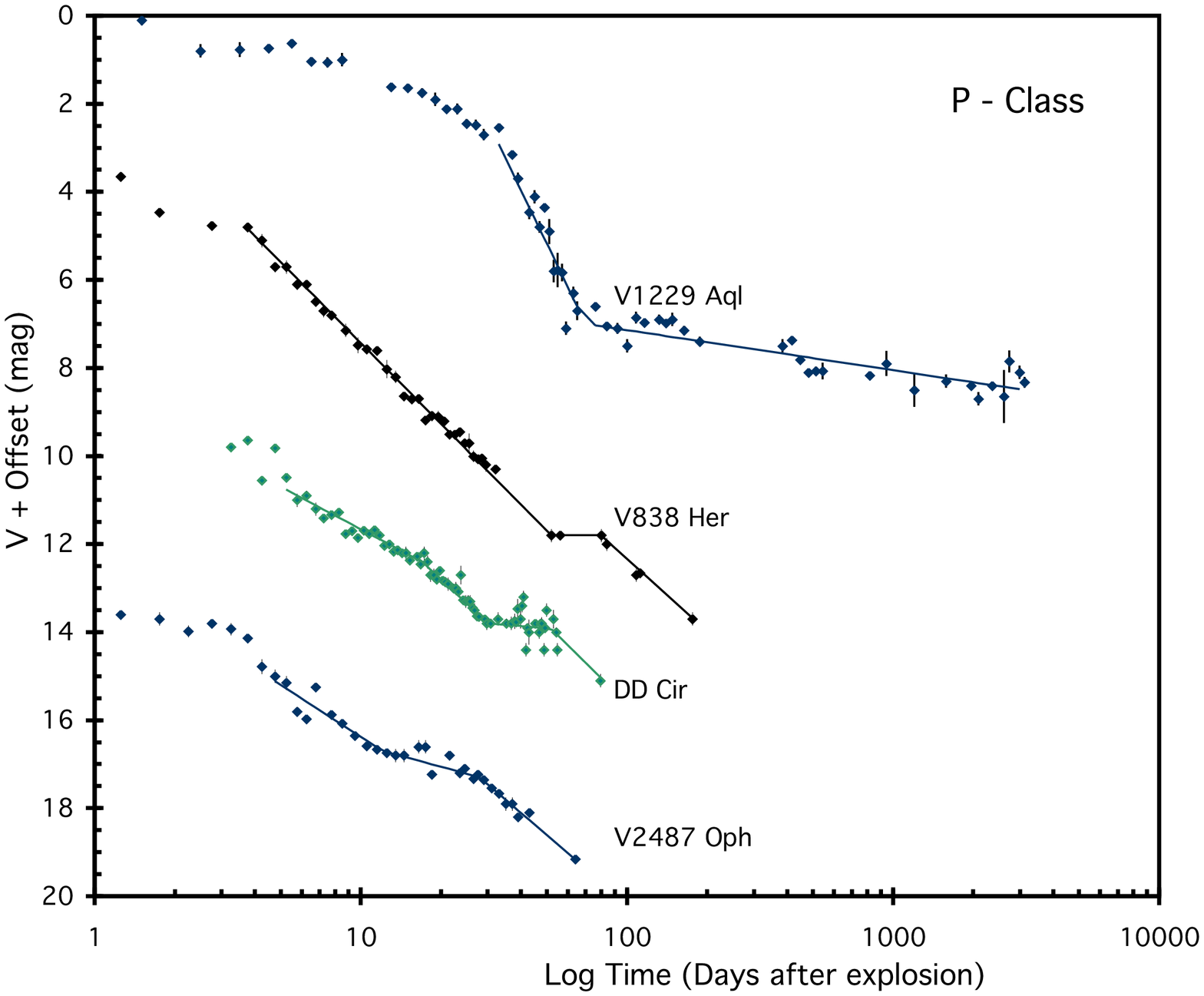}
\caption{
P class light curves on a logarithmic scale.  These four panels show the same data as in Figure 5, except that the log scale allows the late tails to be seen.  It appears that the plateaus are better recognized on a log plot than in a linear plot.}
\end{figure}

\clearpage
\begin{figure}
\epsscale{1.1}
\plottwo{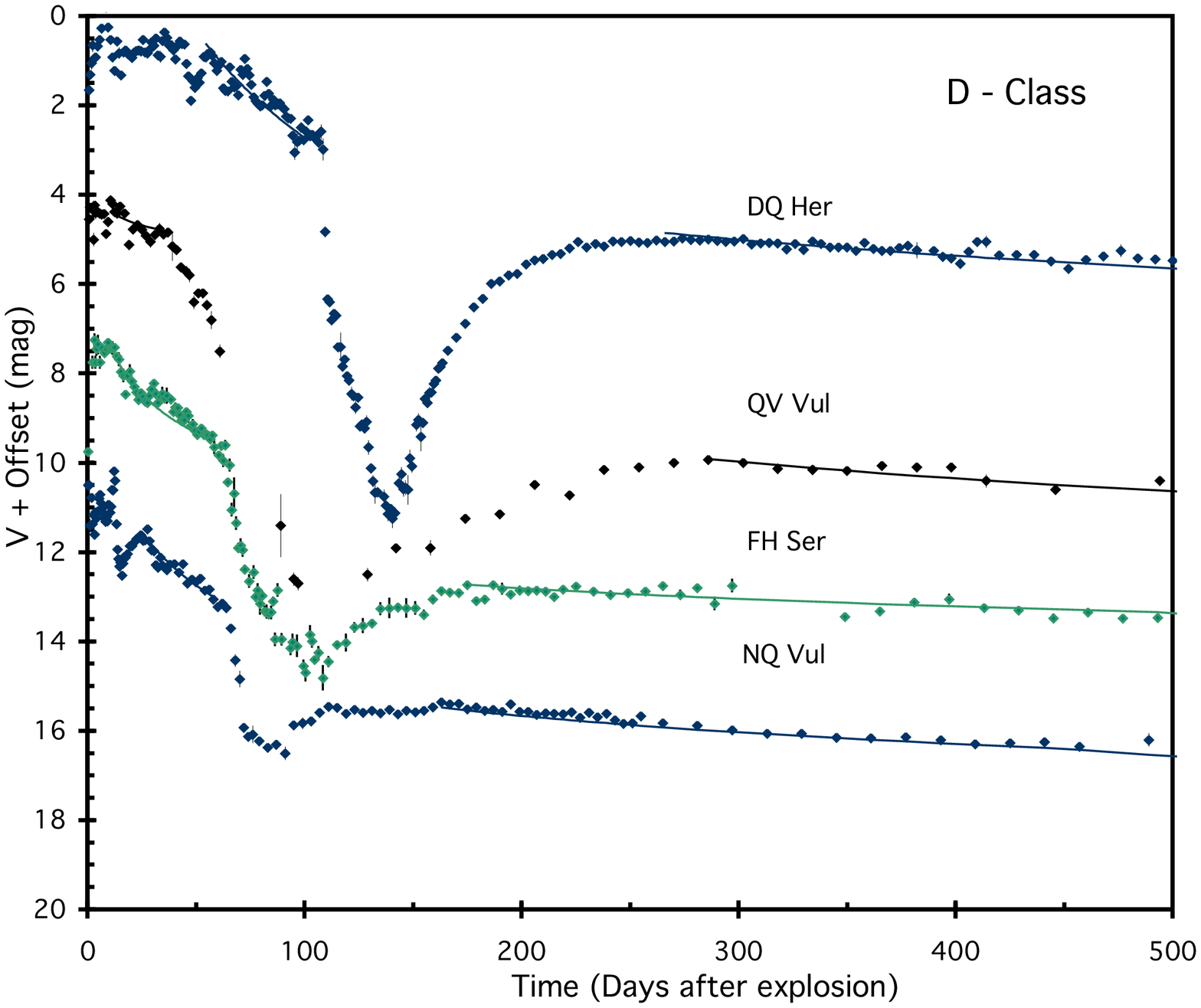}{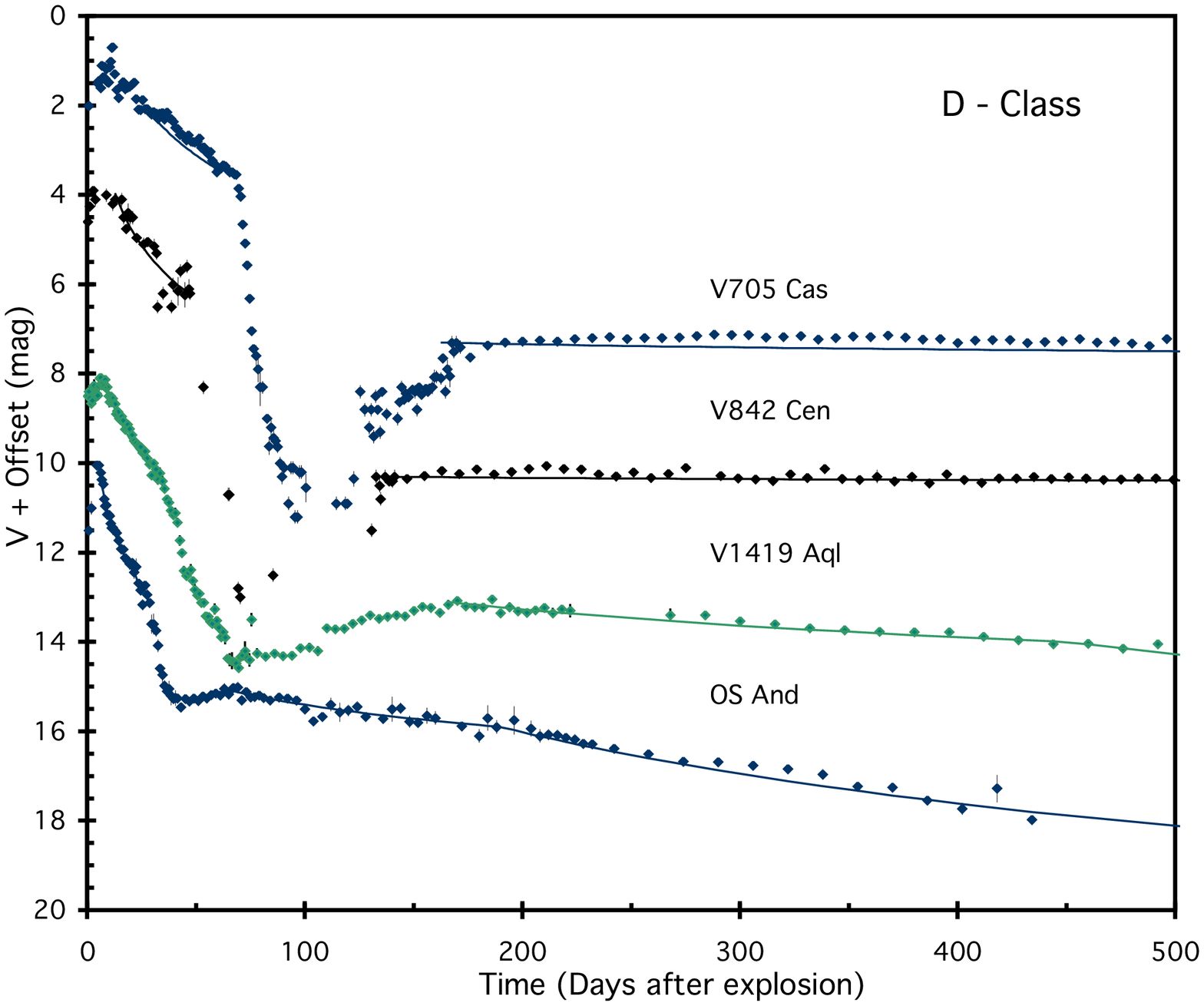}
\plottwo{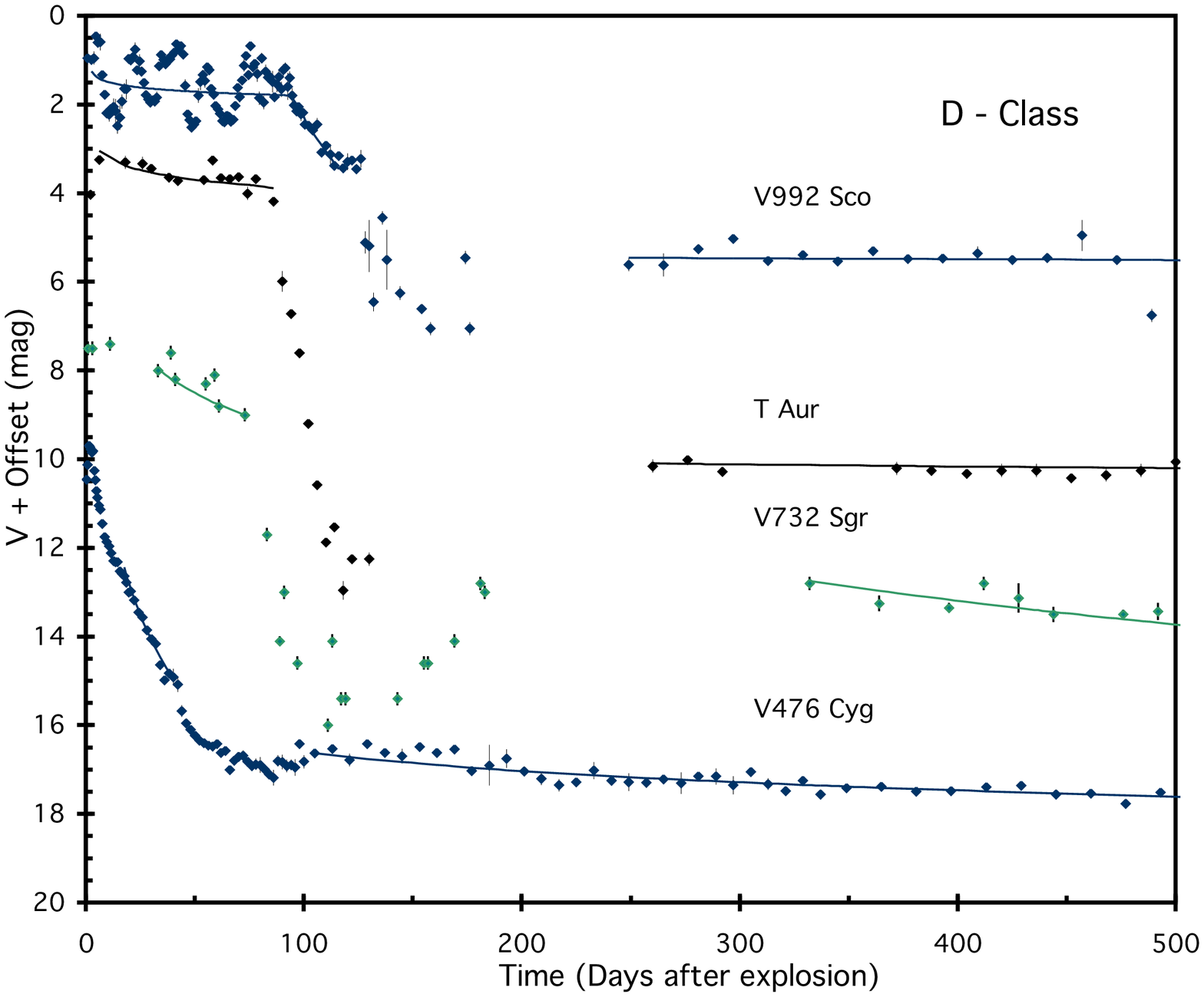}{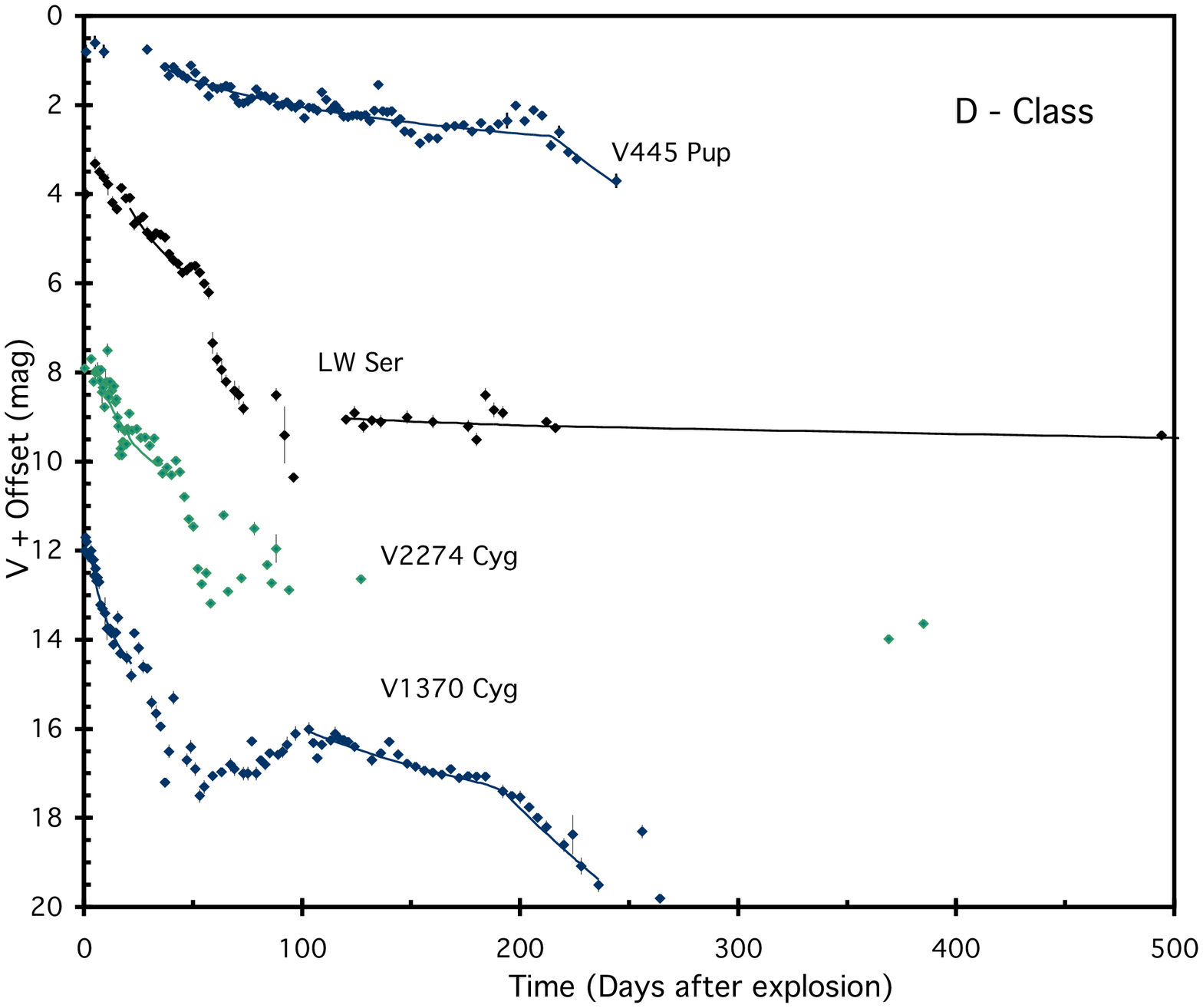}
\caption{
D class light curves.  These four panels show the 16 D class novae with dust dips.  The dust dips come in a wide variety of depths, with the deeper dips coming later in time.}
\end{figure}

\clearpage
\begin{figure}
\epsscale{1.1}
\plottwo{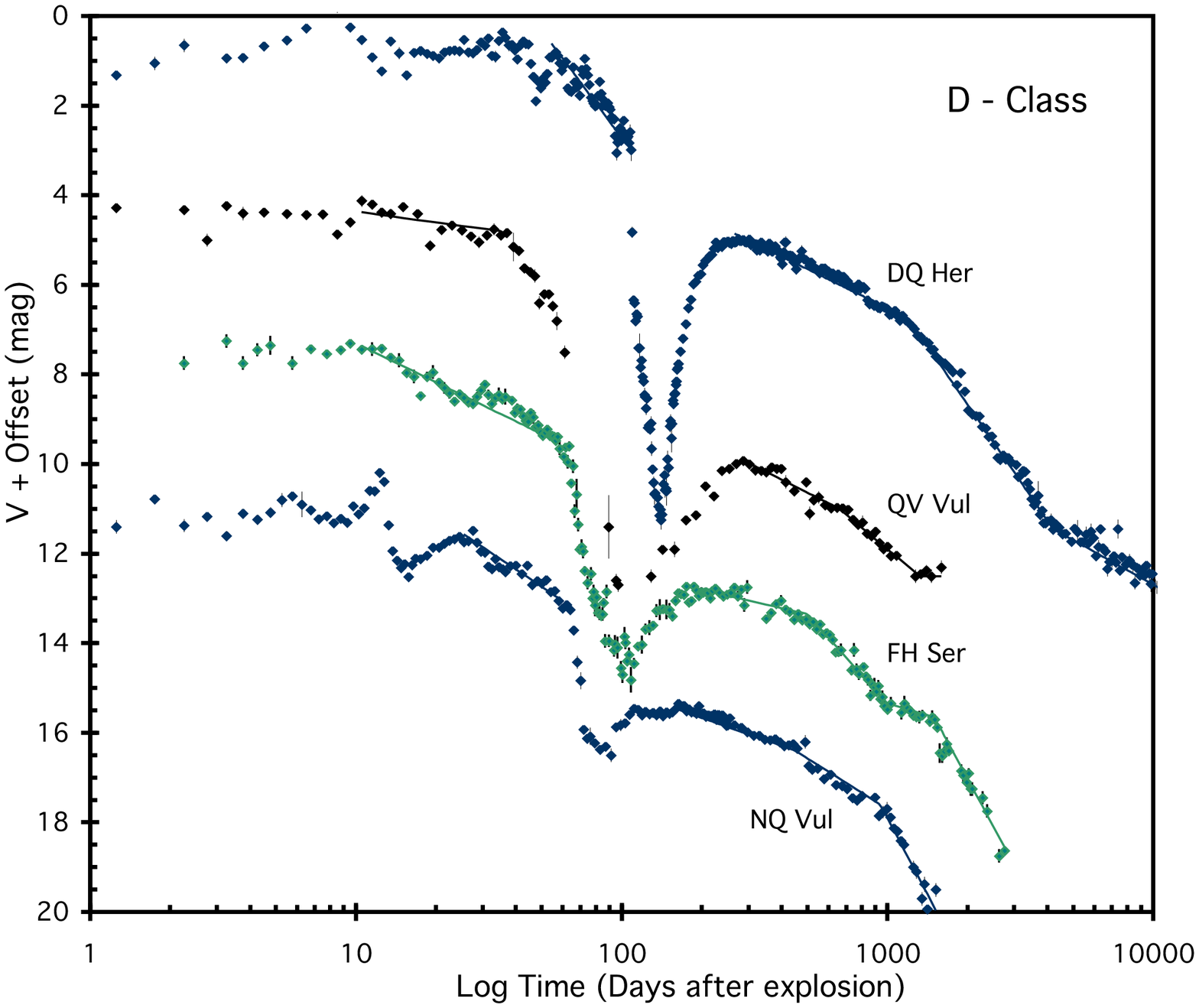}{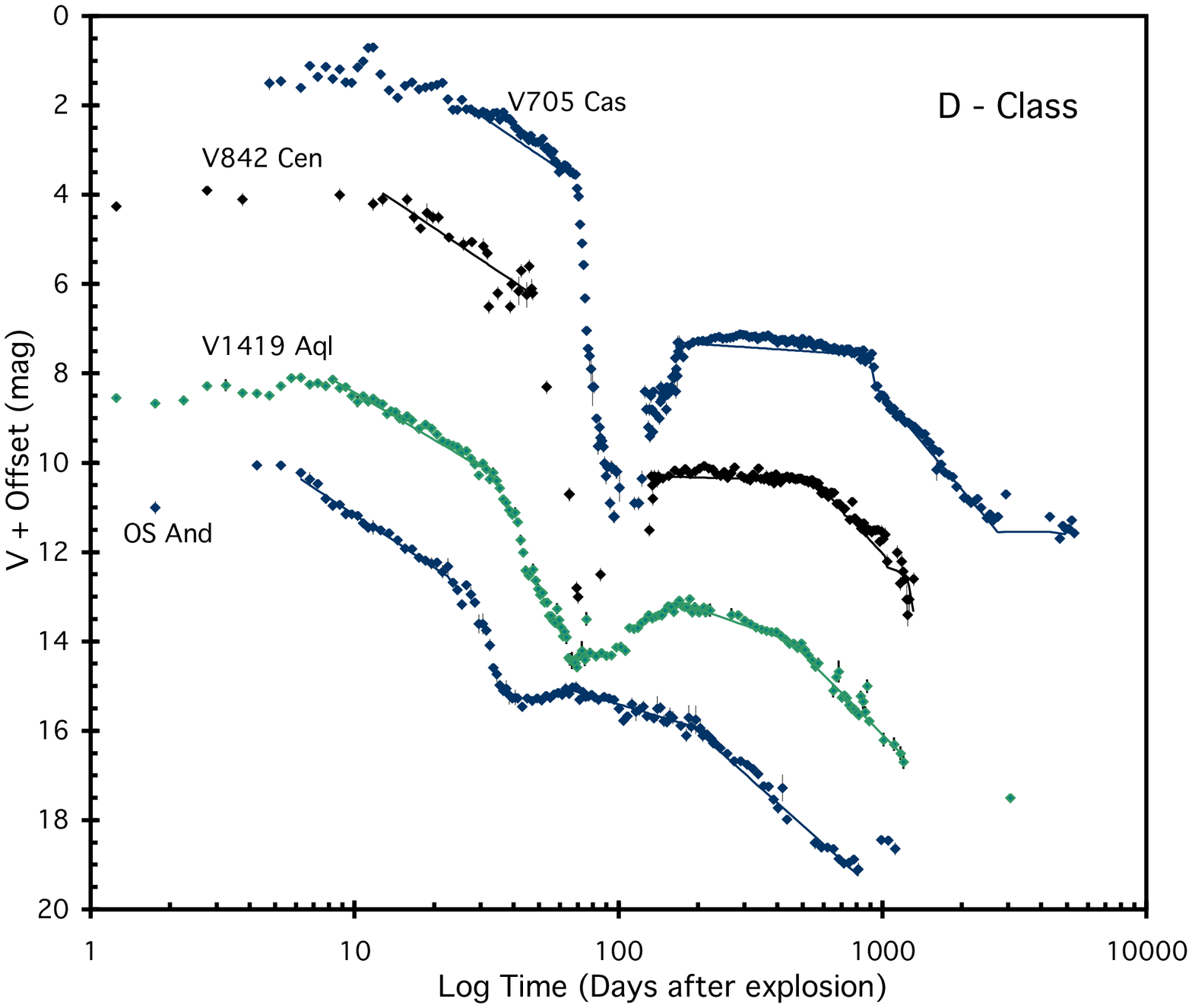}
\plottwo{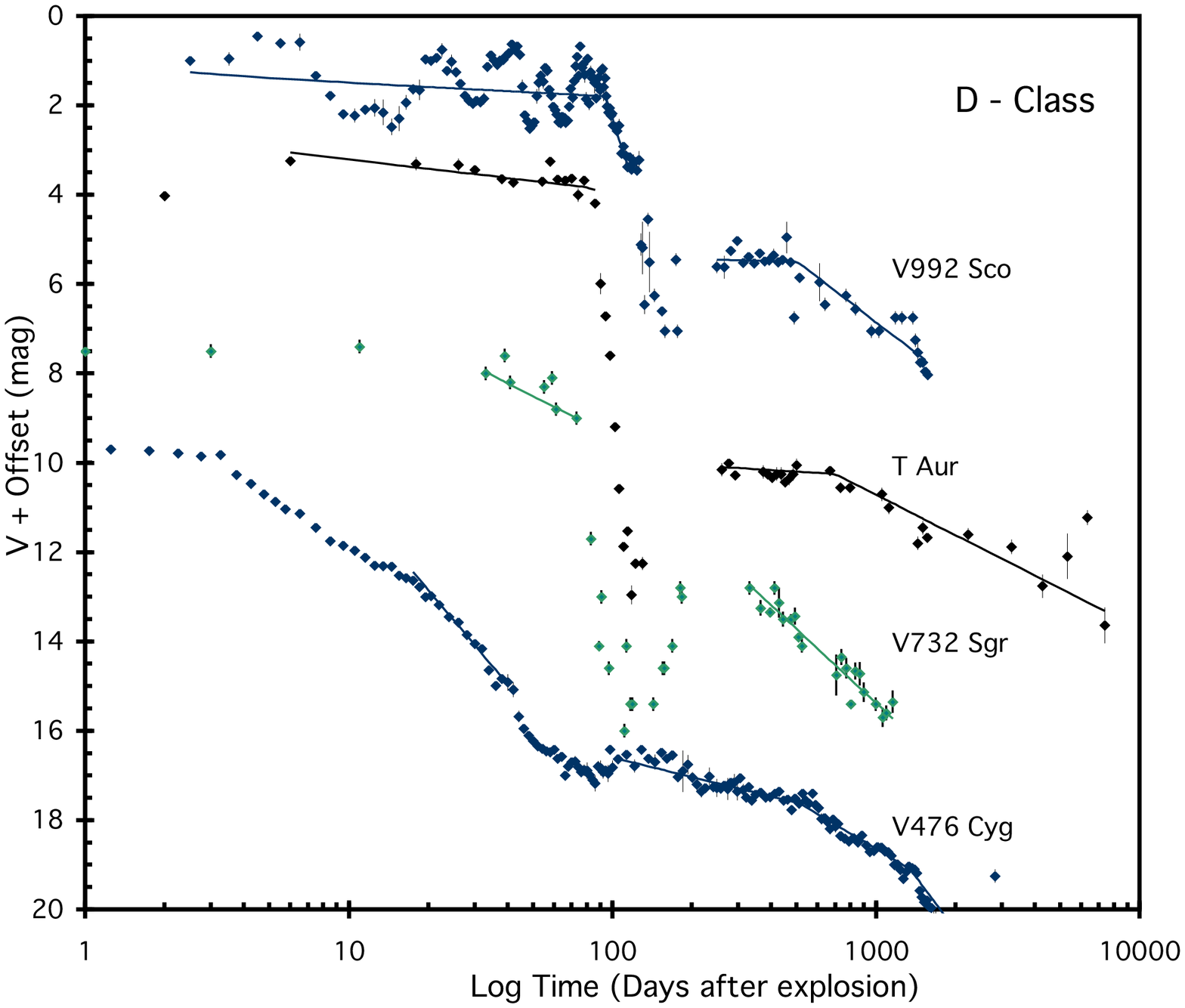}{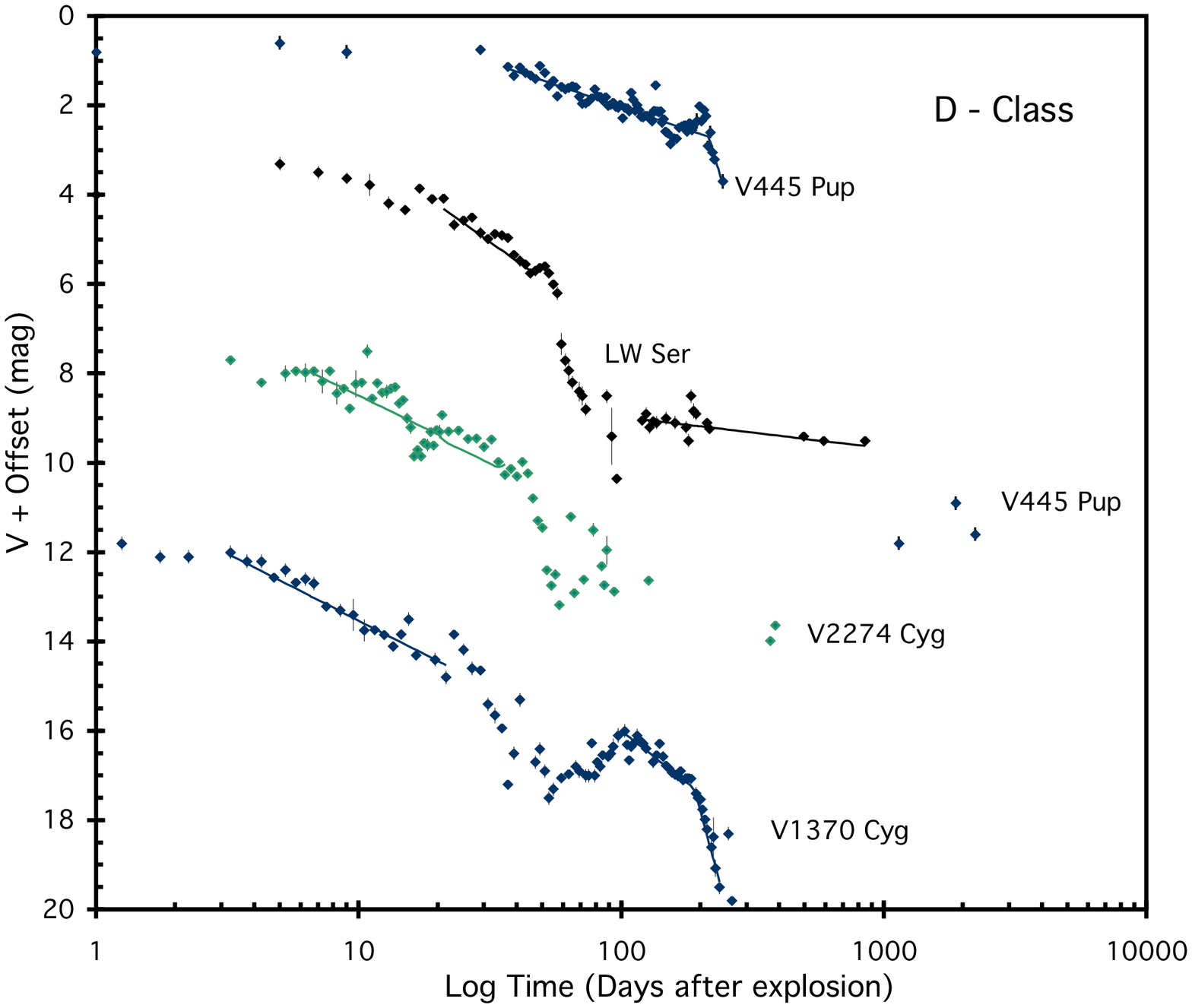}
\caption{
D class light curves on a logarithmic scale.  The initial slopes, lasting for many weeks are much flatter than predicted by the universal decline law, with the flatter slopes generally corresponding to the deeper dips.}
\end{figure}

\clearpage
\begin{figure}
\epsscale{0.7}
\plotone{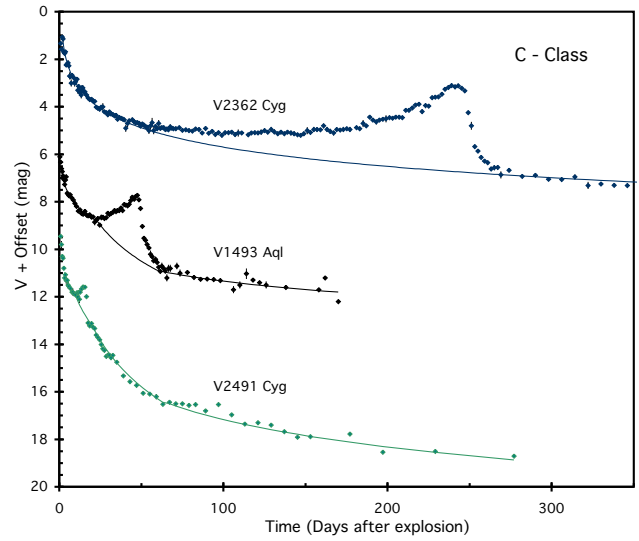}
\caption{
C class light curves.  Only three nova light curves with cusps are known.  On this linear plot, the cusp for V2491 Cyg is inconspicuous, and could easily be mistaken for a jitter or some observational fluctuation.  We suggest that the later the cusp appears, the stronger it is.}
\end{figure}

\clearpage
\begin{figure}
\epsscale{0.7}
\plotone{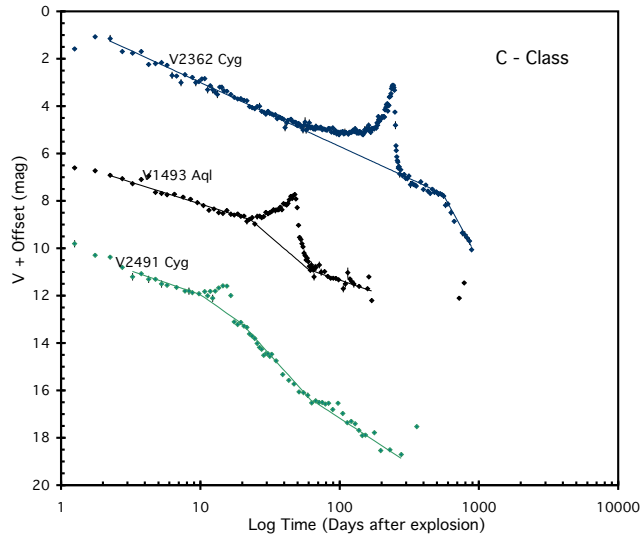}
\caption{
C class light curves on a logarithmic scale.  The V2491 Cyg cusp is readily apparent on this log-scale, whereas it is barely noticeable on a linear scale (see Figure 9).  The broken lines represent the power law segments from Table 9, as fitted to the light curve before and after the cusp.  For V2362 Cyg, the power law fit before and after the cusp is identical, which suggests that the flux from the cusp is something added to the normal universal decline law.  This is not exactly true for the other two cusp novae, but this could easily be due to the usual declines (see Figure 4) having breaks around the times of the cusps.}
\end{figure}

\clearpage
\begin{figure}
\epsscale{0.7}
\plotone{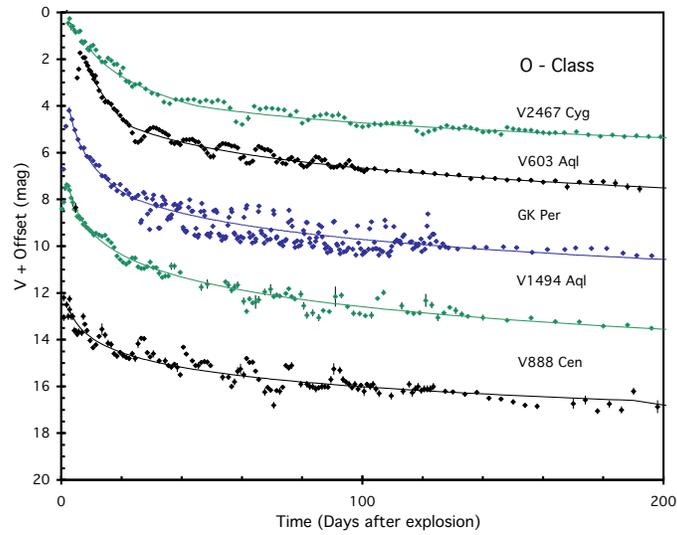}
\caption{
O class light curves.  Only five nova light curves with oscillations are known, for which only those for GK Per and V603 Aql are broadly known.  Oscillations are distinct from jitters (or flares) because (a) the peaks are quasi-periodic, (b) the peaks do not start before or around the nova maximum, and (c) the oscillations extend both above and below the power law segment fitted to the intervals before and after the oscillations.}
\end{figure}

\clearpage
\begin{figure}
\epsscale{0.7}
\plotone{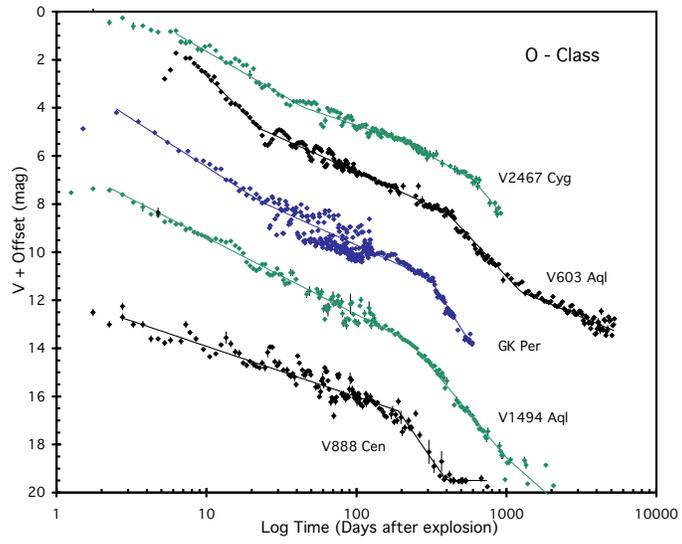}
\caption{
O class light curves on a logarithmic scale.  Overall, the light curves are well fit by the universal decline law.}
\end{figure}

\clearpage
\begin{figure}
\epsscale{0.7}
\plotone{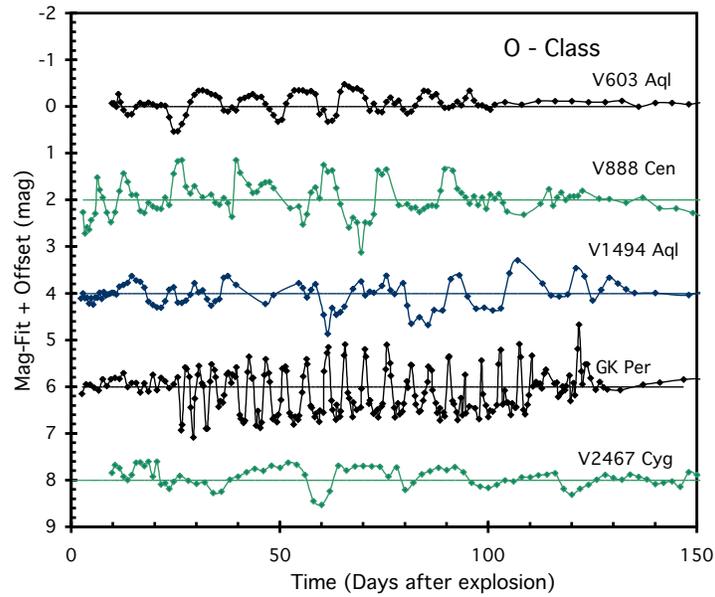}
\caption{
Oscillation close-ups.  This figure displays only the time interval out to 150 days after peak, so as to provide maximum resolution on the oscillations.  In addition, the fitted power law segments have been subtracted out so we are only seeing variations from the oscillations.  We see different oscillation shapes, including fairly sharp spikes while the nova light spends most of its time near the oscillation minimum (like GK Per) and fairly sinusoidal oscillations where a bit more of the time is spent near the oscillation maximum (like V603 Aql).}
\end{figure}

\clearpage
\begin{figure}
\epsscale{0.7}
\plotone{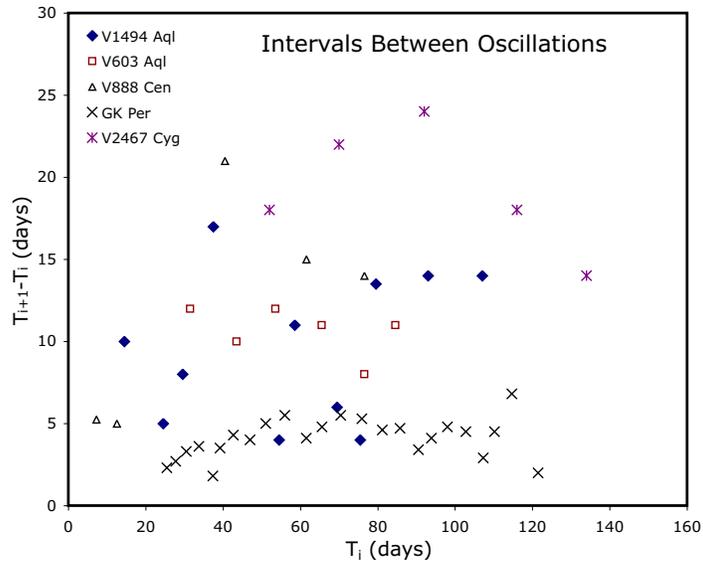}
\caption{
Oscillation spacing.  For the five O class novae, the oscillations are spaced approximately uniformly, with the time between successive peaks being approximately constant.  (V1494 Aql does have substantial scatter about the average, but no systematic trend.)  This is to say that the oscillations are quasi-periodic in {\it linear} time.}
\end{figure}

\clearpage
\begin{figure}
\epsscale{0.7}
\plotone{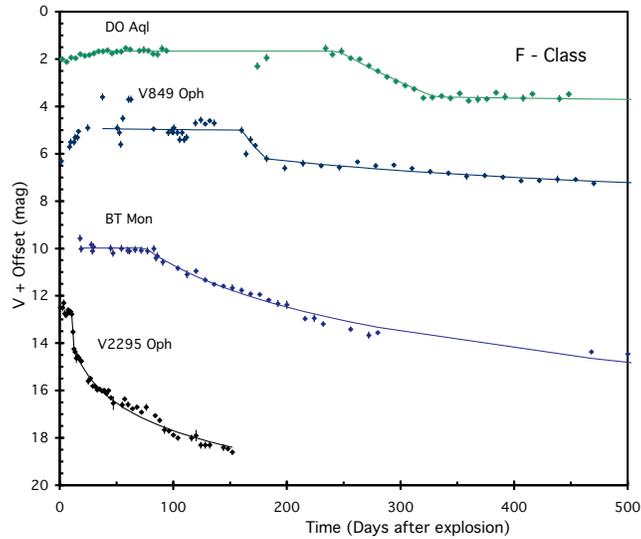}
\caption{
F class light curves.  Four known nova light curves have remarkably flat tops, lasting over 244,  75, 65, and 11 days for DO Aql, BT Mon, V849 Oph, and V2295 Oph respectively.  This is behavior is completely distinct from all other classical novae.}
\end{figure}

\clearpage
\begin{figure}
\epsscale{0.7}
\plotone{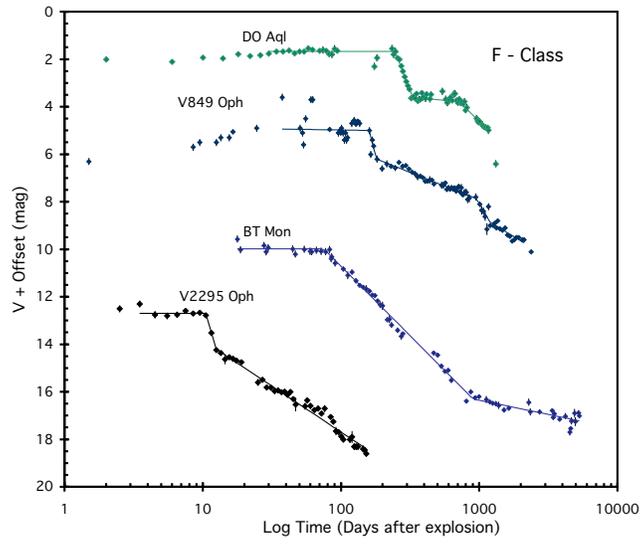}
\caption{
F class light curves on a logarithmic scale.  The flat top is always followed by a steep decline.}
\end{figure}

\clearpage
\begin{figure}
\epsscale{1.1}
\plottwo{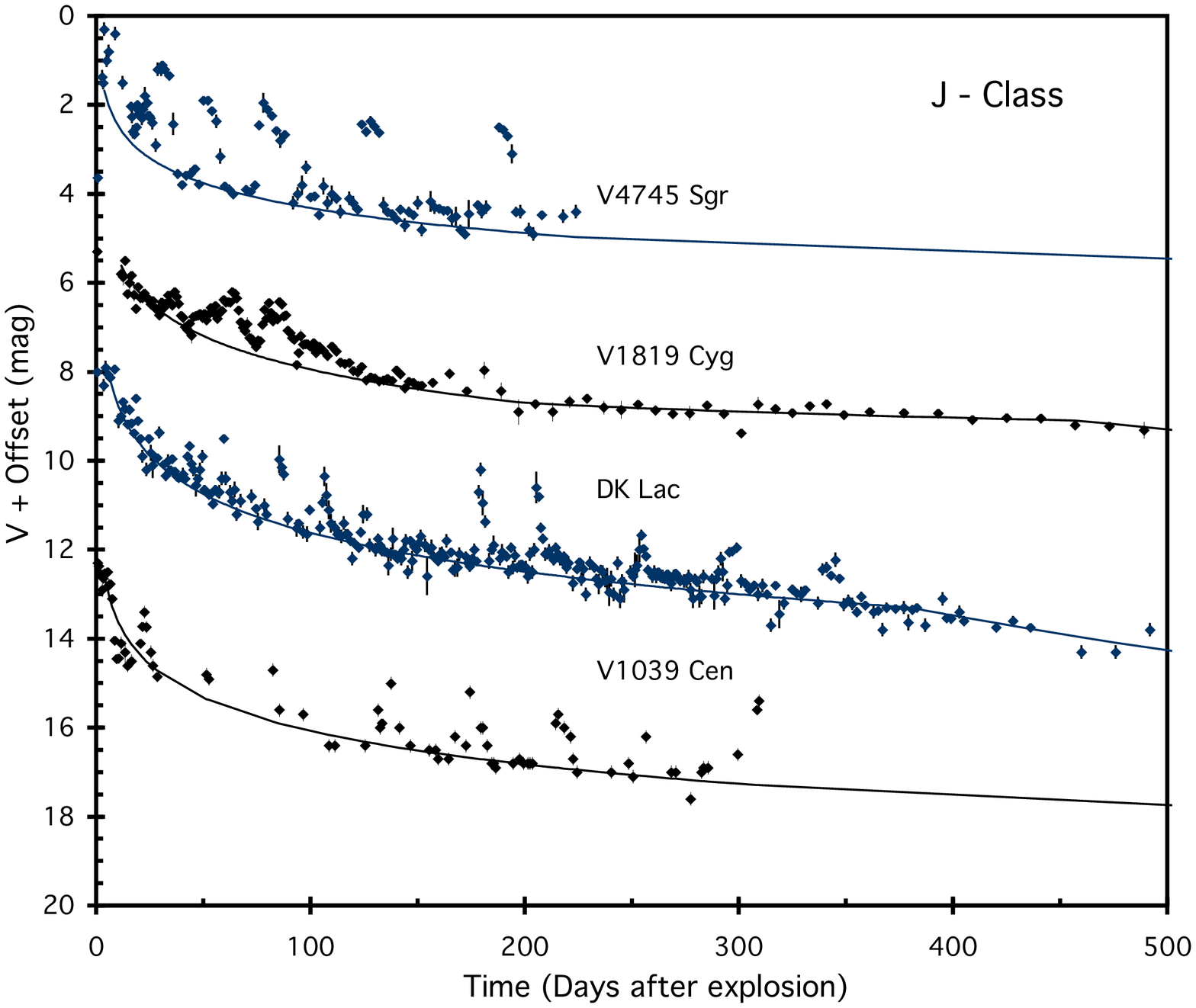}{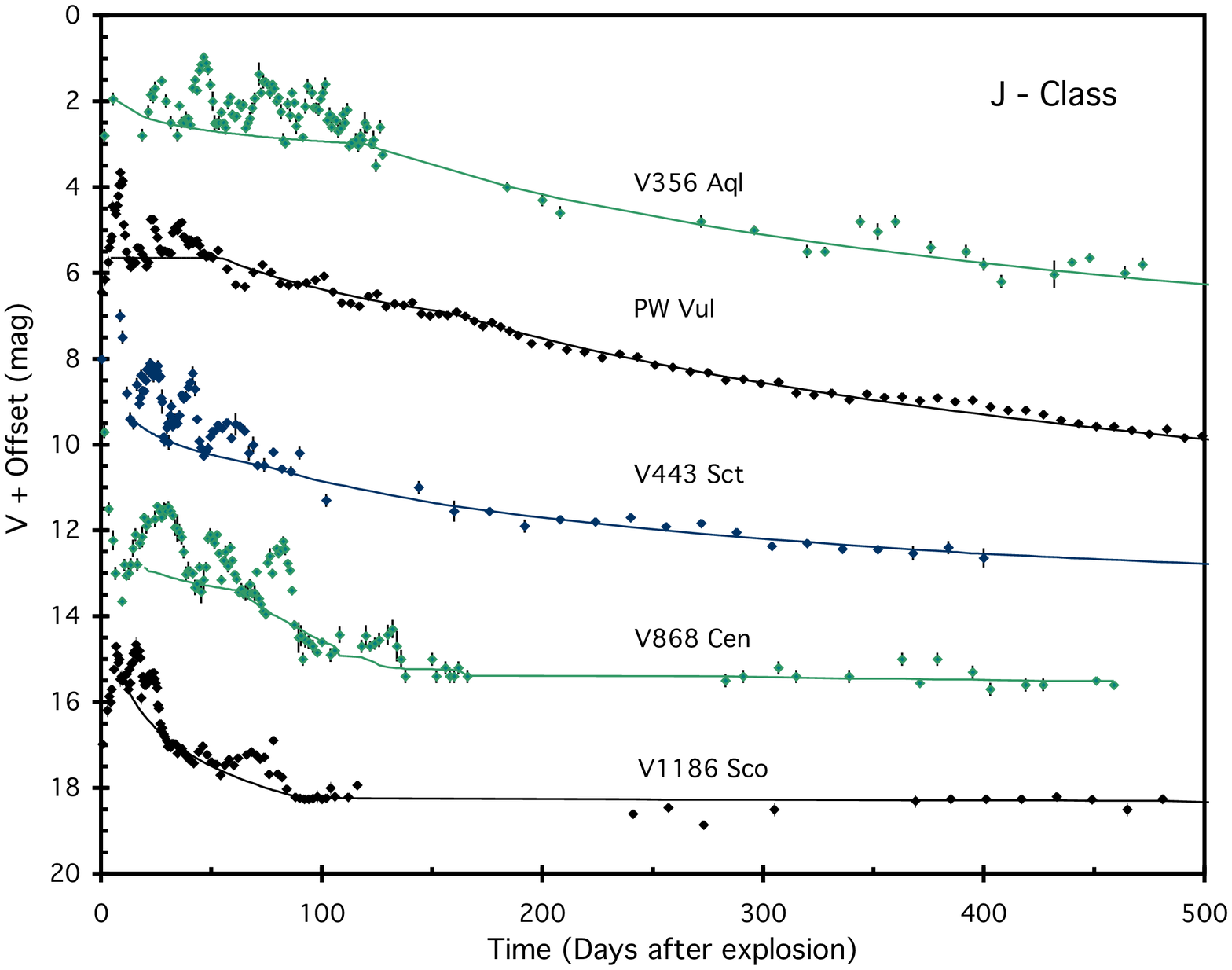}
\epsscale{0.5}
\plotone{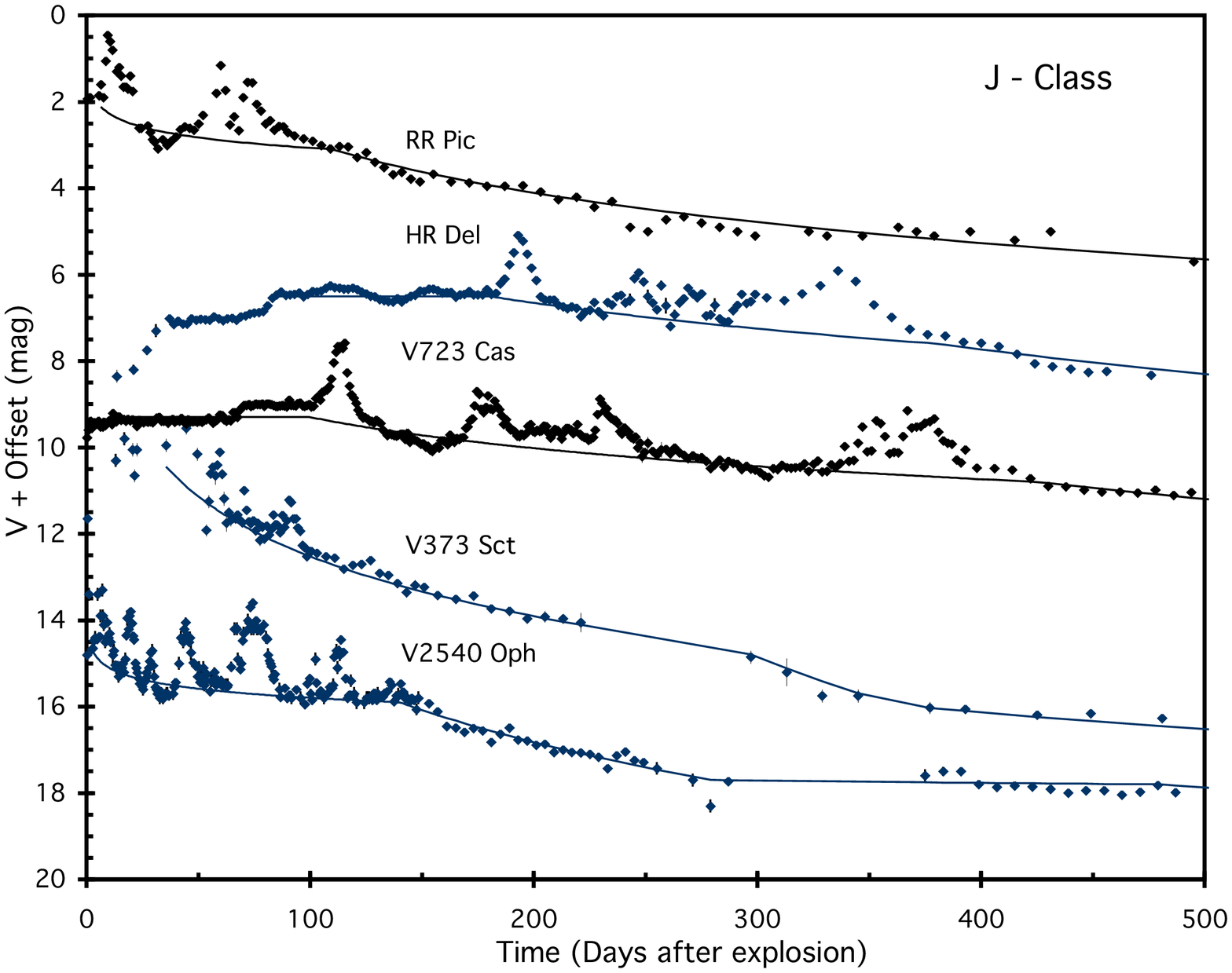}
\caption{
J class light curves.  These three panels show the 14 J class light curves on a linear scale out to 500 days so that the jitters can readily be seen.  The presence of such large jittering raises the question of whether the maximum of the light curve should be taken as the time when the light is brightest or the time when the fitted power law segment meets with the rise.  Similarly, it is an open question as to whether the $t_3$ time should be evaluated with the observed light curve (either as the first or the last time that the light curve passes through 3 mags below peak) or the fitted smooth curve.  In this paper, we take the simple observed peak magnitude for $V_{peak}$ and the last time for which the observed light curve is within 3 mag of that peak as the time for $t_3$.}
\end{figure}

\clearpage
\begin{figure}
\epsscale{1.1}
\plottwo{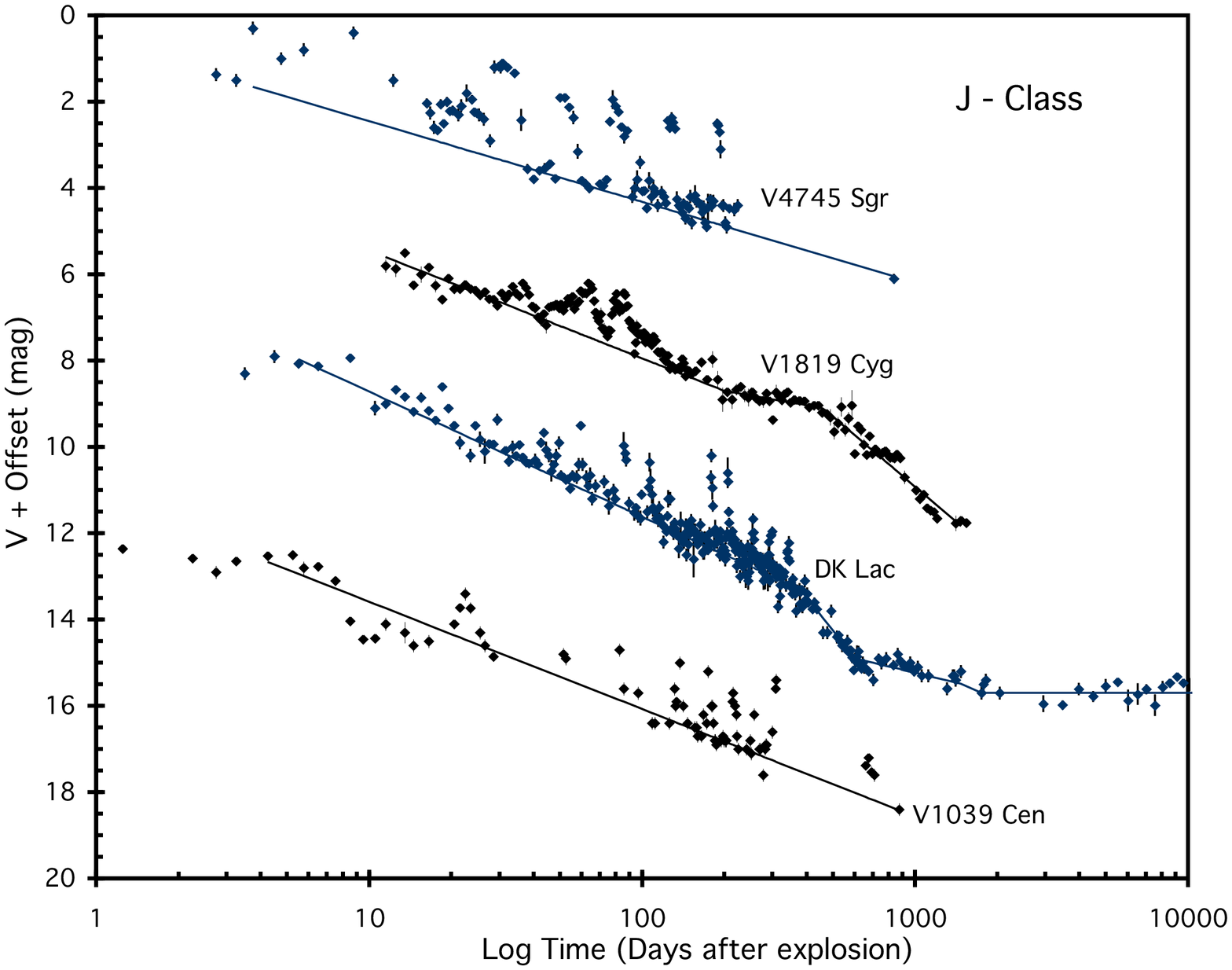}{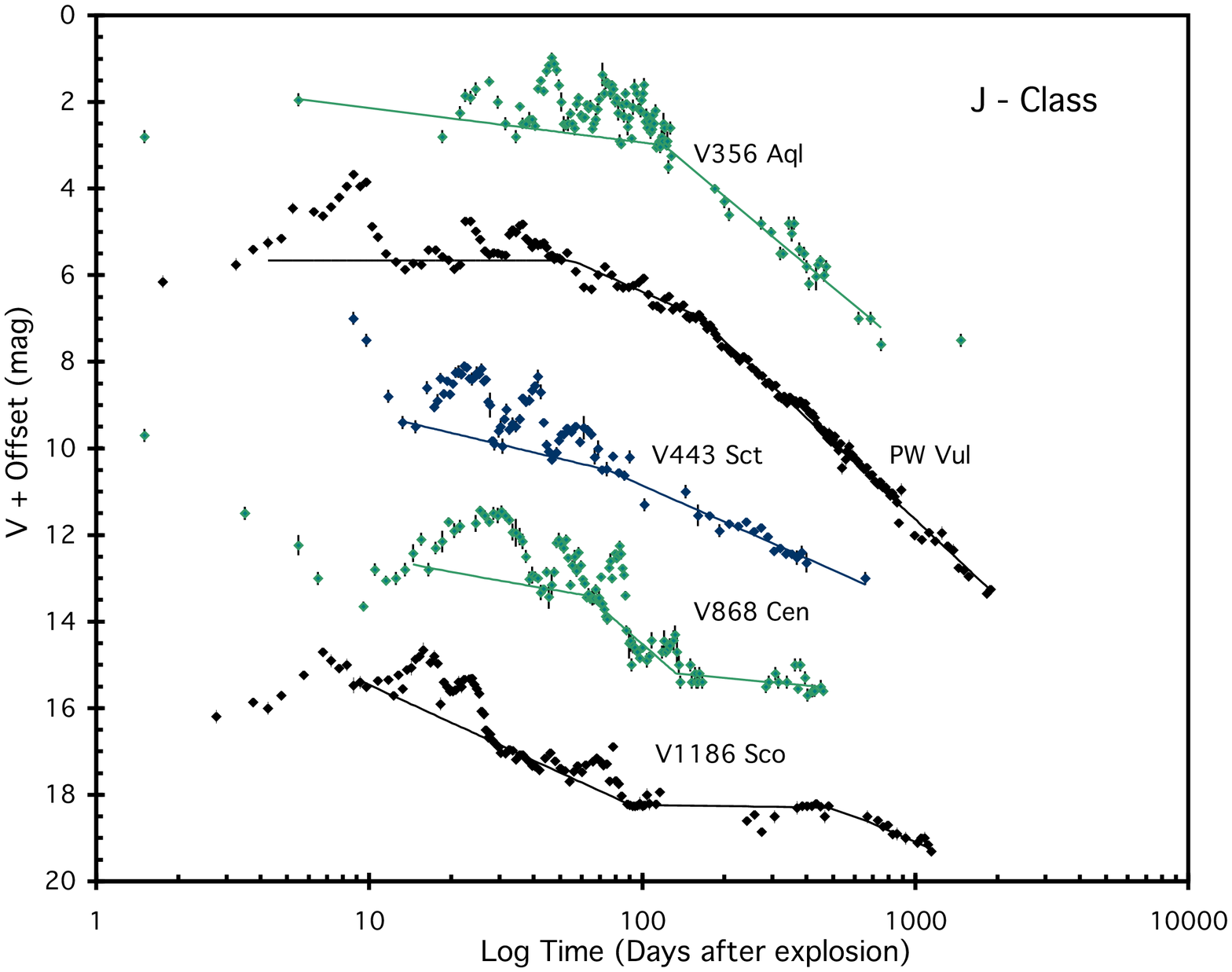}
\epsscale{0.5}
\plotone{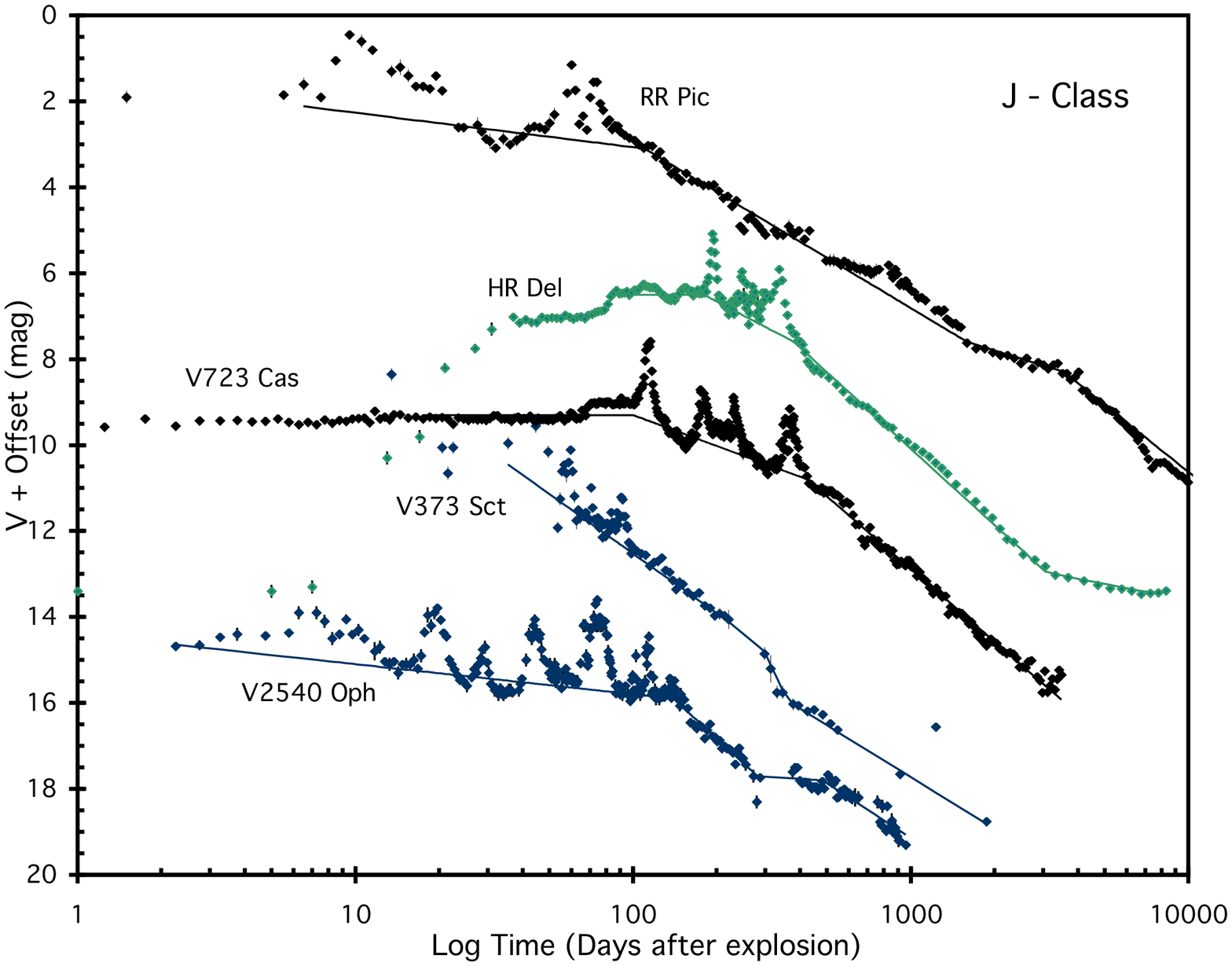}
\caption{
J class light curves on a logarithmic scale.  The jitters always appear to be extra light added onto the universal decline law.}
\end{figure}

\clearpage
\begin{figure}
\epsscale{0.7}
\plotone{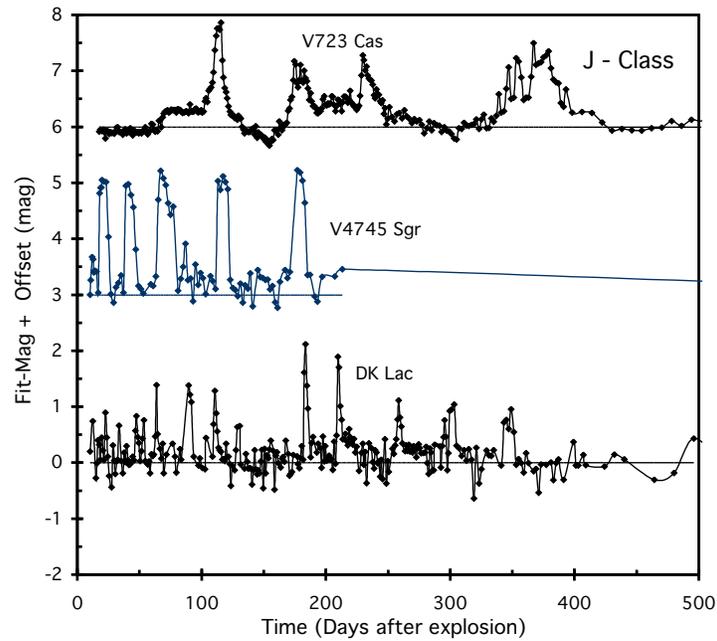}
\caption{
Jitter close-ups.  Here are three good examples of jitters.  The jitters appear to be random in time, generally sharp peaks, the majority of the time spent near the minimum, and the minimum lies close to the power law segments.  We wonder about the apparent regular increase in the interval between peaks for V4745 Sgr.  However, the small number of peaks, the wide variety of possible patterns that we might find interesting, and the lack of such patterns in all other J class light curves, all indicate that this pattern in the V4745 Sgr jitters is likely not significant.}
\end{figure}

\end{document}